\documentclass[useAMS,usenatbib]{mn2e}
\usepackage{graphicx}
\usepackage{subfigure}
\bibpunct{(}{)}{,}{a}{}{;}

\newcommand{\be}{\begin{equation}}
\newcommand{\ee}{\end{equation}}

 \title[Atmospheric refraction in a single mode instrument]{Study of the atmospheric refraction in a single mode instrument - Application to AMBER/VLTI}
\author[S.~Robbe-Dubois et al.]{S.~Robbe-Dubois$^{1}$\thanks{E-mail: Sylvie.Robbe-Dubois@unice.fr}, S.~Lagarde$^{1}$, Y.~Bresson$^{1}$, R.G.~Petrov$^{1}$, M.~Carbillet$^{1}$, 
	\newauthor E.~LeCoarer$^{2}$, F.~Rantakyr\"o$^{3,4}$, I.~Tallon-Bosc$^{5}$, M.~Vannier$^{1}$, P.~Antonelli$^{1}$,
	\newauthor G.~Martinot-Lagarde$^{6,7}$, A.~Roussel$^{1}$ and D.~Tasso$^{1}$\\
$^{1}$Laboratoire Fizeau, Universit\'{e} de Nice Sophia-Antipolis, Observatoire de la C\^{o}te d'Azur, CNRS, UMR 6525, Parc Valrose,\\ 06108 Nice Cedex 2, France\\
$^{2}$Laboratoire d'Astrophysique Observatoire de Grenoble, Universit\'{e} J. Fourier, CNRS, UMR 5571, 414, rue de la piscine,\\ 38041 Grenoble cedex 9, France\\
$^{3}$Gemini Observatory Southern Operations Center, c/o AURA, Casilla 603, La Serena, Chile \\
$^{4}$European Southern Observatory, Casilla 19001, Santiago 19, Chile\\
$^{5}$Universit\'e de Lyon, 69003 Lyon, France ; Universit\'e Lyon 1, Observatoire de Lyon, 9 Av. Charles Andr\'e, 69230 Saint Genis\\ Laval, France ; CNRS, UMR 5574, Centre de Recherche Astrophysique de Lyon ; Ecole Normale\\ Sup\'erieure de Lyon, 69007, Lyon, France  \\
	$^{6}$Laboratoire G\'eosciences Azur, Universit\'{e} de Nice Sophia-Antipolis, Observatoire de la C\^{o}te d'Azur, CNRS, UMR 6526,\\ Av. Nicolas Copernic, 06130 Grasse, France \\
	$^{7}$Division Technique INSU/CNRS UPS 855, 1 place Aristide Briand, 92195 Meudon cedex, France }
\begin{document}
\date{Accepted  . Received ; in original form }

\pagerange{\pageref{firstpage}--\pageref{lastpage}} \pubyear{}
\maketitle
\label{firstpage}
\begin{abstract}
This paper presents a study of the atmospheric refraction and its effect on the light coupling efficiency in an instrument using single-mode optical fibers.
We show the analytical approach which allowed us to assess the need to correct the refraction in J- and H-bands while observing with an 8-m Unit Telescope. We then developed numerical simulations to go further in calculations. 
The hypotheses on the instrumental characteristics are those of AMBER (Astronomical Multi BEam combineR), the near infrared focal beam combiner of the Very Large Telescope Interferometric mode (VLTI), but most of the conclusions can be generalized to other single-mode instruments. We used the software package {\sc caos} (Code for Adaptive Optics Systems) to take into account the atmospheric turbulence effect after correction by the ESO system MACAO (Multi-Application Curvature Adaptive Optics). The opto-mechanical study and design of the system correcting the atmospheric refraction on AMBER is then detailed.
We showed that the atmospheric refraction becomes predominant over the atmospheric turbulence for some zenith angles $z$ and spectral conditions: for $z$ larger than 30$^\circ$ in J-band for example. The study of the optical system showed that it allows to achieve the required instrumental performance in terms of throughput in J- and H-bands. First observations in J-band of a bright star, $\alpha$~Cir star, at more than 30$^\circ$ from zenith clearly showed the gain to control the atmospheric refraction in a single mode instrument, and validated the operating law.
\end{abstract}
  
\begin{keywords}Instrumentation: high angular resolution - Instrumentation: interferometers - Atmospheric effects - Methods: numerical - Methods: laboratory
\end{keywords}

\section{Introduction}\label{1}
Atmospheric refraction is the deviation of light from a straight line as it passes through the atmosphere due to the variation of the air refractive index $n(\lambda)$ with altitude. The wavelength dependence of $n(\lambda)$ implies wavelength-dependent refraction angles of the light. Therefore, the stellar images appear spectrally dispersed at the focal plane of an instrument.

Single mode instruments spatially filter the incoming wavefront. Using single mode optical fibers allows to reduce all wavefront perturbations to photometric and global OPD fluctuations. The fibers are located at the focal point of the instrument, selecting the central part of the stellar image. The performance of these instruments in terms of magnitude is mostly driven by the coupling efficiency of the fibers and by their capabilities over the largest possible bandwidth. In the case of broadband high-resolution observations, the fiber position respective to the dispersed image can then be optimized only for one wavelength. Therefore, the refraction affects the coupling factor as a function of wavelength. Atmospheric refraction controllers are then commonly used or studied in stellar interferometers, Adaptive Optics (AO) systems or coronagraph benches. Risley prisms have been used in a number of optical/IR interferometers to improve performance \citep{cola, teo, bocca, bedd}. \citet{breck} give construction details for a dispersion corrector suitable for use in the visible. 

To reach the quite severe required performance of AMBER in terms of coupling efficiency and stability over a spectral range from 1.1 to 2.4 $\mu$m \citep{petrov, robbe}, the system presented here is made of 2 sets of 3 prisms rotating with respect to each other and inserted prior to the J- and H-spatial filters in each interferometric arm. 
This unusual concept corrects the atmospheric refraction in J- (1.1 $\mu$m to 1.4 $\mu$m) and H- (1.475~$\mu$m to 1.825~$\mu$m) bands.

In Sect.\,\ref{2} we present a numerical study of the atmospheric refraction and its effect on the fiber coupling efficiency. We consider here the instrumental characteristics of AMBER, but the conclusion can be generalized for any other instrument using optical fibers. The fluctuations of the atmospheric turbulence and the subsequent AO correction performed by the MACAO (Multi-Application Curvature Adaptive Optics) system $-$ the ESO 60-actuators curvature sensing system \citep{arseno} $-$ are used to compute the coupling performance in Sect.\,\ref{3}. We show that the atmospheric refraction becomes predominant beyond a value of the zenith angles depending on the spectral band of observation. The optical system used on AMBER to correct the atmospheric refraction is detailed in Sect.\,\ref{4}, i.e. the opto-mechanical description and the tolerance analysis. Sect.\,\ref{5} describes the operations and Sect.\,\ref{6} gives the first results in laboratory and on sky.

\section{Effect of the atmospheric refraction on the fiber coupling}\label{2}
\subsection{The atmospheric refraction}\label{21}
The refraction angle $R(\lambda)$, defined from the apparent zenith angle $z$ of the observed object, and from the atmospheric and geographical conditions of the site (air pressure $P$, air temperature $T$, air density $\delta$, air refraction index $n(\lambda)$ on the ground, 
terrestrial radius $r_{t}$), is a function of the wavelength \citep{danjon}:
\begin{eqnarray}
\frac{R(\lambda)}{n(\lambda)-1} = \hspace{4.5cm}\nonumber \\
\left( 1 - \frac{P}{r_{t}\delta} \right)\tan(z) - \left( \frac{P}{r_{t}\delta} - \frac{n(\lambda)-1}{2}\right)\tan^3(z)  
\label{angleR}
\end{eqnarray}
The Sellmeier equation gives an empirical relationship between $n$ and the vacuum wavelength $\lambda$ for a particular transparent medium. The usual form of the equation is:
\begin{equation}
n^{2}(\lambda)-1 = \frac{B_{1} \lambda^{2}}{\lambda^{2}-C_{1}}+ \frac{B_{2} \lambda^{2}}{\lambda^{2}-C_{2}}+\frac{B_{3} \lambda^{2}}{\lambda^{2}-C_{3}}
\label{nn}
\end{equation}
where $B_{1,2,3}$ and $C_{1,2,3}$ are experimentally-determined Sellmeier coefficients, usually quoted for $\lambda$ measured in micrometers. 

\subsection{Expression of the coupling efficiency versus a tilt - Analytical approach}
\label{22}
An analytical approach to express the coupling efficiency versus a tilt was developed by \citet{tallon}. This tilt represents the inclination of the wavefront reaching an optics feeding the light into a single mode fiber which performs a spatial filtering of the incoming wavefront. 
A classical estimator for the fiber coupling efficiency $C_{\rm\it eff}$ is defined by \citet{ruil} and \citet{cass}: it is the squared modulus of the normalized overlap integral between the
electric field distribution in the focal plane of the telescope and the transmission of the fiber.
Using the Fourier Transform (FT) properties, it can be expressed in the pupil plane, where the
calculation is easier:
\begin{eqnarray}
C_{\rm\it eff} = \frac{\left\langle E_{p}(u,v) | E_{f}(u,v) \right\rangle^{2}} {\left\| E_{p}(u,v)\right\|^{2} \left\|E_{f}(u,v)\right\|^{2}} = \hspace{2cm}
\nonumber \\ 
{ \frac{\left( \int\!\int du dv\: E_{p}(u,v) E_{f}^*(u,v)\right)^{2} }{{ \int\!\int du dv\: E_{p}(u,v) E_{p}^*(u,v)\;\int\!\int du dv\: E_{f}(u,v) E_{f}^*(u,v)}}}
\label{rho}
\end{eqnarray}
where $E_{p}(u,v)=P(u,v) e^{-i\phi(u,v)}$ is the complex pupil function, with $\phi(u,v)$ the phase, and $E_{f}(u,v)$ is the FT of the field propagating through the fiber. 

Considering a single-mode fiber, the field can be approximated by a Gaussian function
$E_{f}(u,v)\propto e^{-\pi^2\omega_{\rm\it 0F}^2r^2}$, with $r=\sqrt{u^2+v^2}$ the FT conjugate variable and $\omega_{\rm\it 0F}=\omega_{\it 0}/F$. $\omega_{\it 0}$ is the radius of the fiber fundamental mode and $F$ is the focal length of the optics injecting the light in the fiber. The squared norm of $E_{f}(u,v)$ can then be expressed by:
\begin{equation}
\left\|E_{f}(u,v)\right\|^{2} = \frac{1}{2\pi\omega_{\rm\it 0F}^2}
\label{Ef}
\end{equation}
If $\alpha,\beta$ denote the angular deviations of the wavefront in the $u,v$ direction, the shift of the image at the fiber entrance is expressed by a convolution of the image function with $\delta(x-\alpha F,y-\beta F)$. Then, the pupil function is:

\noindent $E_{p}(u,v)=P(u,v) e^{-i\phi(u)}e^{-2i\pi (u\alpha+v\beta)}$. If $P(u,v)=1$ inside the pupil and $0$ elsewhere, the squared norm of $E_{p}(u,v)$ can then be expressed by:
\begin{equation}
\left\|E_{p}(u,v)\right\|^{2} = 2\pi \int^{D/2\lambda}_{kD/2\lambda}dr\: r  = \frac{\pi D^2(1-k^2)}{4\lambda^2}
\label{Ep}
\end{equation}
with $D$ the pupil diameter and $k$ the relative coefficient of the central obstruction of the telescope, i.e. the ratio of the diameter of the central obstruction to the primary mirror diameter. 

The scalar product of Eq.\,\ref{rho} can be written as the following function of $\alpha$ and $\beta$:
\begin{eqnarray}
\left\langle E_{p}(u,v) | E_{f}(u,v) \right\rangle = \hspace{4cm} \nonumber \\ 
{{{\scriptsize \infty} \atop{\displaystyle \int\int}}\atop {\scriptstyle -\infty}}
dudvP(u,v)e^{-i\phi(u,v)}e^{-\pi^2\omega_{\rm\it\scriptstyle{0F}}^2(u^2+v^2)}e^{-2i\pi(u\alpha+v\beta)}
\label{equa}
\end{eqnarray}
Considering the absence of aberration ($\phi(u,v)=0$), Eq.\,\ref{equa} is the FT in $\alpha$ and $\beta$ of the radial function $ P(u,v)e^{-\pi^2\omega_{\rm\it 0F}^2(u^2+v^2)}$. This two-dimensional function can then be expressed by the Hankel transform equal to $2\pi\int\limits_{\scriptstyle{0}}^{\scriptstyle{\infty}}r \left[P(r)e^{-\pi^2\omega_{\rm\it 0F}^2r^2}\right]J_{0}(2\pi q r)dr$, where $J_{0}$ is the Bessel function of zero order and $q=\sqrt{\alpha^2+\beta^2}$. Reporting Eq.\,\ref{Ef}, Eq.\,\ref{Ep} and Eq.\,\ref{equa} to Eq.\,\ref{rho} gives the coupling efficiency as a function of the wavefront inclination $q$:
\begin{eqnarray}
C_{\rm\it eff}(q)= \hspace{5cm}\nonumber\\
\frac{32\pi^2}{1-k^2}\frac{\omega_{\rm\it 0F}^2\lambda^2}{D^2}\left(\int^{D/2\lambda}_{kD/2\lambda} dr \, e^{-\pi^2\omega_{\rm\it 0F}^2r^2}r J_{0}(2\pi q r)\right)^2
\label{rho_f}
\end{eqnarray}
In the case of no tilt ($\alpha=\beta=0$):\newline
$C_{\rm\it eff}(0)= \frac{2}{1-k^2} \left(e^{-b^2}-e^{-k^2b^2}\right)^2 /b^2$, with\newline
$b=\pi\!D\omega_{\rm\it 0F}/2\lambda$, consistent with \citet{ruil}. 

Note that the radius $\omega_{\it 0}$ of the fiber fundamental mode is also chromatic:
\begin{equation}
\omega_{\it 0}=a \left( 0.65+\frac{1.619}{V^{3/2}} +\frac{2.879}{V^6}\right)
\end{equation}
where $a$ is the fiber core radius and $V$ the normalized frequency equal to $2\pi a \rm\it NA / \lambda$, $\rm\it NA$ being the numerical aperture.

\subsection{Necessity to correct the atmospheric refraction when observing with an instrument of the VLTI}
\label{23}
\subsubsection{Observing in individual bandwidths}
\label{231}

To assess the necessity to correct the atmospheric refraction while observing in J-, H-, and/or K-bands with a single-mode instrument at the VLTI, 
the coupling efficiency was estimated using the Danjon formula (Eq.\,\ref{angleR}) with Eq.\,\ref{rho}, as well as with Eq.\,\ref{rho_f}. The results of the two calculations agreed. This provided a useful check on our methodology, because when atmospheric turbulence is included, Eq.\,\ref{rho_f} is no longer valid as the problem is no longer circularly symmetric.
The origin of the inclination angle is taken at the central wavelength $\lambda_0$ of the spectral band, so that $\alpha~=~R(\lambda)~-~R(\lambda_0)$ represents the differential dispersion. Taking $\lambda_0$ is considered as a good approximation of the wavelength which maximizes the throughput over the full bandwidth.
The parameters taken into account in the simulations are the following:
\begin{itemize}
\item $P$ = 743 mbar (altitude of Paranal 2635 m)
\item $T$ = 10$^\circ$C
\item $\delta$ = 0.0012932 g/cm$^3$
\item $r_{t}$ = 6371 km
\item $k$ = 0.14 for $D$ = 8 m (Unit Telescope, UT) and $D$~=~1.8 m (Auxiliary Telescope, AT)
\item $z$ = 60$^\circ$ and 30$^\circ$
\item $\rm\it NA$ = 0.14 in J-band, 0.15 in H-band, and 0.16 in K-band.
\item $a$ = 2.8 $\mu$m in J-band, 2.5 $\mu$m in H-band, and 4.9 $\mu$m in K-band.
\item $\lambda_c$~=~0.944~$\mu$m in J-band; $\lambda_c$~=~1.15~$\mu$m in H-band; $\lambda_c$~=~1.9~$\mu$m in K-band: cutting wavelengths of the fibers.
\end{itemize}
The focal length $F$ of the light injection optics of AMBER was originally estimated to optimize $C_{\rm\it eff}(0)$ at $\lambda_c$. However, we needed to replace the fibers by new ones after intensive manipulation. As the initial fibers were not on sale anymore at all manufacturers, the characteristics $\rm\it NA$, $\lambda_c$, and $a$ are no longer adapted to the focal length $F$. This explains the maximum coupling of 0.69 in J-band, 0.61 in H-band, and 0.40 in K-band when it should reach 0.78.
This difference affects the global throughput performance of AMBER, but has a negligible effect in this discussion.

\begin{figure}
\hspace{10mm}\includegraphics[width=63mm]{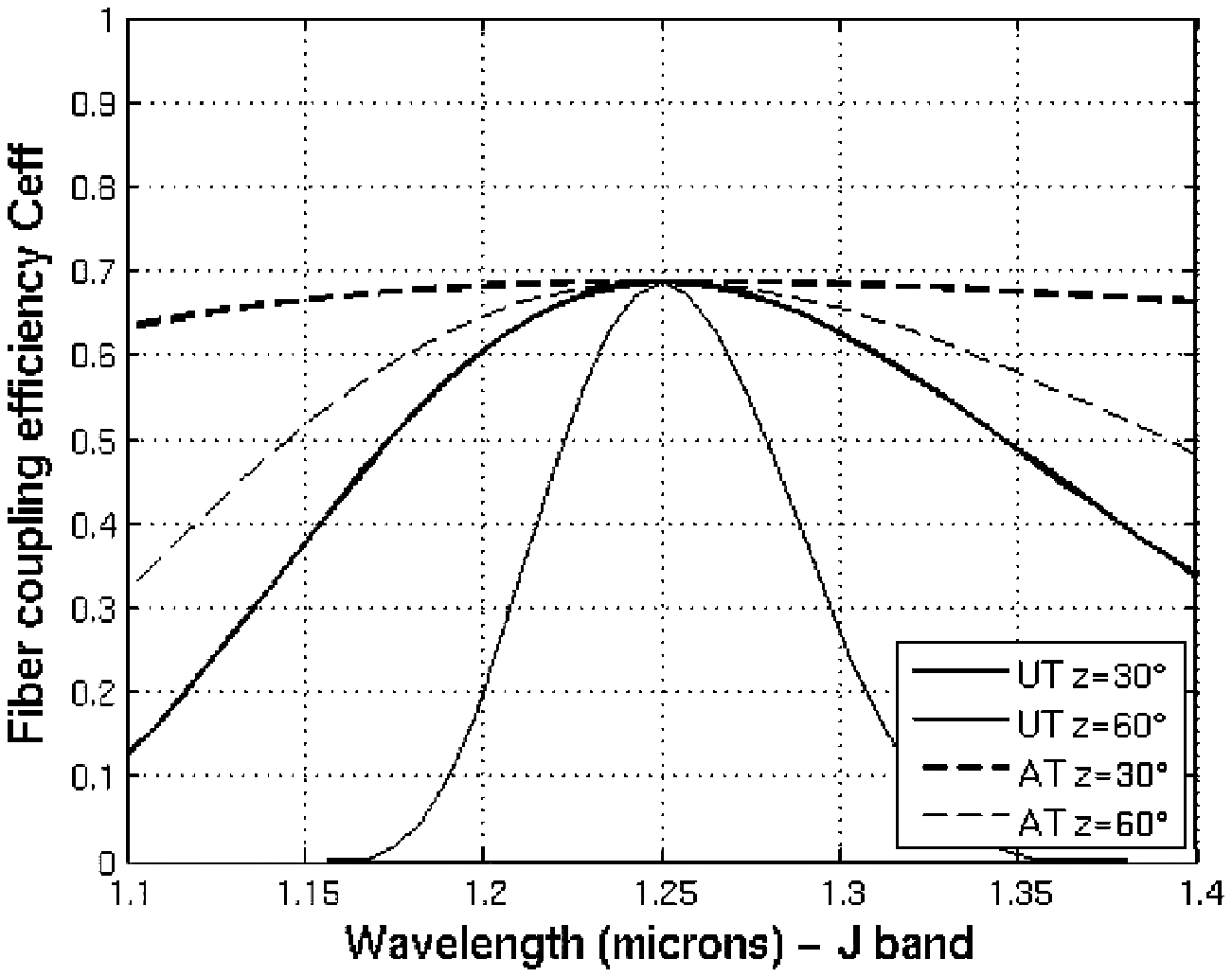}
\begin{center}
\includegraphics[width=70mm]{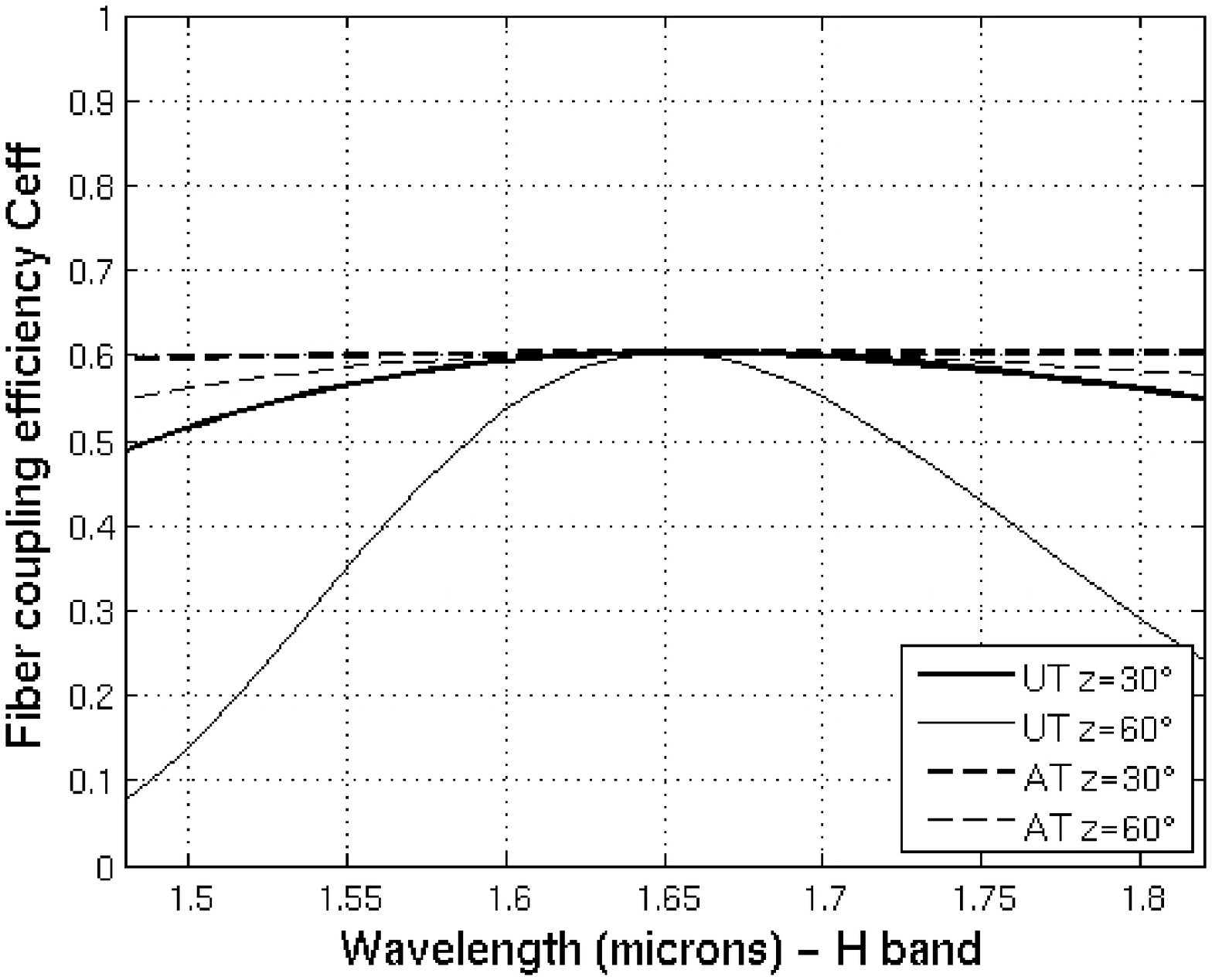}
\includegraphics[width=70mm]{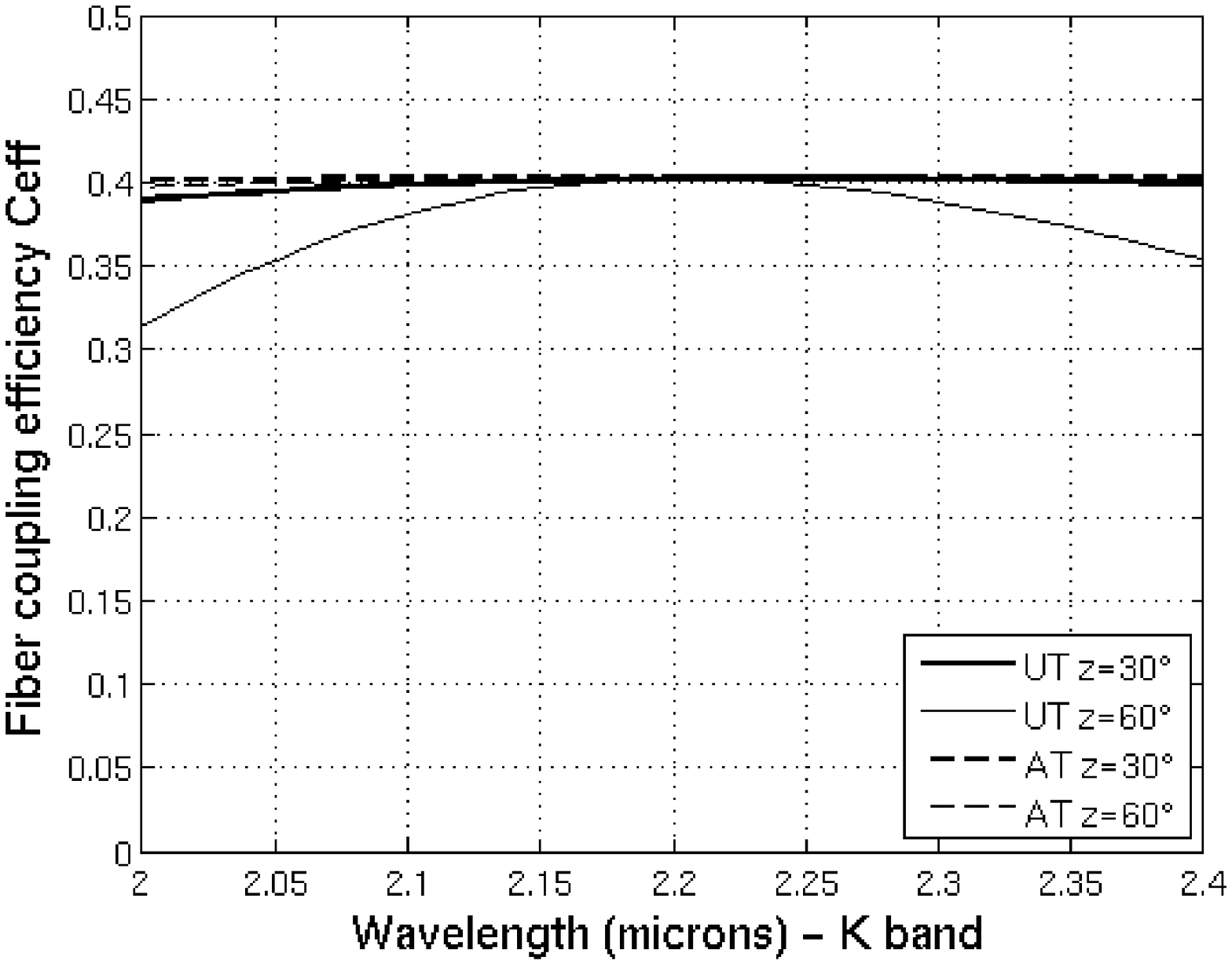}
\end{center}
\caption{Simulated VLTI chromatic variation of the coupling efficiency $C_{\rm\it eff}(\lambda)$ with uncorrected differential atmospheric dispersion $\alpha$ at two zenith angles (60$^\circ$ and 30$^\circ$) in J-, H-, and K-bands. The origin of the refraction angle is taken at the central wavelength of each respective spectral band. Solid lines: UT. Dotted lines: AT.}
\label{fig_rho}
\end{figure}

Figure\,\ref{fig_rho} shows the chromatic variation of the coupling efficiency $C_{\rm\it eff}(\lambda)$ with the differential atmospheric dispersion $\alpha$ at VLTI, observing in J-, H- and K-bands. 

Table\,\ref{obs} gives the corresponding values of the differential dispersion $\alpha$ at a zenith angle of $z$ = 60$^\circ$ (worst case), the coupling efficiencies and the shift of the Airy disk (2.44$\lambda$/D) between the central wavelength $\lambda_0$ and that at the lower edge of the spectral band. The degradation factor $\rho$ is defined as the ratio of the estimated coupling to the ideal one: $\rho~=~C_{\rm\it eff}(\alpha)/C_{\rm\it eff}(0)$.\newline
\begin{table}
	\caption{Differential dispersion $\alpha$ for $z$ = 60$^\circ$, coupling efficiencies $C_{\rm\it eff}$, coupling degradation factor $\rho$, and shift of the Airy disk (2.44$\lambda$/D) between the central wavelength and that at the lower edge of the spectral band.}
	\begin{center}
		\begin{tabular}{c c c c c} \hline\hline
\multicolumn{5}{l} {\scriptsize \textbf{Observing with the UTs, $z$ = 60$^\circ$} }\\	\hline
\scriptsize &\scriptsize $\alpha$ [mas] &\scriptsize	$C_{\rm\it eff}$ &\scriptsize	$\rho$ &\scriptsize	Shift of the Airy disk\\\hline
\scriptsize J  &\scriptsize 150 &\scriptsize	0&\scriptsize	0	&\scriptsize 1	\\
\scriptsize H &\scriptsize 40 &\scriptsize	0.06&\scriptsize 0.10&\scriptsize	1/3$^{rd}$ \\
\scriptsize K &\scriptsize 20 &\scriptsize	0.31&\scriptsize 0.77&\scriptsize	1/7$^{th}$ \\\hline\hline
\multicolumn{5}{l} {\scriptsize \textbf{Observing with the ATs, $z$ = 60$^\circ$ }}\\	\hline
\scriptsize &\scriptsize $\alpha$ [mas] &\scriptsize	$C_{\rm\it eff}$ &\scriptsize	$\rho$ &\scriptsize	Shift of the Airy disk\\\hline
\scriptsize J &\scriptsize 150 &\scriptsize	0.30&\scriptsize 0.44	&\scriptsize 1/4$^{th}$	\\
\scriptsize H &\scriptsize 40 &\scriptsize	0.54&\scriptsize 0.90 &\scriptsize	1/15$^{th}$\\
\scriptsize K &\scriptsize 20 &\scriptsize	0.40&\scriptsize 0.99 &\scriptsize	1/30$^{th}$ \\\hline
		\end{tabular}
			\end{center}
	\label{obs}
\end{table}

If we consider individual spectral bandwidths, and assuming that the origin of the refraction angle is taken at the central wavelengths, let us note that:

- The refraction effect is stronger while observing with UTs.

- The refraction effect is stronger at the shortest wavelengths. In J-band, at zenith angles above 30$^\circ$ the coupling efficiency drops by more than 50\% for wavelengths lower than 1.15~$\mu$m and greater than 1.35~$\mu$m. In H-band, at 60$^\circ$ the coupling efficiency drops by more than 50\% for wavelengths lower than 1.55~$\mu$m and greater than 1.80~$\mu$m. In K-band, the coupling efficiency drops from 0.40 to 0.31 which represents a loss of about 20\% at the shortest wavelength.

The specifications defined on AMBER was to put within 15\% the optimal coupling efficiency in J- and H-bands at the same level than that in K-band. Mainly, the dispersion correction at $z$ of about 60$^\circ$ in J and H must lead to the optimal coupling efficiency of 0.78 in the optimal optical configuration, within 15\%, this error being expected taking into account optical element transmission and manufacturing errors.

\subsubsection{Observing at low spectral resolution in individual bandwidths}
\label{232}
In addition, we can average out the coupling over $\lambda$ to estimate the fiber transmission throughout the overall bandwidths. This results in the estimations presented in Table\,\ref{obs2}. It gives an order of the coupling efficiency while observing in low spectral resolution with AMBER as a function of the zenith angle.
\begin{table}
	\caption{Simulated coupling efficiency averaged over J-band (1.1$-$1.4~$\mu$m), H-band (1.48$-$1.82~$\mu$m), and K-band (2.0$-$2.4~$\mu$m).}
	\begin{center}
	\begin{tabular}{c | c c | c c | c c} \cline{2-7}
	\scriptsize &\multicolumn{2}{l} {\scriptsize \textbf{J}}&\multicolumn{2}{l} {\scriptsize \textbf{H}}&\multicolumn{2}{l} {\scriptsize \textbf{K}}\\\cline{2-7}
	\scriptsize &\scriptsize \textbf{UT}&\scriptsize \textbf{AT}&\scriptsize \textbf{UT}&\scriptsize \textbf{AT}&\scriptsize \textbf{UT}&\scriptsize \textbf{AT}\\\hline\hline
		\scriptsize $z$ = 10$^\circ$&\scriptsize 0.66&\scriptsize 0.69&\scriptsize 0.60&\scriptsize 0.60 &\scriptsize 0.40&\scriptsize	0.40\\\hline
		\scriptsize $z$ = 20$^\circ$&\scriptsize 0.58&\scriptsize 0.68&\scriptsize 0.59&\scriptsize 0.60 &\scriptsize 0.40&\scriptsize	0.40\\\hline
	\scriptsize $z$ = 30$^\circ$&\scriptsize 0.47&\scriptsize 0.67&\scriptsize 0.57&\scriptsize 0.60 &\scriptsize 0.40&\scriptsize	0.40\\\hline
	\scriptsize $z$ = 40$^\circ$&\scriptsize 0.35&\scriptsize 0.65&\scriptsize 0.53&\scriptsize 0.60 &\scriptsize 0.40&\scriptsize	0.40\\\hline
	\scriptsize $z$ = 50$^\circ$&\scriptsize 0.26&\scriptsize 0.62&\scriptsize 0.48&\scriptsize 0.60 &\scriptsize 0.39&\scriptsize	0.40\\\hline
	\scriptsize $z$ = 60$^\circ$&\scriptsize 0.18&\scriptsize 0.56&\scriptsize 0.38&\scriptsize 0.59 &\scriptsize 0.37&\scriptsize	0.40\\\hline
	\end{tabular}
	\end{center}
	\label{obs2}
	\end{table}

The coupling efficiency averaged over the whole bands drops by more than 50\% for zenith angles higher than 40$^\circ$ in J-band with UTs, drops by more than 20\% of its initial value for zenith angles above 50$^\circ$ in H-band with UTs, and is almost constant - from 0.40 to 0.37 - in K-band.

This analytical approach demonstrates that an atmospheric refraction controller is necessary to observe at large zenith angle in J- and H-bands, especially with the UTs for which $C_{\rm\it eff}$ is almost null at $z$ = 60$^\circ$ at extreme wavelengths. No correction is necessary in K-band. 

\subsubsection{Simultaneous observations}
\label{233}
\begin{figure}
\begin{center}
\includegraphics[width=65mm,height=45mm]{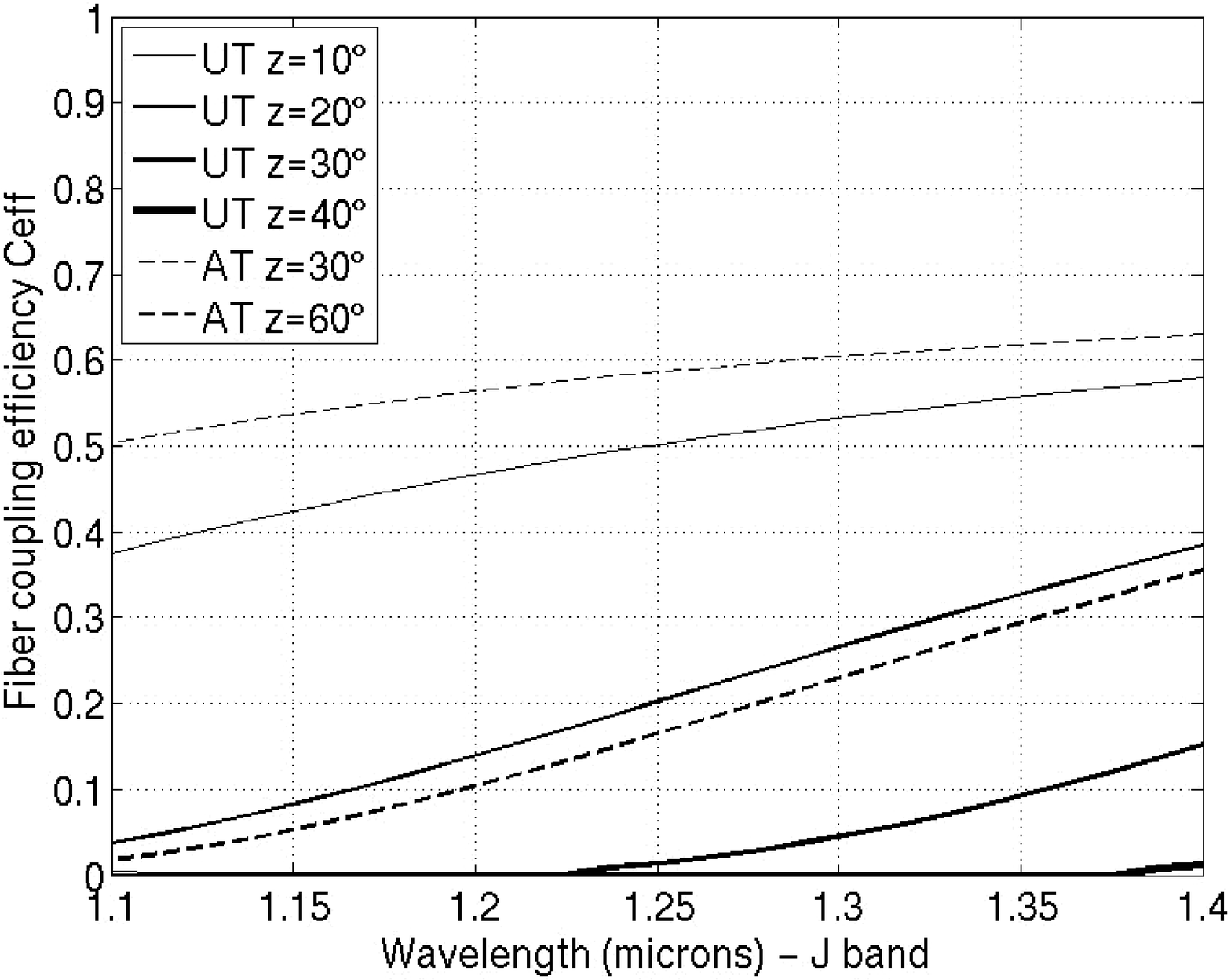}
\includegraphics[width=65mm,height=45mm]{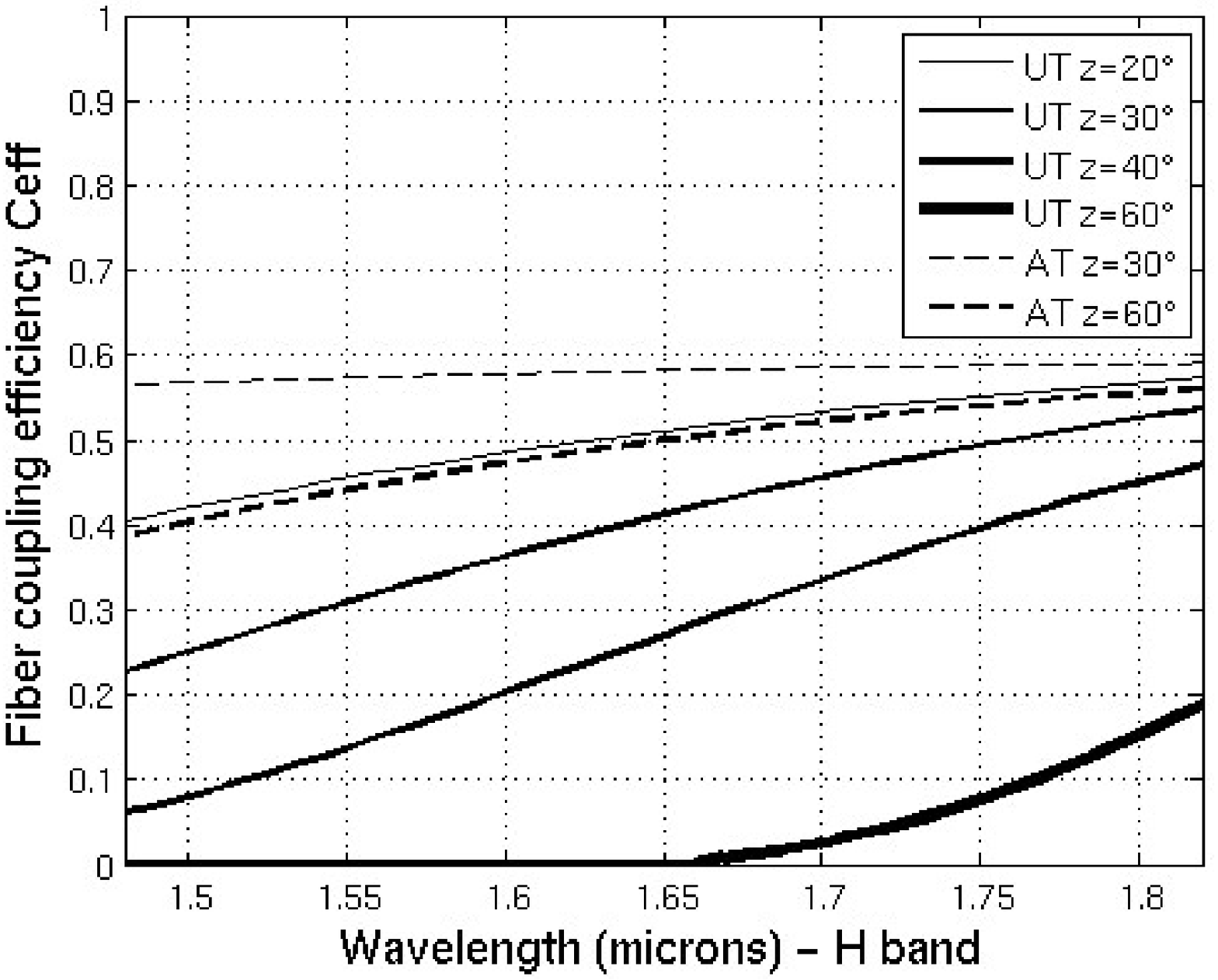}
\end{center}
\caption{VLTI chromatic variation of the coupling efficiency $C_{\rm\it eff}(\lambda)$ with uncorrected differential atmospheric dispersion $\alpha$ at different zenith angles in J- and H-bands while the VLTI is optimized at 2.2~$\mu$m (i.e. $C_{\rm\it eff}$ maximum at 2.2~$\mu$m). Solid lines: UT. Dotted lines: AT.}
\label{fig_rho2}
\end{figure}

The results above consider observations in individual bandwidths. Let us now assume simultaneous observations in the three spectral bands J, H, and K, as it can be performed with the AMBER instrument. There is one spatial filter with single mode fibers per band. They are all aligned together along the same optical axis defined by the artificial sources. 
While observing a scientific target, the light coming from the telescope is optimized in one spatial filter, in K- or H-band, along the previously defined optical axis. Due to the atmospheric refraction, in a direction different from zenith, the optical path differs from the initial axis as a function of wavelength. 
While observing in J-, H- and K-bands, the system must correct the refraction effect in J- and H-bands with respect to the
central wavelength of K-band (2.2~$\mu$m), which is more demanding than inside each respective wavelength. The differential dispersion angle is then
$\alpha = R(\lambda) - R(2.2 \mu m)$.  Figure\,\ref{fig_rho2} shows the chromatic variation of the coupling efficiency $C_{\rm\it eff}(\lambda)$ with the differential atmospheric dispersion $\alpha$ at VLTI, observing in J- and H-bands with the VLTI optimized at 2.2~$\mu$m.

The loss of coupling efficiency becomes more critical:

- In J-band: with UTs, the coupling efficiency drops by more than 70\% of its initial value at zenith angles above 20$^\circ$. 

- In H-band: with UTs, the coupling efficiency drops by more than 50\% of its initial value at zenith angles above 35$^\circ$, and 70\% above 45$^\circ$.

The corresponding averaged coupling efficiencies are shown in Table\,\ref{obs3}. 
\begin{table}
	\caption{Simulated coupling efficiency averaged over J-band (1.1$-$1.4~$\mu$m), and H-band (1.48$-$1.82~$\mu$m), with the VLTI optimized at 2.2~$\mu$m.}
	\begin{center}
	\begin{tabular}{l | l l | l l | l l  } \cline{2-7}
	\scriptsize &\multicolumn{2}{l} {\scriptsize \textbf{J}}&\multicolumn{2}{l} {\scriptsize \textbf{H}}&\multicolumn{2}{l} {\scriptsize \textbf{K}}\\\cline{2-7}
	\scriptsize &\scriptsize \textbf{UT}&\scriptsize \textbf{AT}&\scriptsize \textbf{UT}&\scriptsize \textbf{AT}&\scriptsize \textbf{UT}&\scriptsize \textbf{AT}\\\hline\hline
		\scriptsize $z$ = 10$^\circ$&\scriptsize 0.49&\scriptsize 0.66&\scriptsize 0.56&\scriptsize 0.59&\scriptsize 0.40 &\scriptsize 0.40 \\\hline
		\scriptsize $z$ = 20$^\circ$&\scriptsize 0.20&\scriptsize 0.63&\scriptsize 0.50&\scriptsize 0.59 &\scriptsize 0.40&\scriptsize 0.40\\\hline
	\scriptsize $z$ = 30$^\circ$&\scriptsize 0.04&\scriptsize 0.58&\scriptsize 0.40&\scriptsize 0.58 &\scriptsize 0.40&\scriptsize 0.40\\\hline
	\scriptsize $z$ = 40$^\circ$&\scriptsize - &\scriptsize 0.49&\scriptsize 0.26&\scriptsize 0.57 &\scriptsize 0.40&\scriptsize 0.40\\\hline
	\scriptsize $z$ = 50$^\circ$&\scriptsize - &\scriptsize 0.35&\scriptsize 0.13&\scriptsize 0.54 &\scriptsize 0.39&\scriptsize 0.40\\\hline
	\scriptsize $z$ = 60$^\circ$&\scriptsize - &\scriptsize 0.17&\scriptsize 0.04&\scriptsize 0.49&\scriptsize 0.37&\scriptsize 0.40
\\\hline
	\end{tabular}
	\end{center}
	\label{obs3}
	\end{table}

\begin{figure}
\begin{center}
\includegraphics[width=65mm,height=45mm]{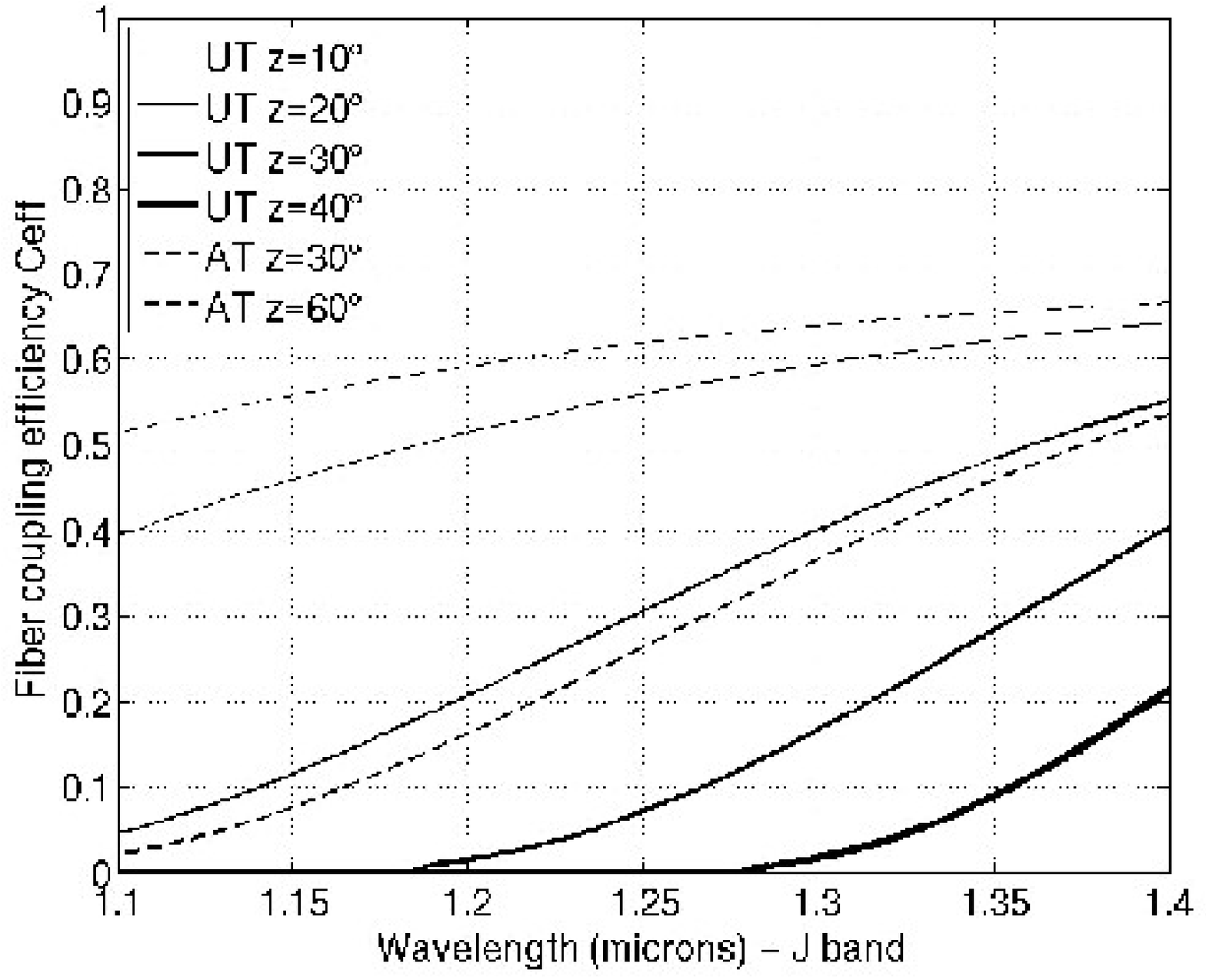}
\includegraphics[width=65mm,height=45mm]{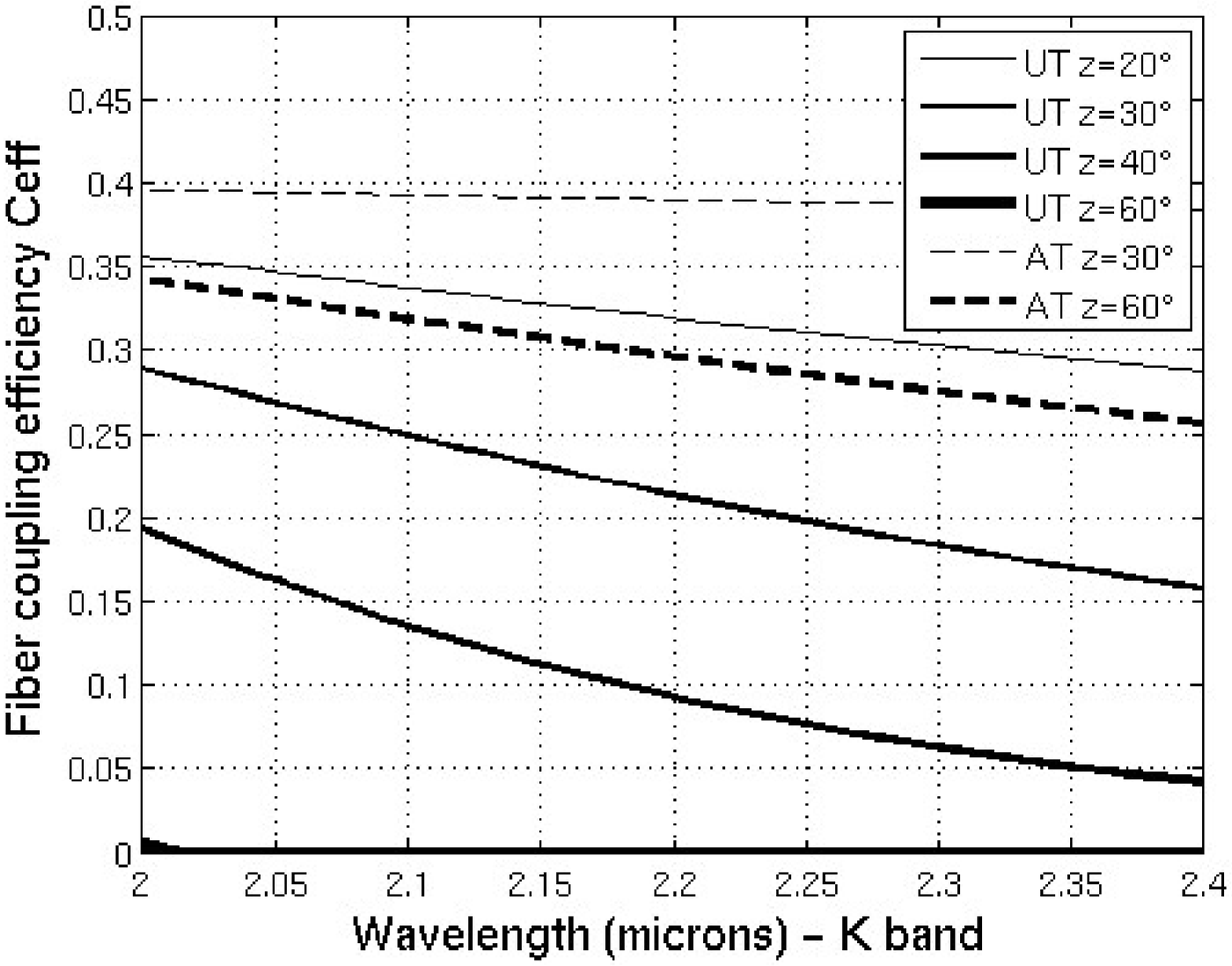}
\end{center}
\caption{VLTI chromatic variation of the coupling efficiency $C_{\rm\it eff}(\lambda)$ with uncorrected differential atmospheric dispersion $\alpha$ at different zenith angles in J- and K-bands with the VLTI optimized at 1.65~$\mu$m (i.e. $C_{\rm\it eff}$ maximum at 1.65~$\mu$m). Solid lines: UT. Dotted lines: AT.}
\label{fig_rho3}
\end{figure}

If we observe in J- and K-bands in the case the VLTI is optimized at 1.65~$\mu$m, the situation is less critical than previously, as seen in Figure\,\ref{fig_rho3}. The loss of coupling efficiency is then:

- In J-band: with UTs, the coupling efficiency drops by more than 70\% of its initial value at zenith angles above 25$^\circ$. 

- In K-band: with UTs, the coupling efficiency drops by more than 20\% of its initial value at zenith angles above 40$^\circ$, and by more than 50\% above 55$^\circ$.

The corresponding averaged coupling efficiency is shown in Table\,\ref{obs4}. 
\begin{table}
	\caption{Simulated coupling efficiency averaged over J-band (1.1$-$1.4~$\mu$m), and K-band (2.0$-$2.4~$\mu$m), with the VLTI optimized at 1.65~$\mu$m.}
	\begin{center}
	\begin{tabular}{c | c c | c c | c c} \cline{2-7}
	\scriptsize &\multicolumn{2}{l} {\scriptsize \textbf{J}}&\multicolumn{2}{l} {\scriptsize \textbf{H}}&\multicolumn{2}{l} {\scriptsize \textbf{K}}\\\cline{2-7}
	\scriptsize &\scriptsize \textbf{UT}&\scriptsize \textbf{AT}&\scriptsize \textbf{UT}&\scriptsize \textbf{AT}&\scriptsize \textbf{UT}&\scriptsize \textbf{AT}\\\hline\hline
		\scriptsize $z$ = 10$^\circ$&\scriptsize 0.54&\scriptsize 0.68&\scriptsize 0.60&\scriptsize 0.60&\scriptsize 0.40&\scriptsize 0.40 \\\hline
		\scriptsize $z$ = 20$^\circ$&\scriptsize 0.30&\scriptsize 0.65&\scriptsize 0.59&\scriptsize 0.60&\scriptsize 0.39&\scriptsize 0.40 \\\hline
	\scriptsize $z$ = 30$^\circ$&\scriptsize 0.12&\scriptsize 0.61&\scriptsize 0.57&\scriptsize 0.60&\scriptsize 0.36&\scriptsize 0.40 \\\hline
	\scriptsize $z$ = 40$^\circ$&\scriptsize 0.03 &\scriptsize 0.54&\scriptsize 0.53&\scriptsize 0.60&\scriptsize 0.31&\scriptsize 0.40 \\\hline
	\scriptsize $z$ = 50$^\circ$&\scriptsize - &\scriptsize 0.42&\scriptsize 0.48&\scriptsize 0.60&\scriptsize 0.25&\scriptsize 0.39 \\\hline
	\scriptsize $z$ = 60$^\circ$&\scriptsize - &\scriptsize 0.26&\scriptsize 0.38&\scriptsize 0.59&\scriptsize 0.17&\scriptsize 0.38
\\\hline
	\end{tabular}
	\end{center}
	\label{obs4}
	\end{table}

\section{Atmospheric fluctuations and MACAO correction}
\label{3}
While $z$ increases, two effects occur. The atmospheric refraction increases, and the atmospheric Strehl ratio decreases. The goal of this paragraph is to estimate which effect is predominant on the coupling degradation, and to determine if the refraction correction by an optical system stays relevant in spite of the Strehl degradation.
\begin{figure}
\begin{center}
\includegraphics[width=42mm]{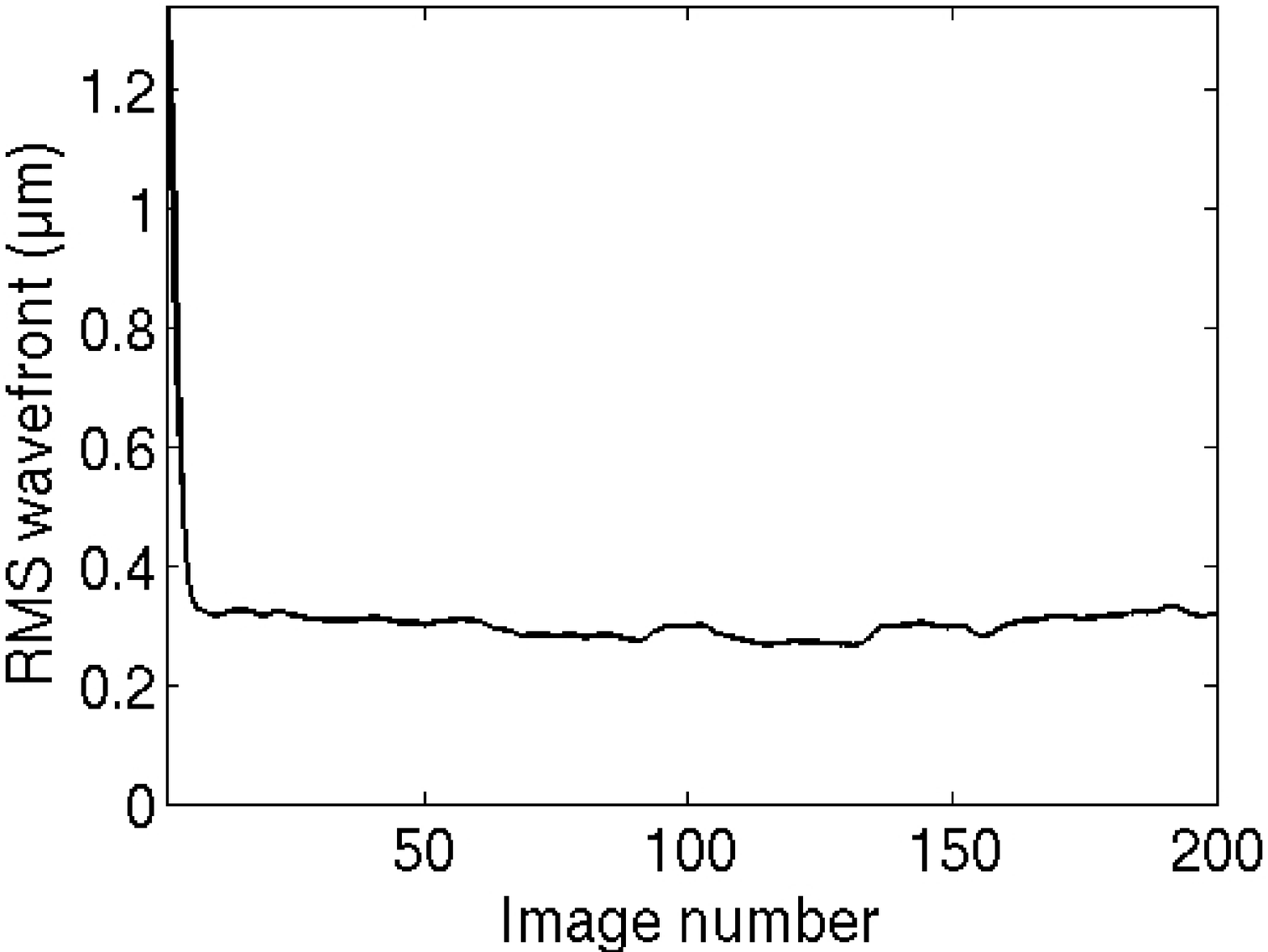}
\includegraphics[width=42mm]{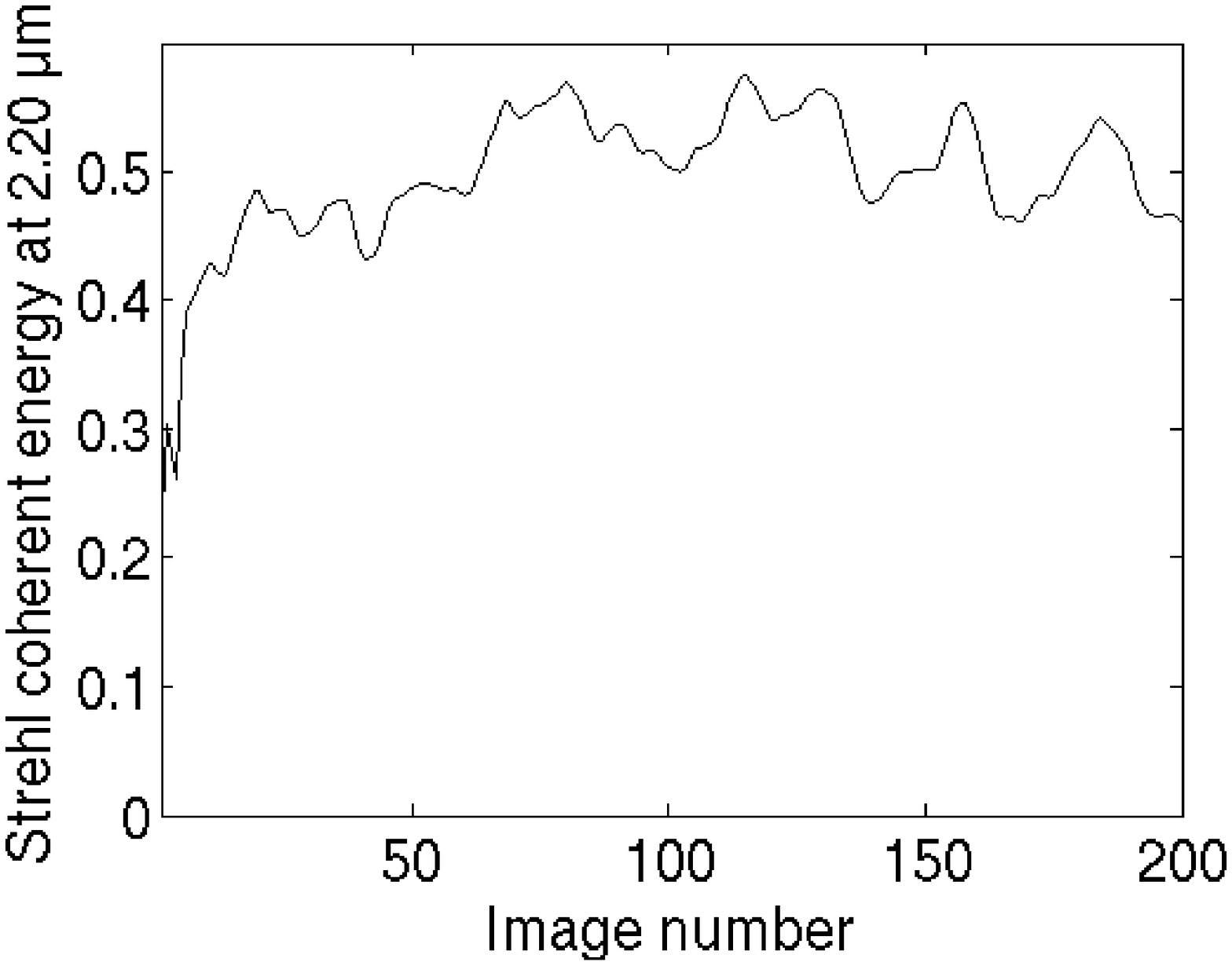}
\end{center}
\caption{Wavefront rms (left) and K-band Strehl ratio (right) as a function of image number, i.e. time, for one simulated time history. Note the stabilization of the performance after $\approx$~10~images.}
\label{rms}
\end{figure}
\subsection{Description of the simulated parameters}
We computed the coupling efficiency while taking into account the atmosphere fluctuations, and the residual wavefront errors after correction by MACAO.
Under median seeing conditions (0.65$\arcsec$) and at 2.2~$\mu$m, MACAO is specified to deliver at least 50\% of Strehl ratio for bright sources (m$_V\leq 8$), and at least 25\% for faint sources (m$_V\leq 15.5$).

In order to simulate in details the optical wavefront coming from the propagation of a star light through the atmosphere, and at the exit of the 60-actuators MACAO system, we made use of simulations performed with the software package {\sc caos} \citep{caos}, developed within the {\sc caos} problem-solving environment \citep{carbill} and permitting a detailed physical modeling of both the atmospheric optical turbulence and MACAO. We hence simulated a 3-layers Paranal median-seeing atmosphere (see Table~\ref{tab:param} for details on the physical parameters) and, assuming the usual Taylor hypothesis and a time sampling of 3\,ms, made this atmosphere evolve over a total evolution time history of 2\,s.
\begin{table}
\begin{center}
\caption{Main parameters of the end-to-end {\sc caos} simulations.}
\begin{tabular}{lrr}
\hline\hline
\multicolumn{2}{l}{\bf Turbulent atmosphere parameters}							\\
\scriptsize Fried parameter $r_0 @$ 500\,nm				&   \scriptsize 12\,cm				\\
\scriptsize Number of turbulent layers							&   \scriptsize 3					\\
\scriptsize Layers altitudes												&   \scriptsize 10\,m, 1\,km, 10\,km	\\
\scriptsize Layers $C_N^2$ profile relative percentage	&  \scriptsize  20\%, 60\%, 20\%	\\
\scriptsize Wind velocities													&   \scriptsize $\sim$6\,m/s, $\sim$6\,m/s, 32\,m/s \\
\scriptsize Wind directions													&   \scriptsize 0\degr, 90\degr, 0\degr \\
\scriptsize Wavefront outer-scale ${\cal L}_0$			&   \scriptsize 25\,m				\\ \hline\hline
\multicolumn{2}{l}{\bf  Telescope parameters}												\\
\scriptsize Diameter																&  \scriptsize  8\,m				\\
\scriptsize Obstruction ratio												&  \scriptsize 14\%				\\ \hline\hline
\multicolumn{2}{l}{\bf  AO system main parameters}										\\
\scriptsize Number of sensing elements							&  \scriptsize 60				\\
\scriptsize Number of actuators											&  \scriptsize  60				\\
\scriptsize Number of reconstructed modes 					&  \scriptsize up to 48			\\
\scriptsize Time-filter type												&  \scriptsize  pure integration		\\
\scriptsize Closed-loop gain												&  \scriptsize  0.4				\\
\scriptsize Exposure time of the wavefront sensor		&  \scriptsize  3\,ms					\\  \hline
\label{tab:param}
\end{tabular}
\end{center}
\end{table}

\begin{figure*}
\begin{center}
\includegraphics[width=50mm]{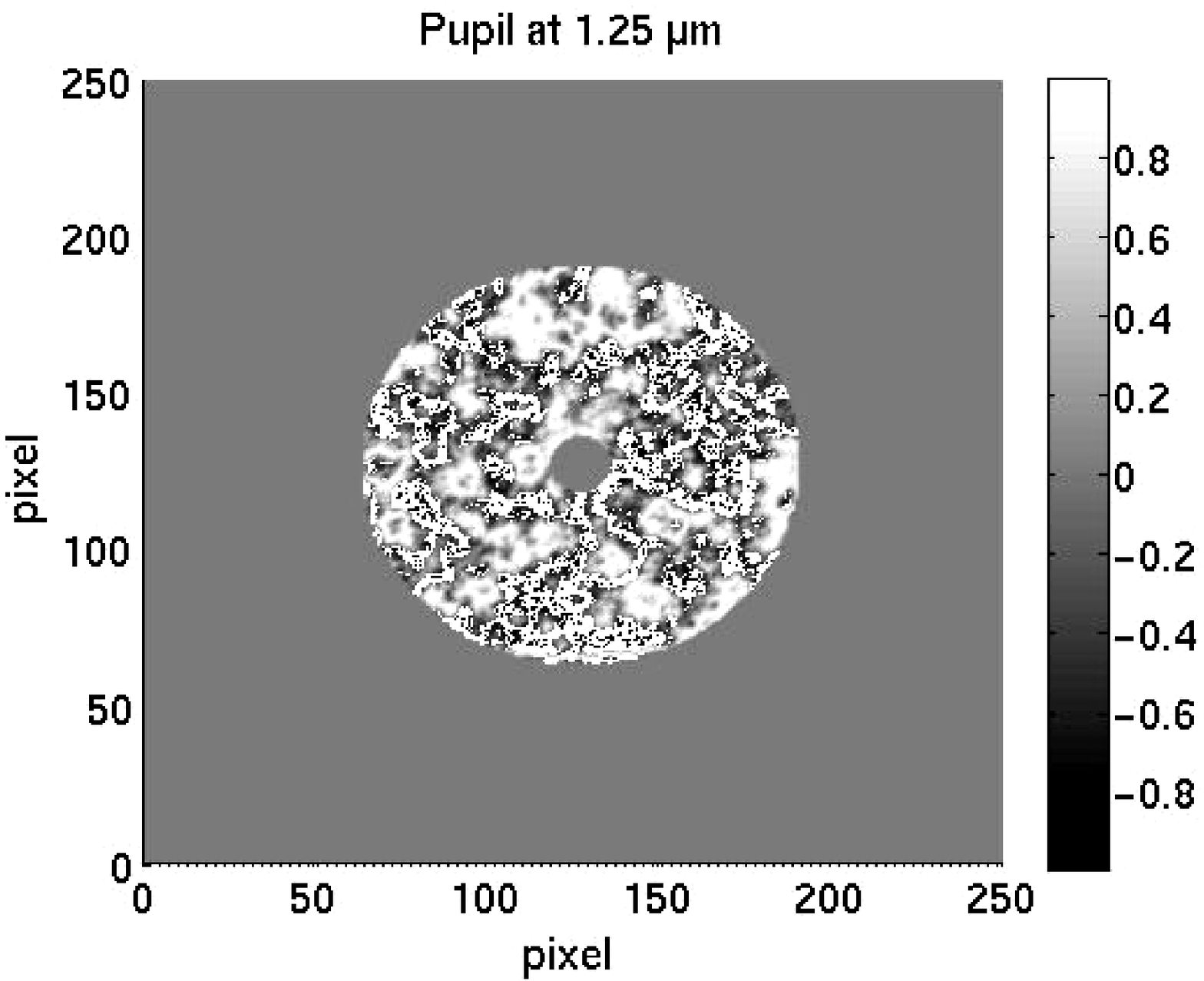}
\includegraphics[width=50mm]{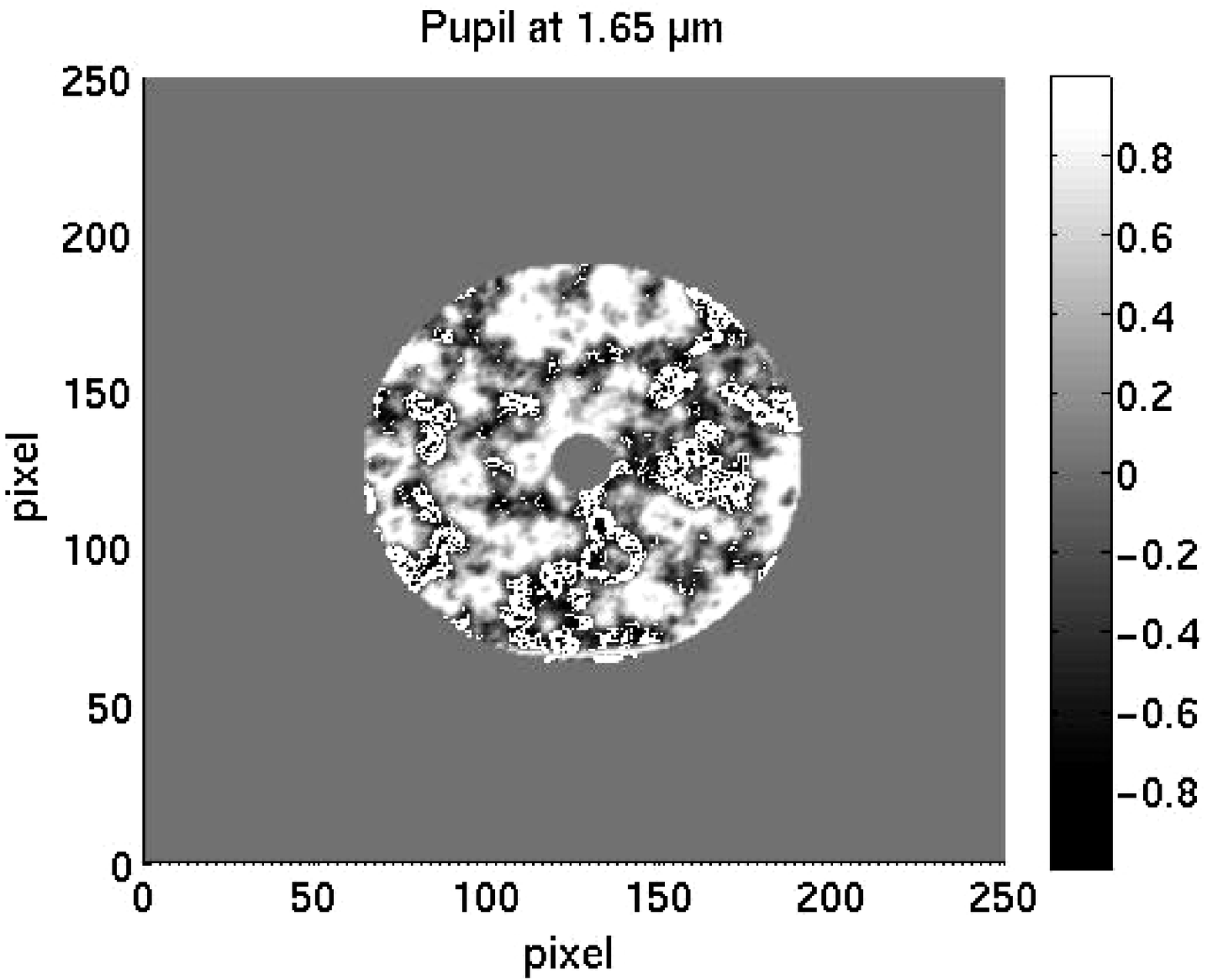}
\includegraphics[width=50mm]{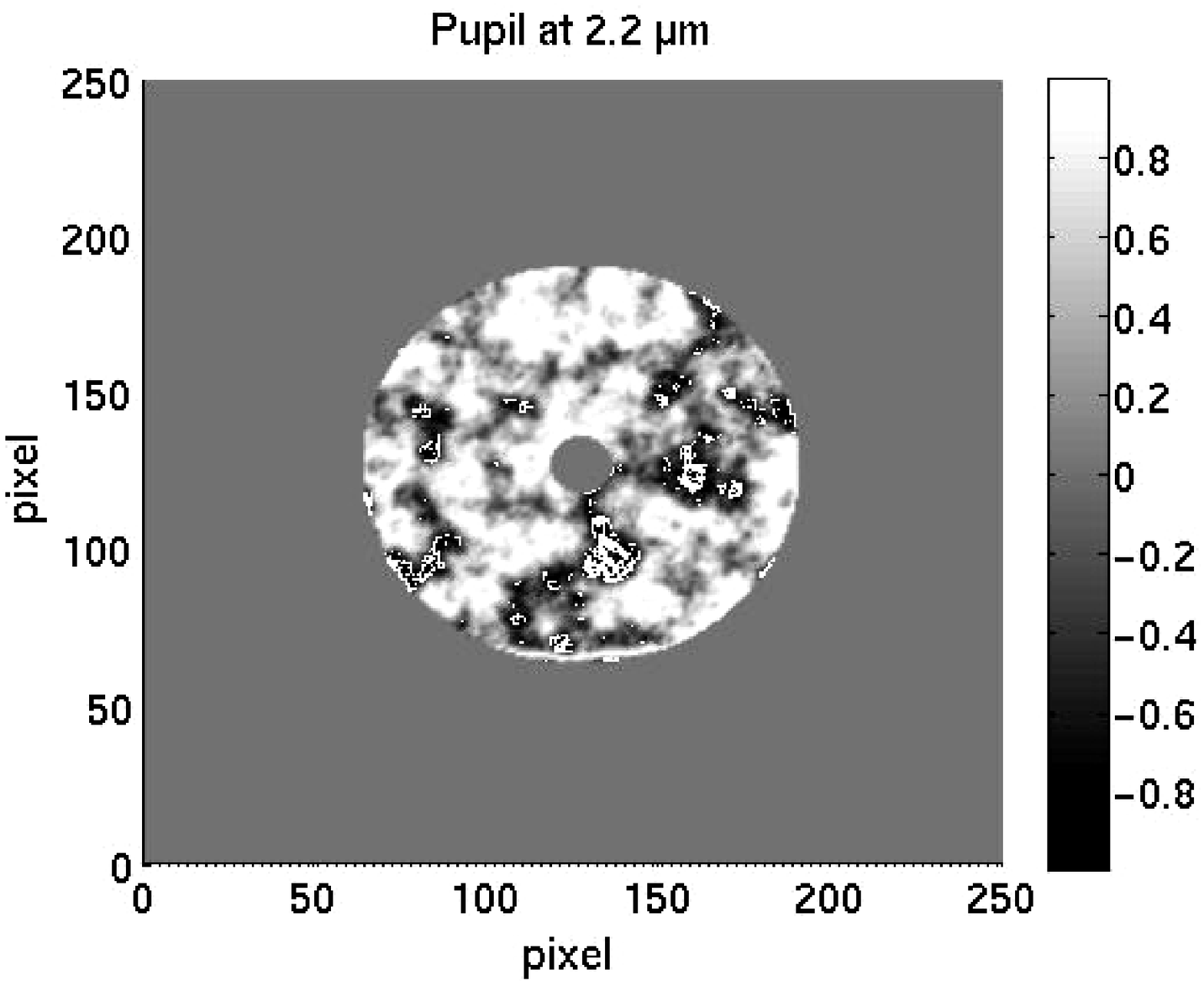}
\end{center}
\begin{center}
\includegraphics[width=50mm]{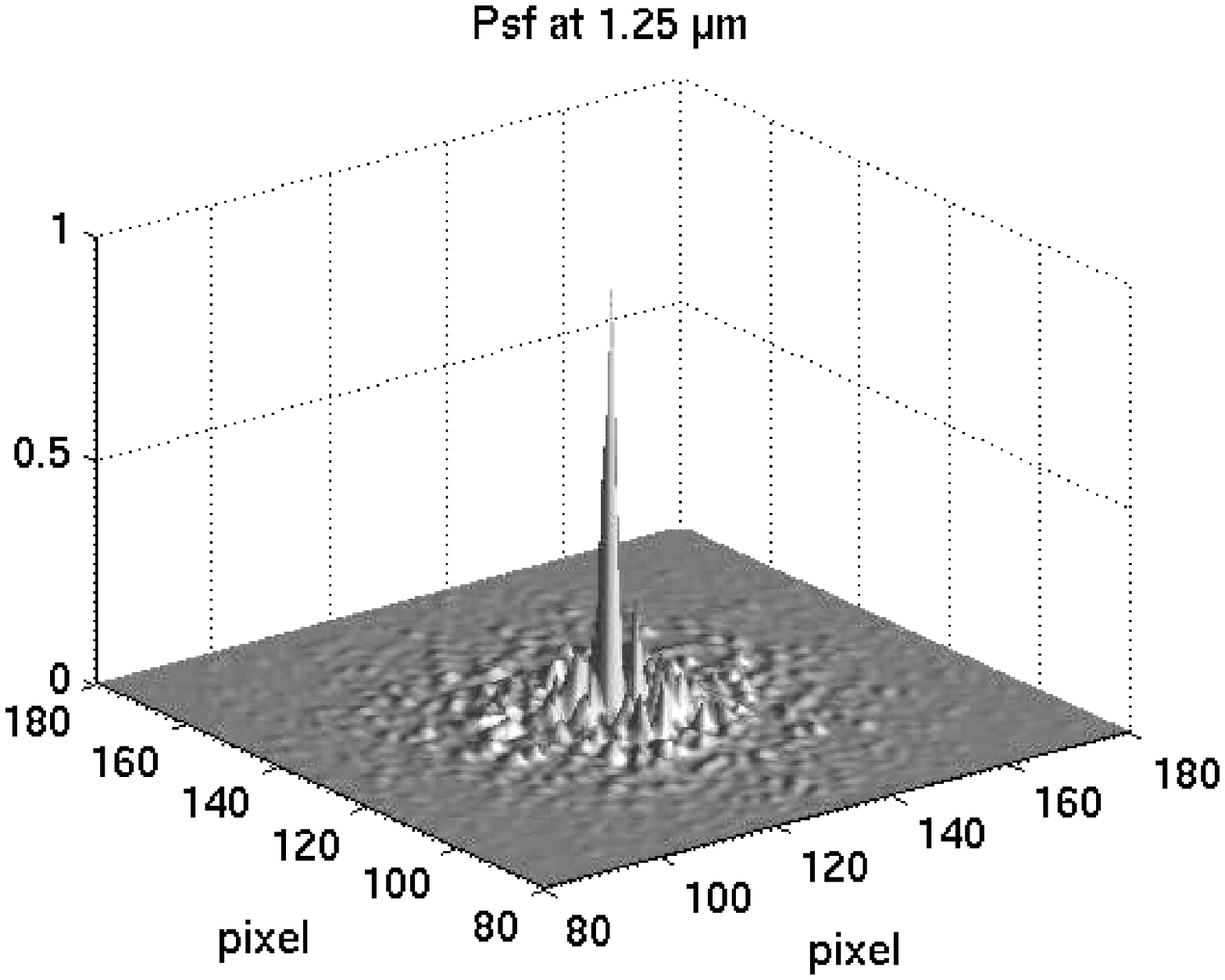}
\includegraphics[width=50mm]{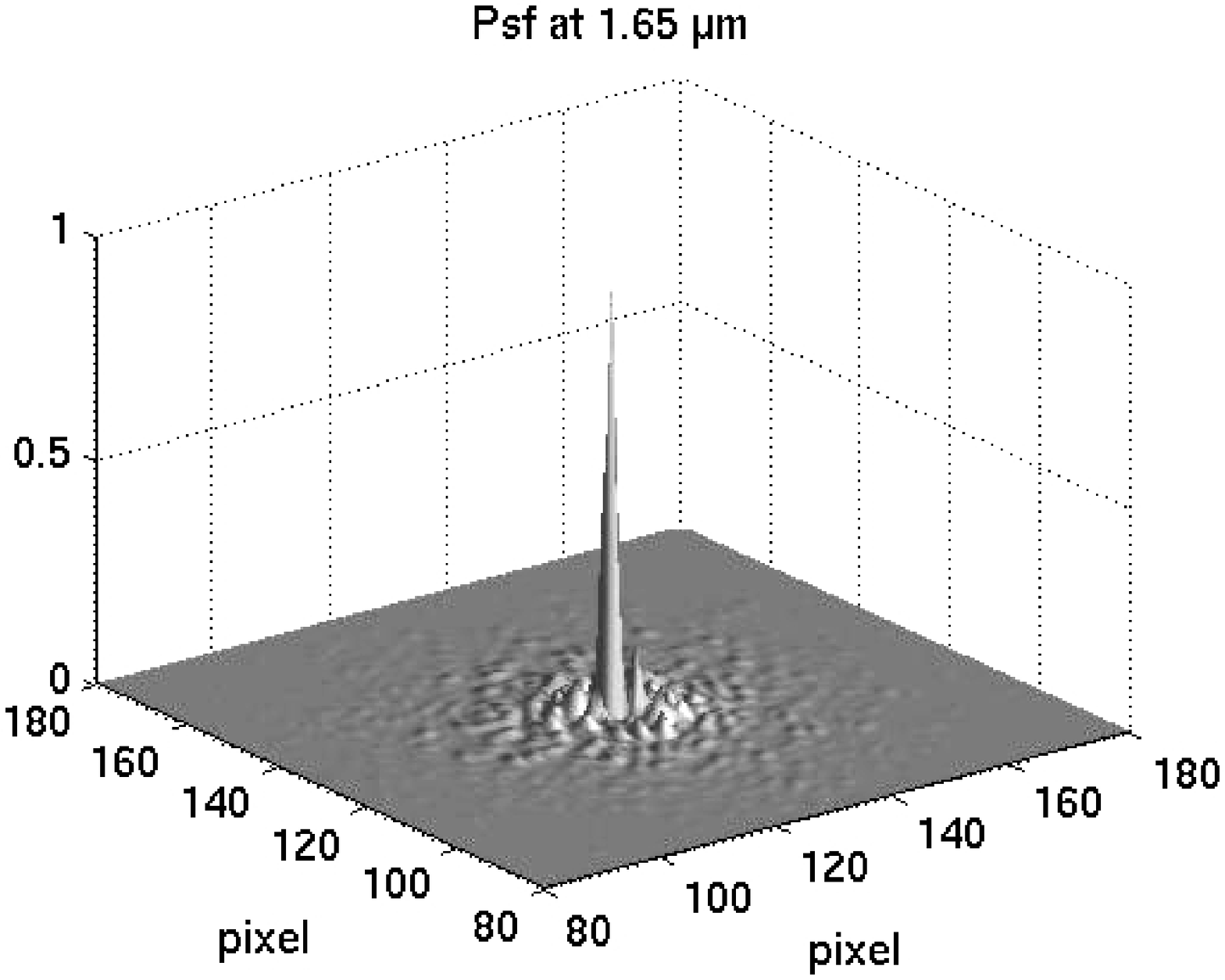}
\includegraphics[width=50mm]{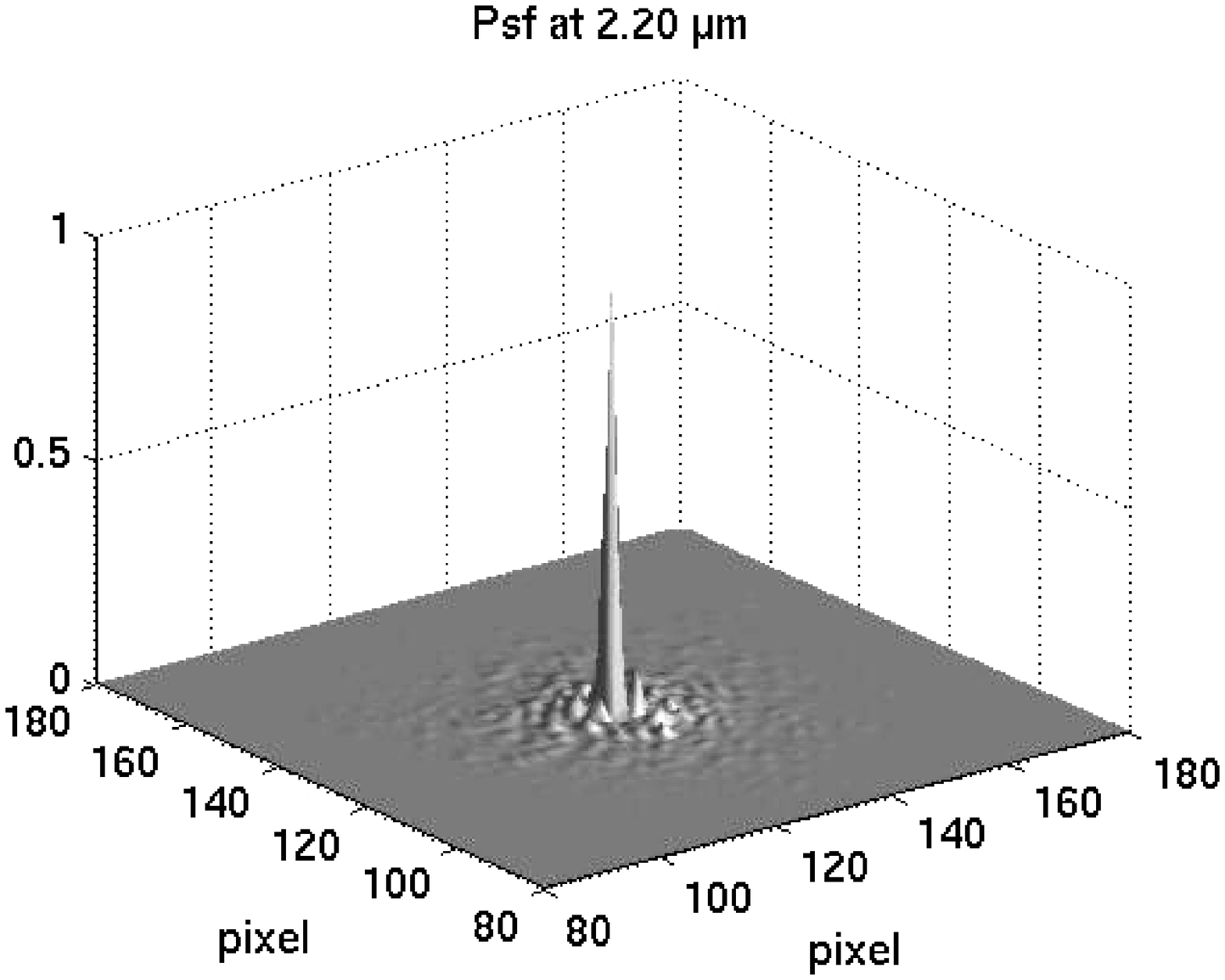}
\end{center}
\caption{Top: example of a simulated post-MACAO wavefront, with a Strehl ratio of 10\%, 30\%, and 50\% at 1.25~$\mu$m, 1.65~$\mu$m, and 2.2~$\mu$m, respectively. The phase value is indicated by the scale in radians. Bottom: associated normalized PSF.} 
\label{atmos}
\end{figure*}

A 60-elements wavefront sensor and a 60-actuators deformable mirror were then modeled, considering the correction of up to 48 Zernike modes with a simple pure integrator for time-filtering of the mirror commands. The performance obtained, in terms of resulting K-band Strehl ratio, are identical to the performance given for MACAO by \citet{arseno}. Moreover, and in order to have some statistics for our results, we made a series of ten different time evolution histories of 2\,s, starting from 10 different sets of random seeds (and hence 10 independent realizations) for both the turbulent atmosphere and the wavefront sensor noise. In the analysis below, we studied all ten histories and, for all purposes, the conclusions are independent of which exact file is used.

The AO system performance stabilizing after a few tens iterations, we considered for our resulting data the 630 last 3-ms wavefronts (and hence the data over the last 1.89\,s). Figure\,\ref{rms} shows the wavefront rms and the K-band Strehl ratio for the first 200 iterations of one simulated history. 

Concerning tip-tilt errors, they stay below 10~mas rms on sky over 2\,s for the UTs, corresponding to the performance range of the InfraRed Image Sensor (IRIS) of the VLTI \citep{gitton}. Figure\,\ref{atmos} shows an example of a typical wavefront and the associated normalized Point Spread Function (PSF) with a Strehl ratio of 10\%, 30\%, and 50\% at 1.25~$\mu$m, 1.65~$\mu$m, and 2.2~$\mu$m, respectively. 

\subsection{Computation of $C_{\rm\it eff}$ corrupted by the turbulence phase and correlation with the Strehl}
\begin{figure}
\begin{center}
\includegraphics[width=60mm,height=40mm]{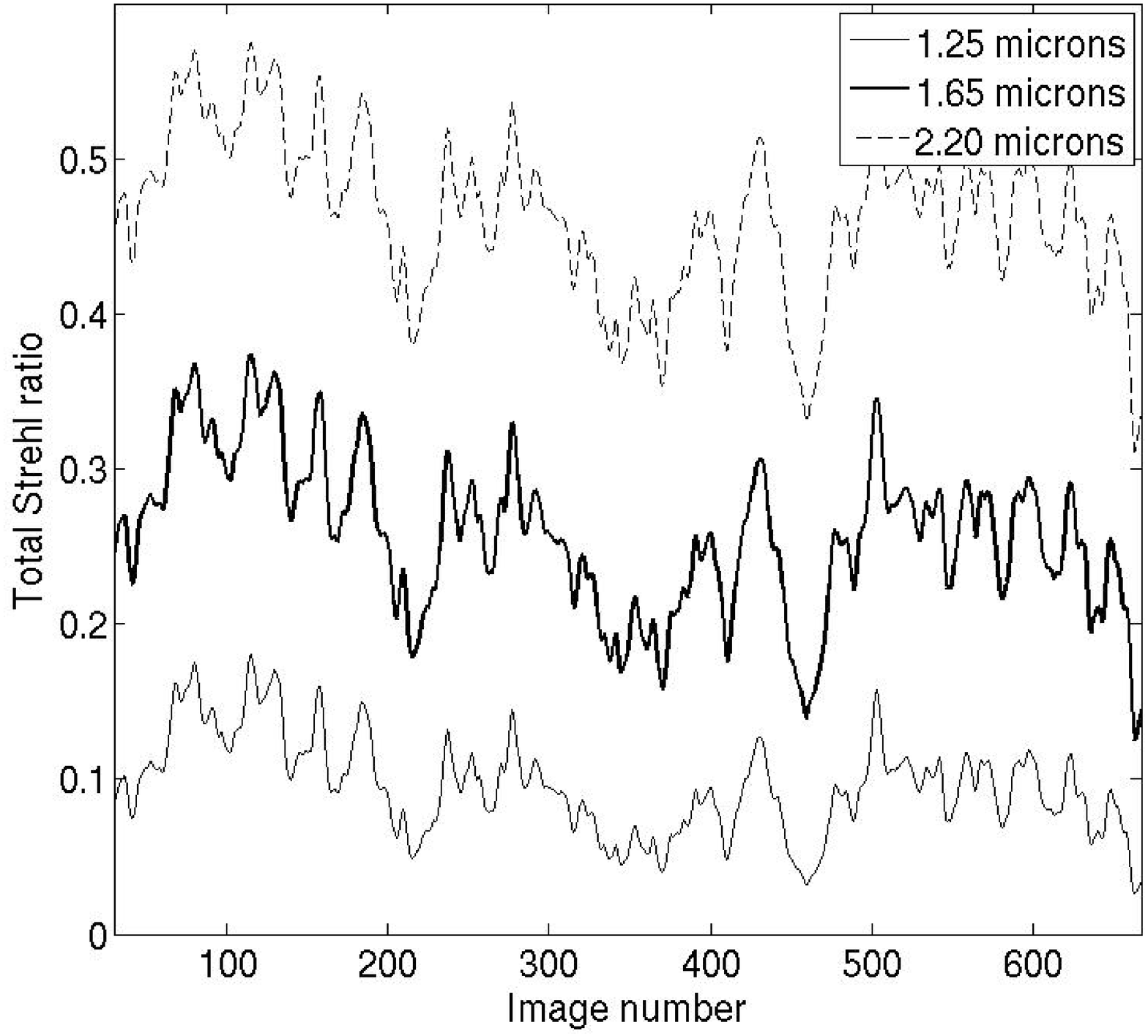} 
\includegraphics[width=60mm,height=40mm]{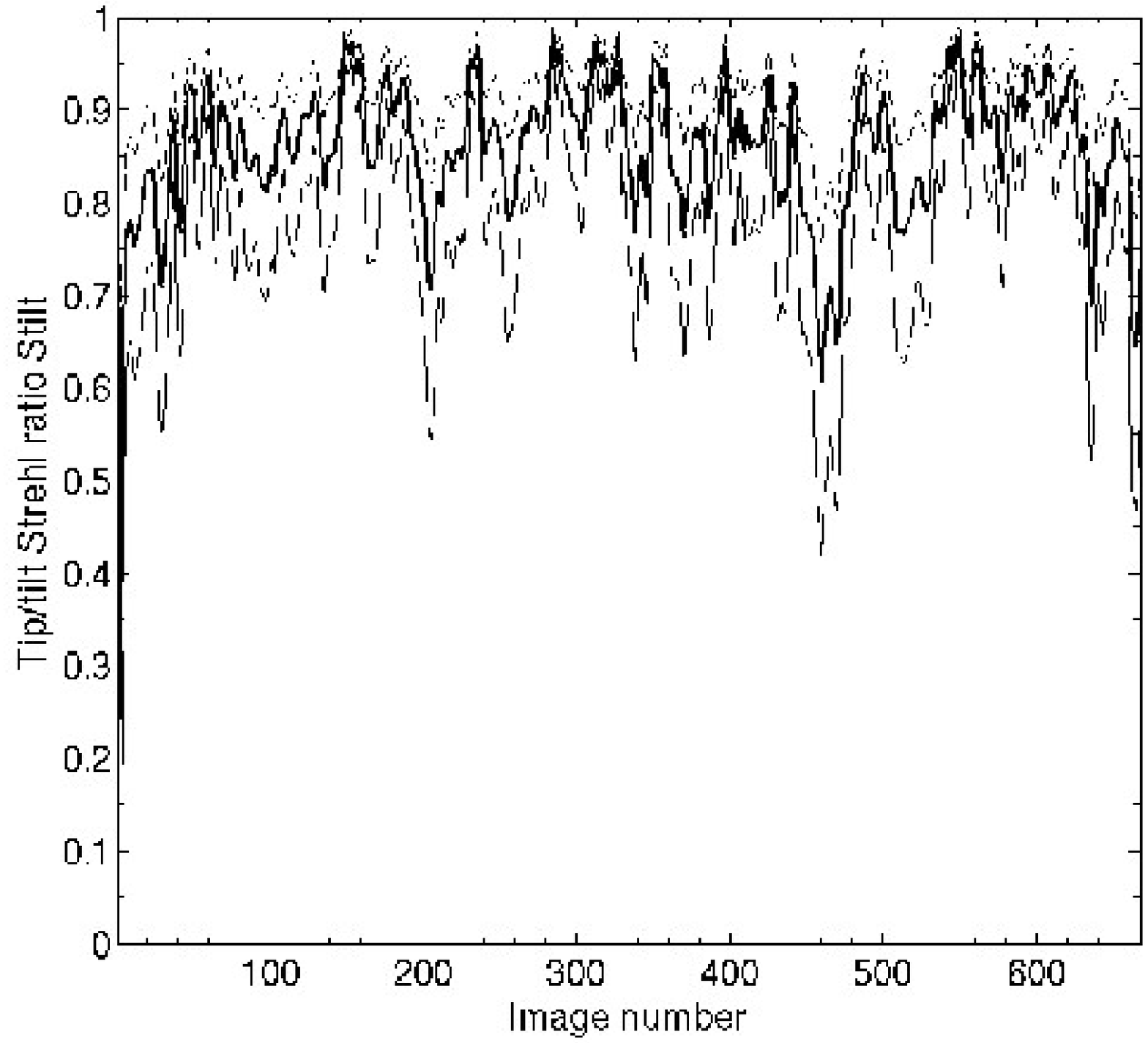}
\includegraphics[width=60mm,height=40mm]{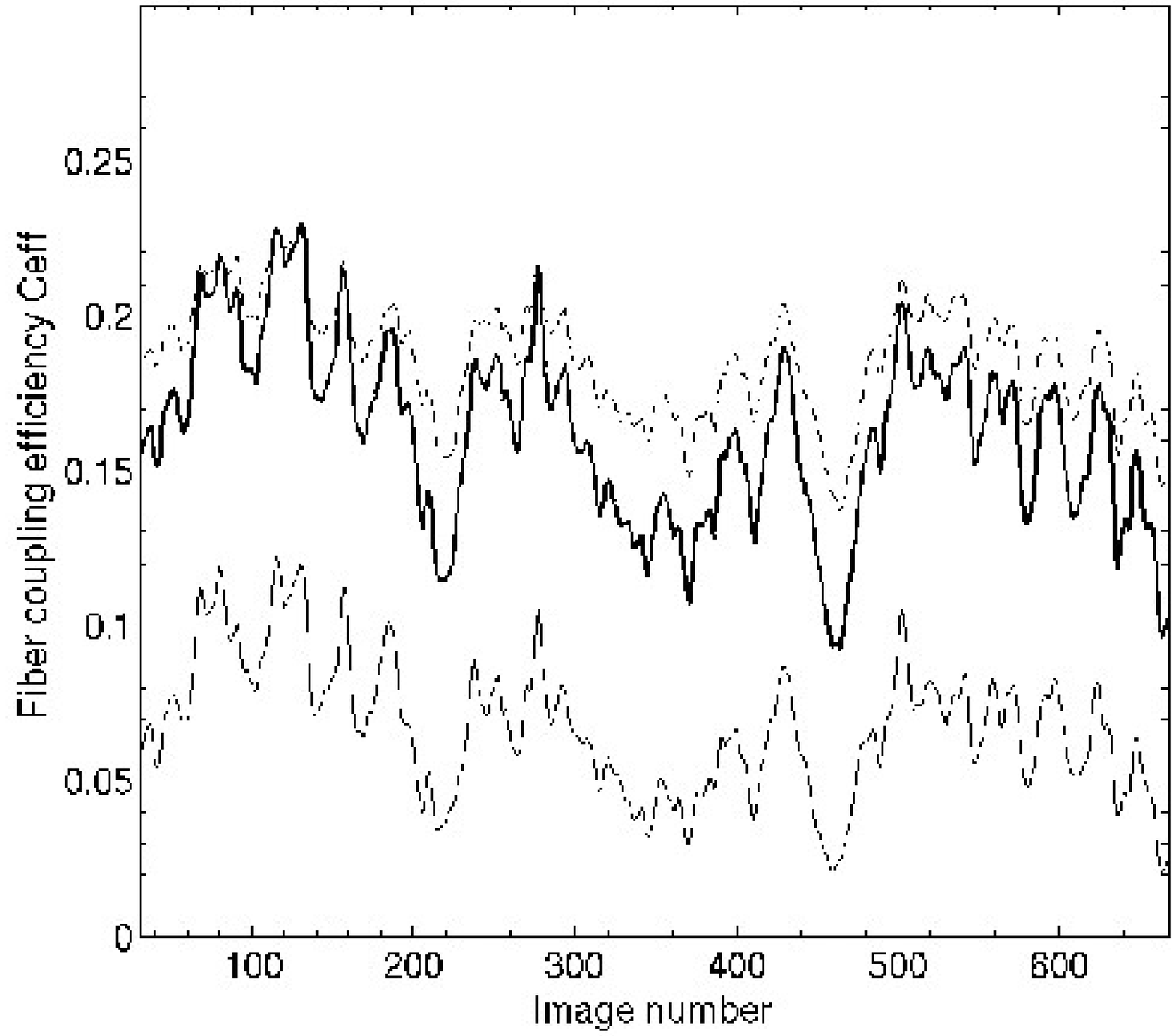}
\end{center}
\caption{One simulation of the atmosphere at Paranal for $z~=~0^\circ$. Evolution of $\rm\it SR$ (top) and of $S_{tilt}$ (middle). Bottom: Evolution of the coupling efficiency at 1.25 $\mu$m (solid line), 1.65 $\mu$m (thick-solid line), and 2.2 $\mu$m (dashed line).}
\label{atmos_ceff1}
\end{figure}

We computed the coupling efficiency by introducing the phase in Eq.\,\ref{rho}, and searched for correlations with the Strehl ratio variation and also with the first orders of aberration, i.e. the tip-tilt, for several values of the zenith angle $z$. Assuming that the tip-tilt and high-order aberrations are independent \citep{sandler}, the total Strehl ratio $\rm\it SR$ is then the product of the tip-tilt Strehl ratio $S_{tilt}$ with the higher-order aberrations Strehl ratio $S_{high}$. The simulated data providing $\rm\it SR$ and $S_{high}$ (see Fig.\,\ref{atmos_ceff1} at $z~=~0^\circ$), it is easy to deduce $S_{tilt}$, by simply computing:
\begin{eqnarray}
\lefteqn{\hspace{1cm} \rm\it SR = \frac{\rm\it PSF_{AO}[0,0]}{\rm\it PSF_{id}[0,0]}} \nonumber\\
\lefteqn{\hspace{1cm} \rm\it S_{high} = \frac{max(\rm\it PSF_{AO} )}{\rm\it PSF_{id}[0,0]}} \nonumber\\
\lefteqn{\hspace{1cm} \rm\it S_{tilt} = \frac{\rm\it SR}{\rm\it S_{high}}}
\label{Stt}
\end{eqnarray}
where $\rm\it{PSF_{AO}}$ is the PSF obtained from the post-MACAO wavefront, and $ \rm\it PSF_{id}$ that obtained from the ideal wavefront. The Strehl ratios are estimated for each 3-ms wavefront.

\begin{figure*}
\begin{center}
\includegraphics[width=50mm,height=40mm]{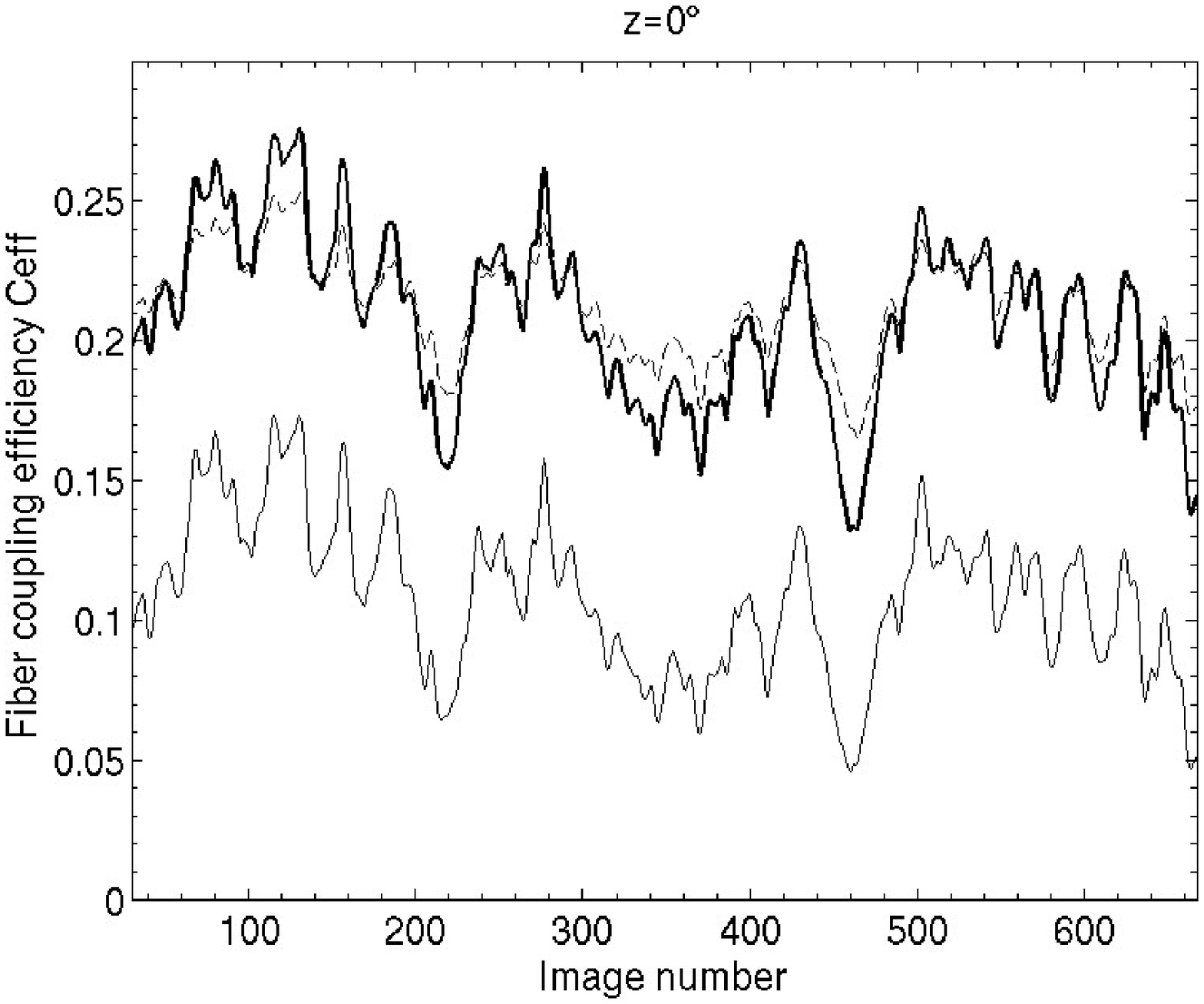}
\includegraphics[width=50mm,height=40mm]{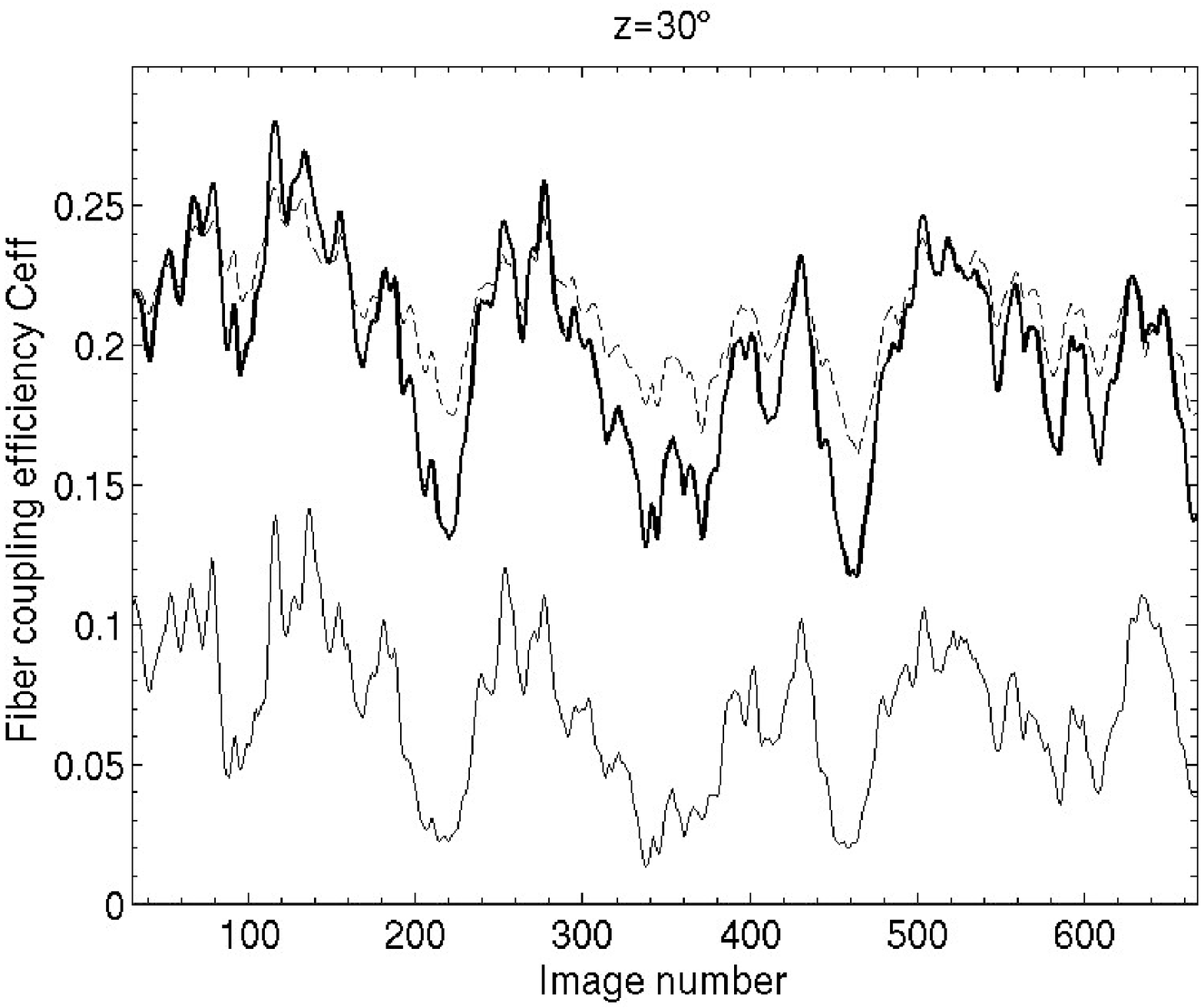}
\includegraphics[width=50mm,height=40mm]{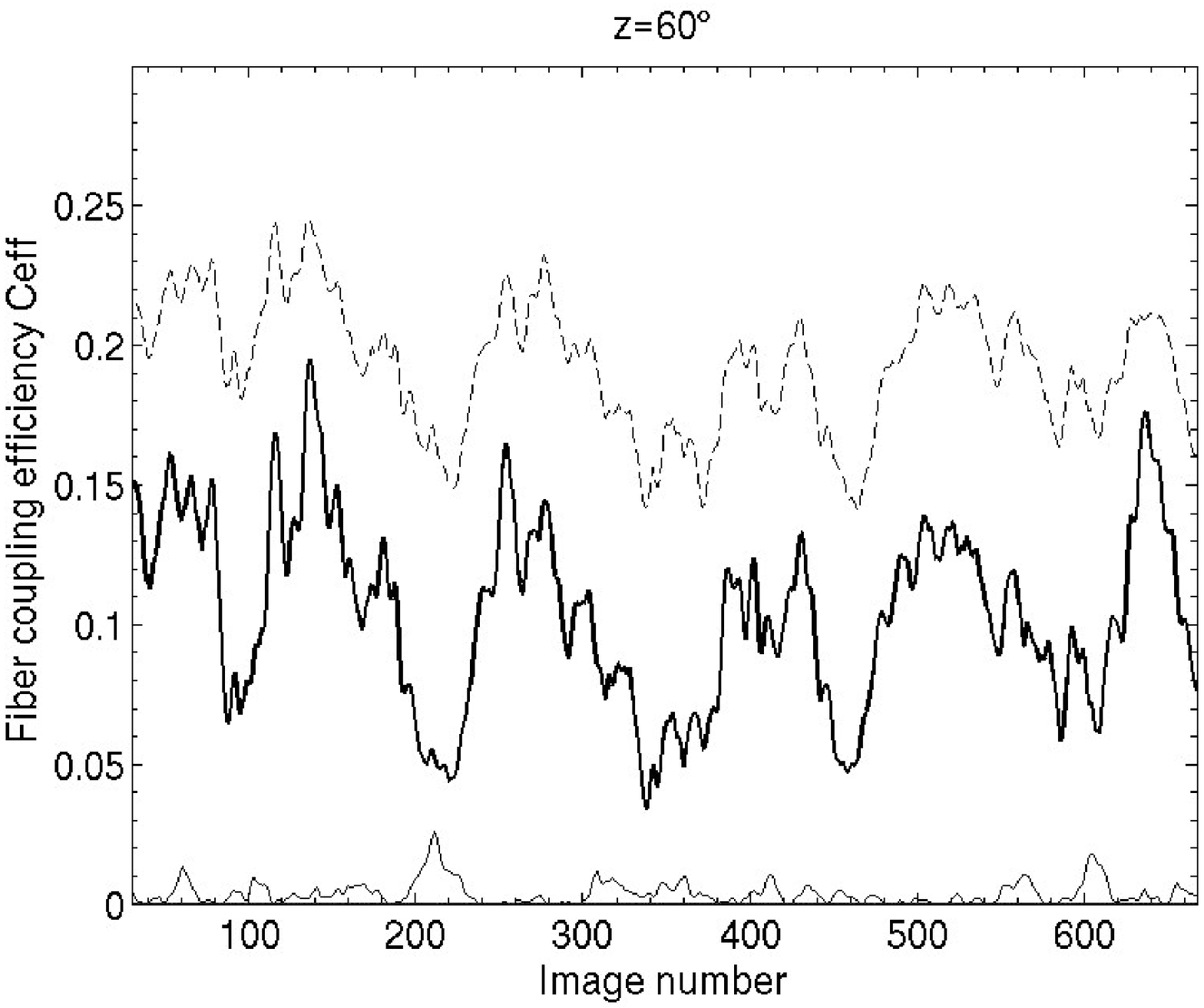}
\end{center}
\begin{center}
\includegraphics[width=50mm,height=40mm]{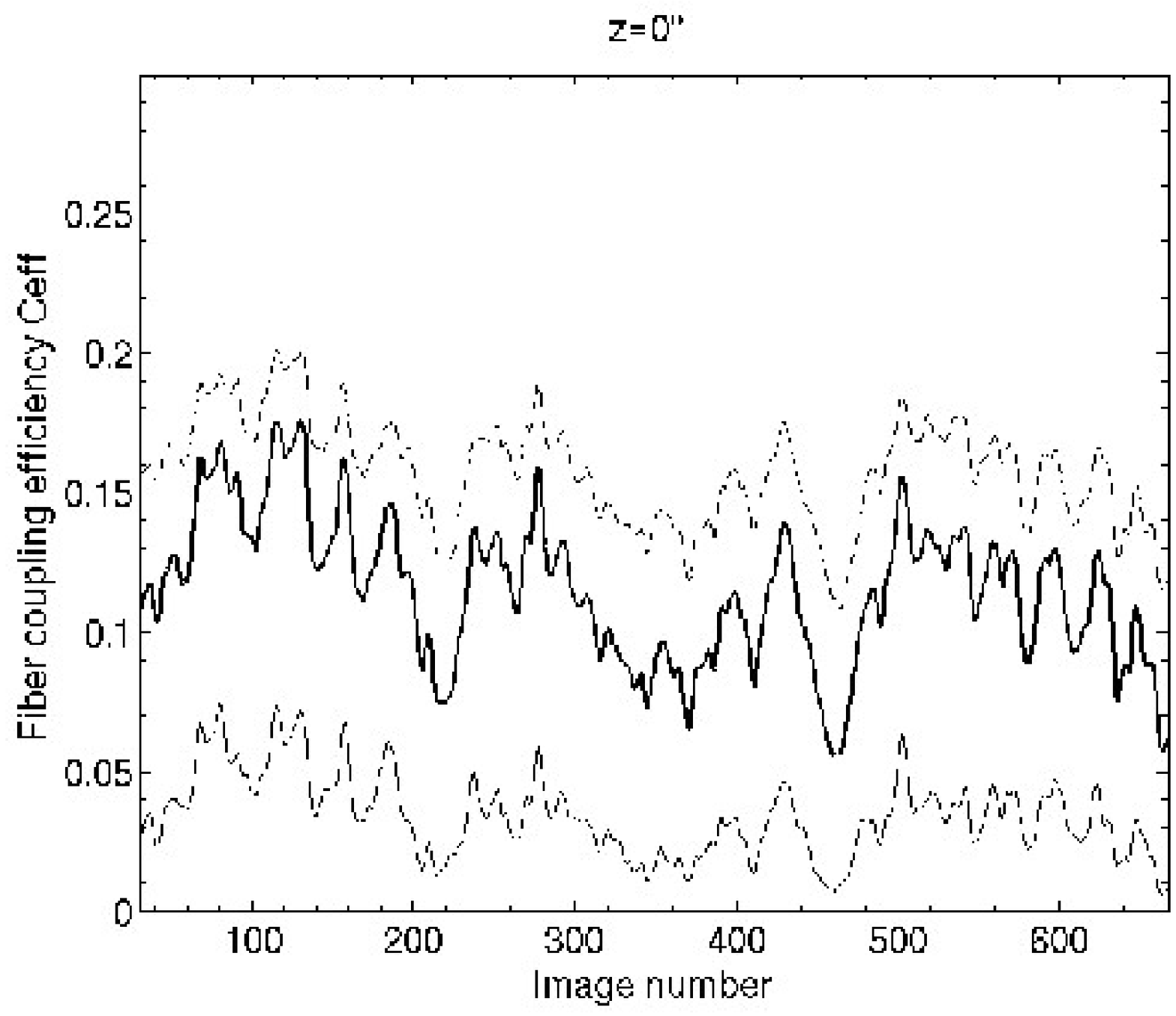}
\includegraphics[width=50mm,height=40mm]{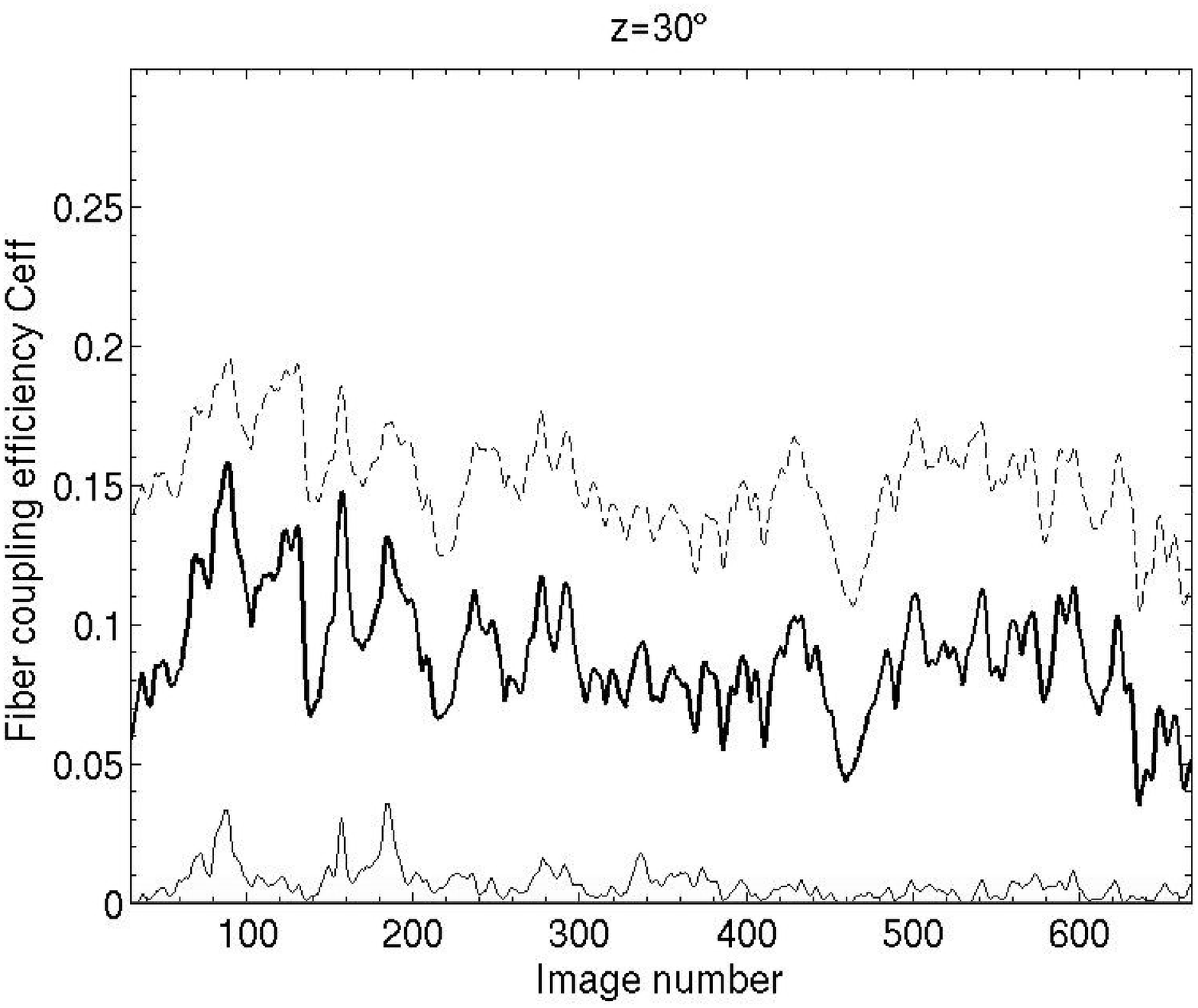}
\includegraphics[width=50mm,height=40mm]{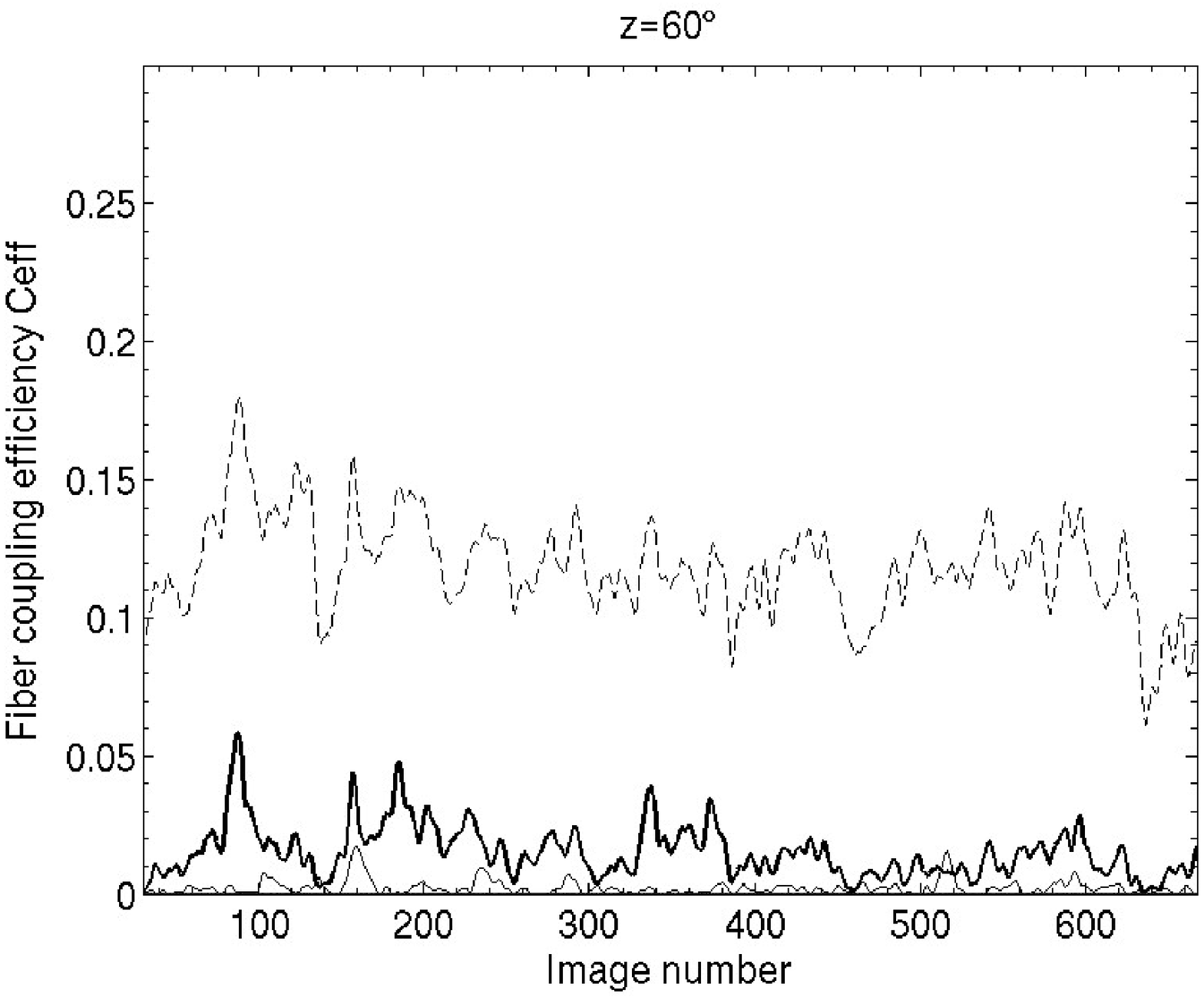}
\end{center}
\caption{Top: evolution of the coupling efficiency at the maximal wavelength of the three spectral bands at a zenith angle of 0$^\circ$, 30$^\circ$ and 60$^\circ$. Maximal wavelengths: 1.4~$\mu$m (solid line), 1.82~$\mu$m (thick solid line), and 2.4~$\mu$m (dashed line). Bottom: evolution of the coupling efficiency at the minimal wavelength of the three spectral bands at a zenith angle of 0$^\circ$, 30$^\circ$ and 60$^\circ$. Minimal wavelengths: 1.1~$\mu$m (solid line), 1.48~$\mu$m (thick solid line), and 2.0~$\mu$m (dashed line).}
\label{atmos_ceff2}
\end{figure*}

The evolution of $C_{\rm\it eff}$ for three values of the zenith angle $z$, and at the maximal (upper) and the minimal (lower) wavelength of the three spectral bands, is shown on Fig.\,\ref{atmos_ceff2}. The conclusions are:

- As expected, the coupling efficiency in each wavelength is the product of the maximal coupling by the Strehl ratio. For example, a Strehl ratio of 50\% at 2.4~$\mu$m gives a coupling efficiency of 0.20.

- The photometry in the AMBER instrument is about the same in the H-band and in the K-band, a lower Strehl ratio in H than in K compensating for the higher maximal coupling efficiency due to the presently used fibers.

- The conjugated effects of the atmospheric aberrations and refraction is the highest at the shortest wavelengths: J-band is very affected. 

In order to quantify the dependency between the fiber coupling and the $\rm\it SR$, we computed the correlation coefficients between these two parameters. We studied the whole J- and H-bands as the effects are stronger than in K-band. We also calculated the correlation coefficients between the fiber coupling and $S_{high}$. From the results shown in Table\,\ref{correl}, we can deduce that:

- The correlation between the fiber coupling and $\rm\it SR$ decreases while $z$ increases. The atmospheric refraction is predominant.

- Aberrations of higher orders than tip-tilt are predominant in the effect of the atmospheric turbulence on the fiber coupling, at least as far as the tip-tilt stays in the IRIS performance.

\begin{table}
\caption{Correlation coefficients between the fiber coupling and $\rm\it SR$, $S_{high}$, and $S_{tilt}$, in J- and H-bands, and for three different zenith angle values.}
	\begin{center}
		\begin{tabular}{c c c c } \hline
\multicolumn{4}{l} {Correlation with $\rm\it SR$}	\\\hline
\scriptsize  &\scriptsize $z$ = 0$^\circ$ &\scriptsize $z$ = 30$^\circ$ &\scriptsize	$z$ = 60$^\circ$ 	\\
\scriptsize J-band &\scriptsize 0.984 $\pm$ 0.003 &\scriptsize 0.64 $\pm$ 0.06 &\scriptsize	0.20 $\pm$ 0.03\\
\scriptsize H-band &\scriptsize 0.947 $\pm$ 0.004 &\scriptsize 0.82 $\pm$ 0.04 &\scriptsize	0.50 $\pm$ 0.07\\\hline
\multicolumn{4}{l} {Correlation with $S_{high}$} 	\\\hline
\scriptsize  &\scriptsize $z$ = 0$^\circ$ &\scriptsize $z$ = 30$^\circ$ &\scriptsize	$z$ = 60$^\circ$ 	\\
\scriptsize J-band &\scriptsize 0.835 $\pm$ 0.001 &\scriptsize 0.57 $\pm$ 0.04 &\scriptsize	0.20 $\pm$ 0.04\\
\scriptsize H-band &\scriptsize 0.813 $\pm$ 0.001 &\scriptsize 0.71 $\pm$ 0.04 &\scriptsize	0.45 $\pm$ 0.05\\\hline
\multicolumn{4}{l} {Correlation with $S_{tilt}$} 	\\\hline
\scriptsize  &\scriptsize $z$ = 0$^\circ$ &\scriptsize $z$ = 30$^\circ$ &\scriptsize	$z$ = 60$^\circ$ 	\\
\scriptsize J-band &\scriptsize -0.403 $\pm$ 0.006 &\scriptsize -0.21 $\pm$ 0.03 &\scriptsize	-0.01 $\pm$ 0.03\\
\scriptsize H-band &\scriptsize -0.383 $\pm$ 0.001 &\scriptsize -0.33 $\pm$ 0.02 &\scriptsize	-0.16 $\pm$ 0.03\\\hline
		\end{tabular}
		\end{center}
	\label{correl}
\end{table}

\begin{figure}
\begin{center}
\includegraphics[width=50mm,height=40mm]{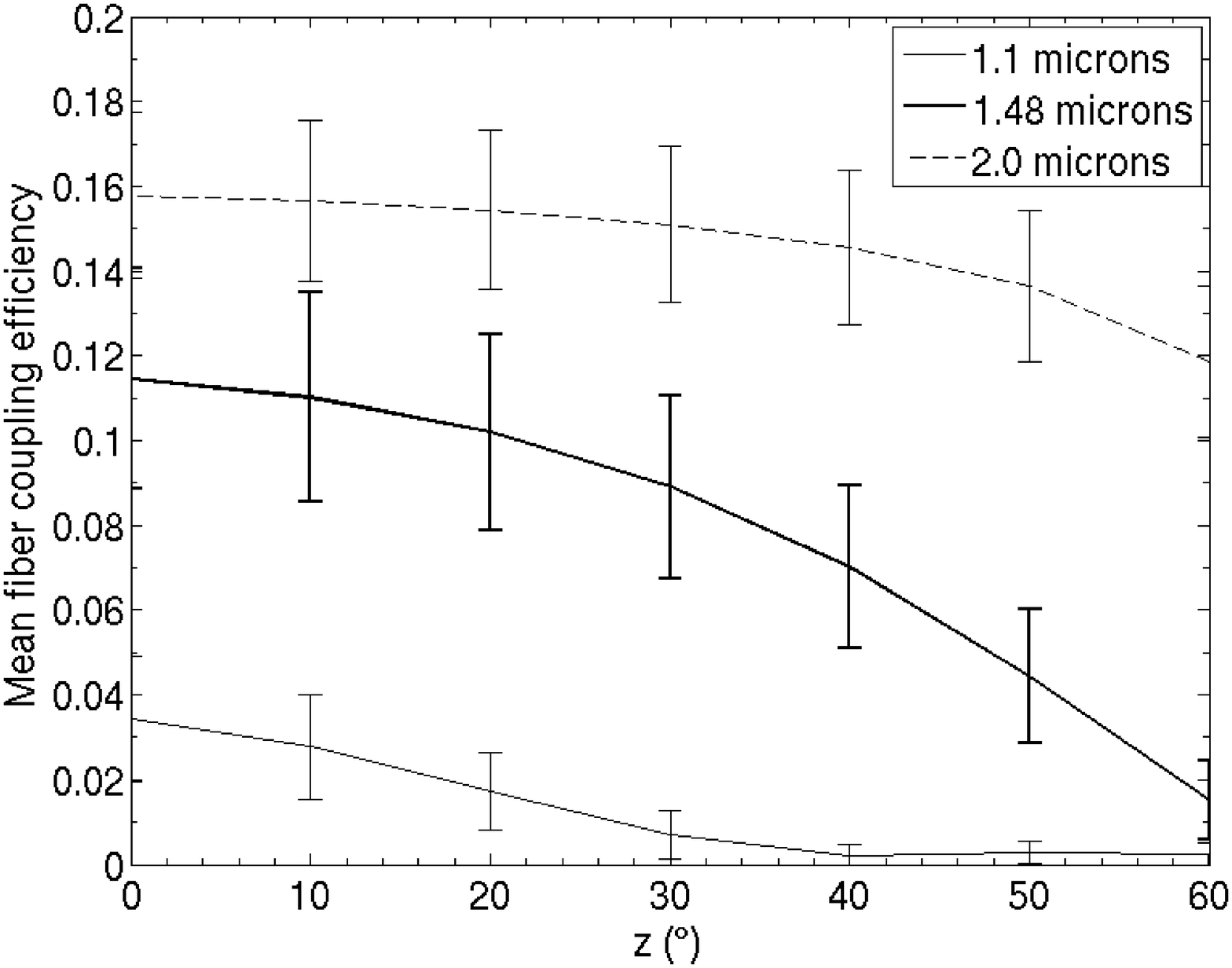}
\includegraphics[width=50mm,height=40mm]{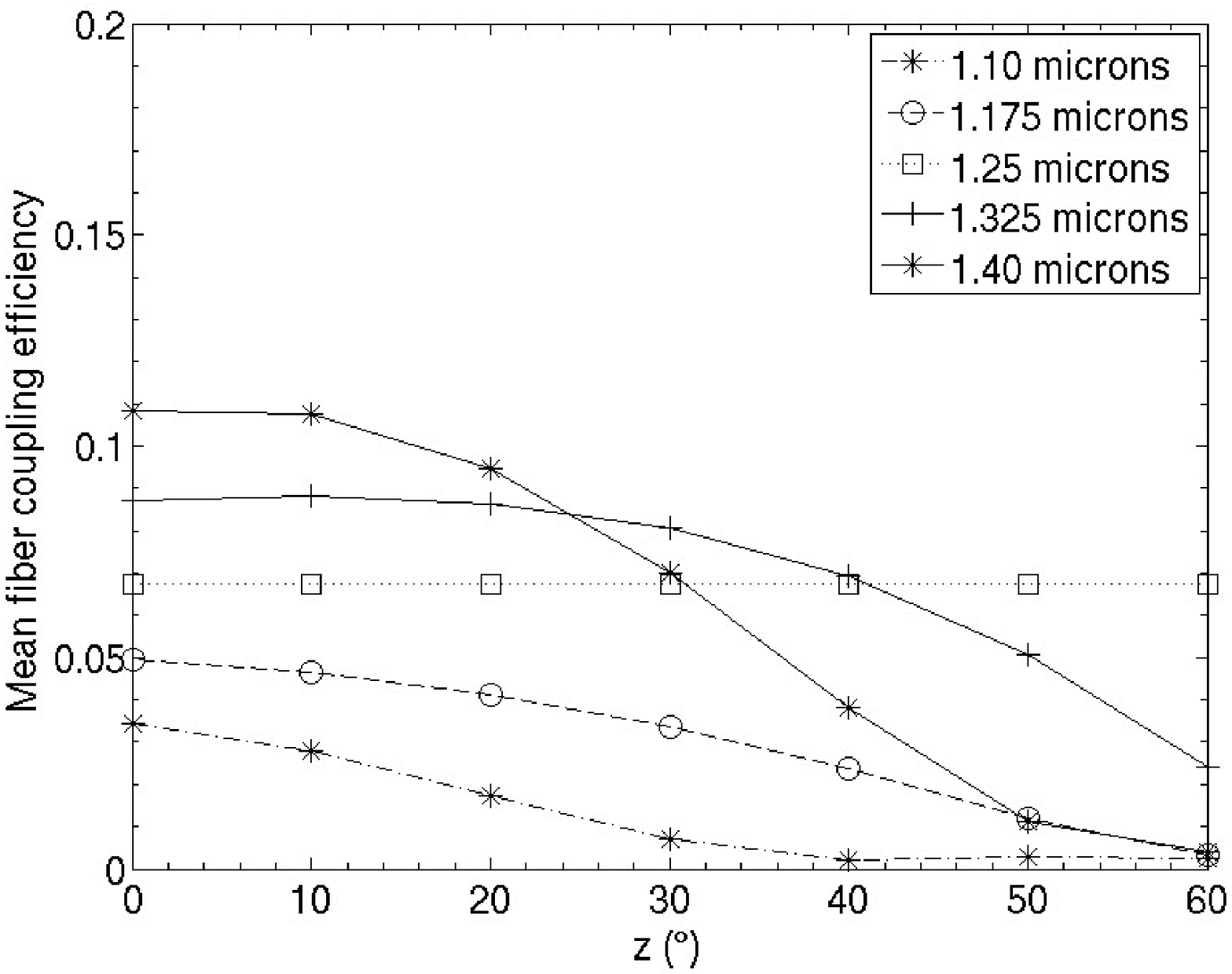}
\includegraphics[width=50mm,height=40mm]{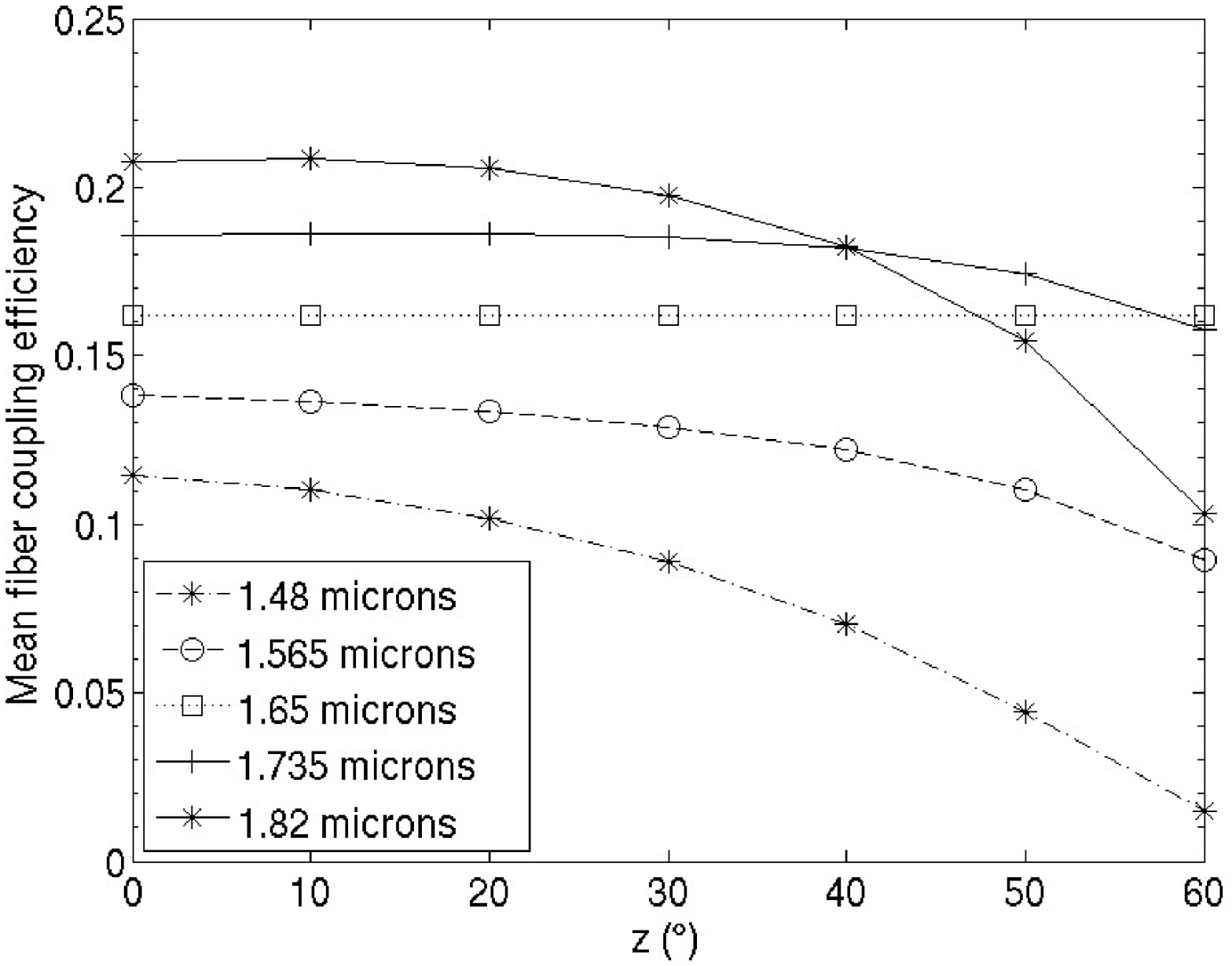}
\end{center}
\caption{Mean coupling efficiency as a function of the zenith angle (top). Mean fiber coupling efficiency as a function of $z$ for five wavelengths equally sampled in J-band (middle), and in H-band (bottom).}
\label{atmos2}
\end{figure}

Maximum AMBER time exposures in low spectral resolution mode are 25 to 100 ms. For high spectral resolution, for which integration times are longer, the VLTI is optimized at the central wavelength of the observed spectral band, and the dispersion effect is not so significant. Nevertheless, in the case of an instrument observing with exposures larger than 1~s, we plotted the mean coupling efficiency as a function of the zenith angle. Figure\,\ref{atmos2} shows the results at the minimal wavelengths of J-, H-, and K-bands where the effect is the most critical. We note that 50\% of the coupling efficiency is lost for $z$ larger than 45$^\circ$ at 1.48~$\mu$m, and for $z$ larger than 22$^\circ$ at 1.1~$\mu$m. 
We then plotted the mean fiber coupling efficiency as a function of $z$ for five wavelengths equally sampled in J-band, and in H-band. We would almost loose half of J-band flux if no atmospheric refraction correction was performed. As stated in the paragraphs above, let us note the decrease of $C_{\rm\it eff}$ with wavelength, at given values of $z$ lower than 30$^\circ$ for which the atmospheric turbulence is predominant. The predominance of the refraction is then established for larger values of $z$ for which $C_{\rm\it eff}$ decreases with wavelengths deviating from the central wavelength.

\subsection{Taking into account the dependence on $z$ of $\rm\it SR$}
\begin{figure}
\begin{center}
\includegraphics[width=50mm,height=40mm]{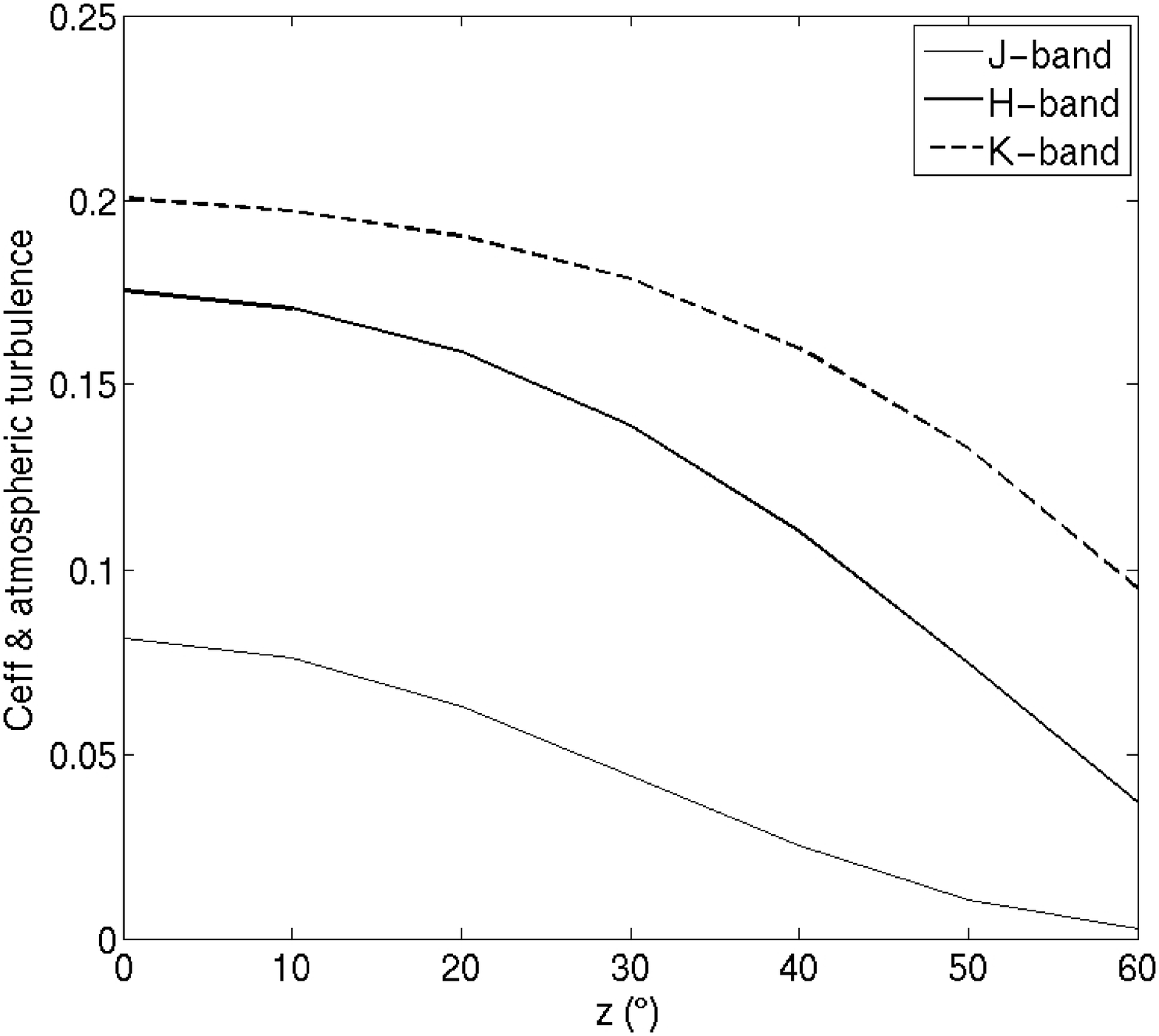}
\includegraphics[width=50mm,height=40mm]{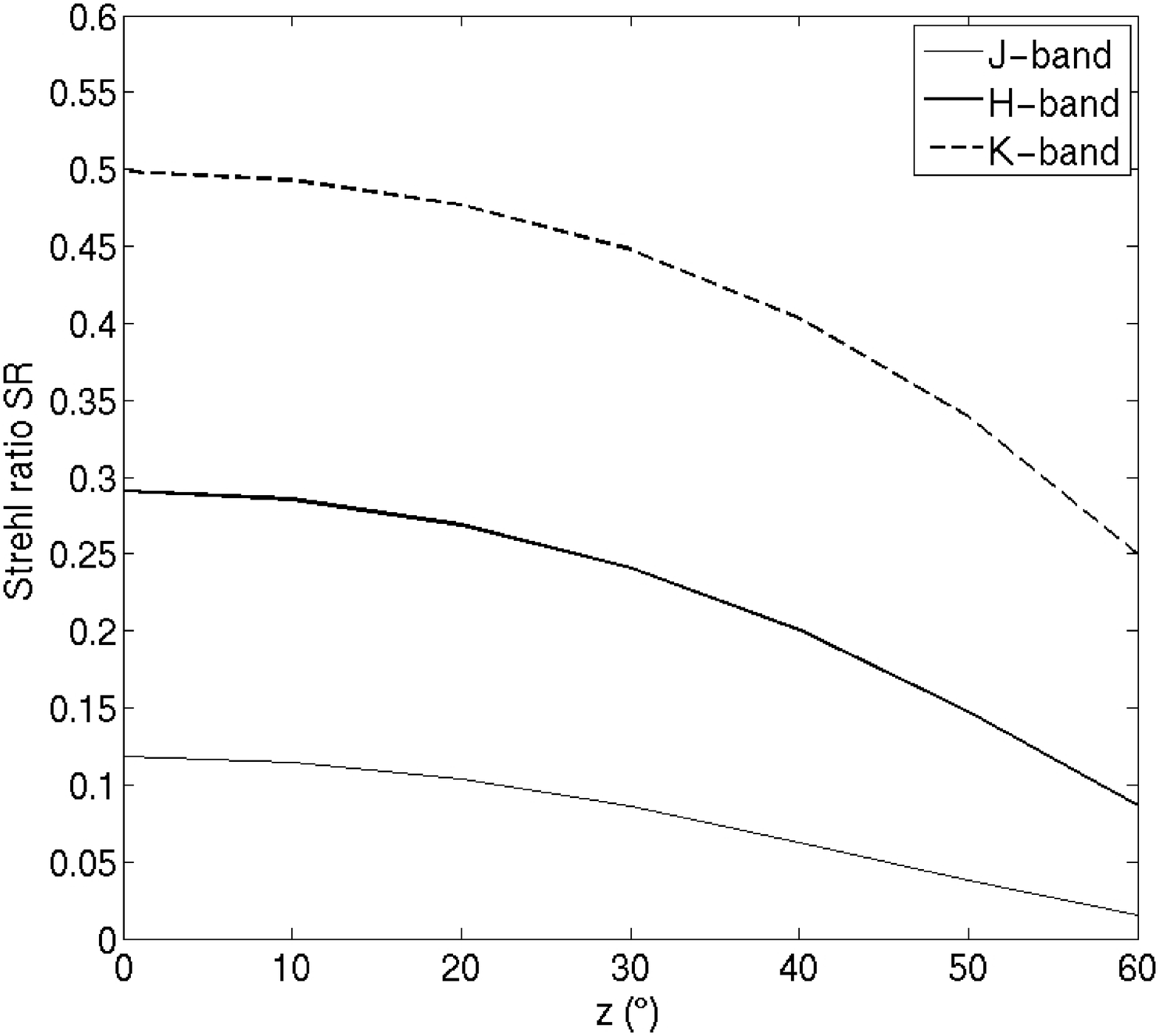}
\end{center}
\caption{Evolution of the coupling efficiency (top) as a function of $z$, taking into account the evolution of $\rm\it SR$ (bottom) and the atmospheric refraction. Averages are computed over the whole individual bandwidths. The origin of the refraction angle is taken at the central wavelength of each spectral band.}
\label{SR-z}
\end{figure}

Up to this Section, we supposed a constant $\rm\it SR$ whatever the zenith angle $z$ was. The evolution of $C_{\rm\it eff}$ versus $z$ was estimated only considering the effect of the atmospheric refraction. In fact, the Fried parameter $r_0$ depends on $z$, as {$\rm\bf (cosz)^{(3/5)}$} \citep{roddier}, so does the Strehl, as {$\rm\bf e^{-0.134(D/r_0)^{5/3}}$} \citep{noll}. We here computed the evolution of $\rm\it SR$ as a function of $z$, taken the initial value at $0^\circ$ of 50\% at 2.2 $\mu$m. The results were similar, within 1\%, to those given by ESO in \citet{puech}. We then considered the degradation of $C_{\rm\it eff}$ due to the atmospheric refraction and plotted the complete coupling efficiency versus $z$, taking into account both evolutions, of the refraction and of the Strehl. Figure\,\ref{SR-z} shows this evolution in the three spectral bands, while $C_{\rm\it eff}$ is the average value over the whole individual bandwiths. The origin of the refraction angle is here again taken at the central wavelength of each spectral band, as in Sect.\,\ref{231} assuming observations in individual bandwidths.

Before doing this study, we could suspect the Strehl degradation to be so important for large $z$ that the fiber injection would be lost. In this case, a correction of the refraction would be useless. But Figure\,\ref{SR-z} shows that this is not the case and that the refraction correction by an optical system stays relevant in J- and H-bands in spite of the Strehl degradation.

Having demonstrated the importance to correct for the atmospheric refraction in J- and H-bands while observing with UTs, we describe, in the next sections, the correction system installed on the AMBER instrument.

\section{The atmospheric refraction controller: opto-mechanical description} 
\label{4}
This section describes the Atmospheric transverse Dispersion Controller (ADC) specifically studied for the instrument AMBER. Sect.\,\ref{41} gives the characteristics of the optical elements. Sect.\,\ref{42} presents the results of the tolerance analysis. This analysis, described in the Appendix~A, takes into account errors on the prism angle due to manufacturing and to cementing operation, errors on the ADC rotation axis relative to the AMBER optical axis, and mechanical control of the optical elements. The ray-tracing Software ZEMAX is then used to analyze additional parameters, as surface irregularities and positioning errors of the optics and axes (see Sect.\,\ref{43}). The final expected performance of the VLTI in terms of fiber coupling efficiency is given in Sect.\,\ref{44}. Sec.\,\ref{45} gives a short view of the mechanics of the ADC.

\subsection{Characteristics of the optics}
\label{41}
\begin{figure*}
\begin{center}
\includegraphics[width=120mm]{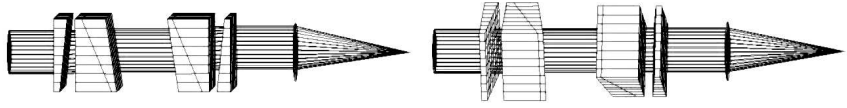}
\end{center}
\caption{Relative axial rotation of one ADC system relative to the other one at z=0$^\circ$ and at z=60$^\circ$.}
\label{adc}
\end{figure*}

The system has to reduce the dispersion (making almost null the image spread) while maintaining the overall compact Airy disk at the fiber entrance. To optimize the coupling efficiency, the resulting optimized system is composed of 
two sets of 3 prisms (see Fig.\,\ref{adc}). The two sets, rotating with respect to each other, are inserted in each interferometric arm prior to the J- and H-spatial filters. Each system is composed of a BK7 wedge and a pair of wedges made respectively in F2 and SF14 and cemented together.  

The transmission of the materials for the 6 prisms is 0.88 in J-band and 0.86 in H-band. Each element has a central thickness of about 5 mm (for information, the transmission in K-band is 0.62 $-$ another choice of materials would likely be more efficient if the system had to be used in K-band as well).

The wedges were manufactured, and the cementing carried out, by SEOP (Sud Est Optique de Pr\'ecision, Lorgues, France). 
The prism angles are respectively: 6.285$^\circ$ for the BK7, 22.500$^\circ$ for the F2, and 14.005$^\circ$ for the SF14. The angle accuracy is $\pm$0.01$^\circ$.

The surface flatness is 73~nm peak-to-valley PV for the first prism and 70~nm PV for the two others. The resulting wavefront optical quality at the exit of a system is less than 90~nm PV on a surface of 23~mm.

The theoretical parallelism of the edges after the doublet cementing is 0.5$^\circ$.

The three ADC systems located in the three interferometric beams have the same thickness with a $\pm$20~$\mu$m accuracy in order to respect the requirement on the chromatic optical path difference \citep{robbe}.

Let us analyze how the performance of the instrument is affected by the non-ideal manufacturing of the optical pieces, mainly the prism angles, the cementing, and the positioning errors.

\subsection{Manufacturing errors}
\label{42}
\begin{figure}
\begin{center}
\includegraphics[width=70mm]{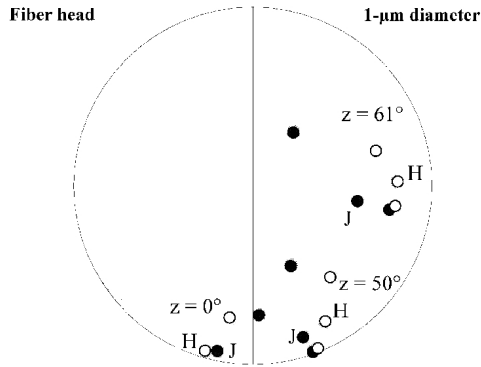}
\end{center}
\caption{Combined optical and positioning errors of the ADC. Three values of the zenith angle $z$ are considered and the following errors are taken into account: prism angles manufacturing, error on the prism cementing, sensitivity of the axial rotation of the two cemented prisms respective to the first one, tip-tilt of this rotation axis, tip-tilt of the 3-prisms assembly, sensitivity of the 360$^\circ$-rotation of the two ADC prism system, inclination induced during this rotation. The spots represent the dispersion of the beam from 1.1~$\mu$m (dark spots for the J-band) to 1.82~$\mu$m (clear spots for the H-band). The worse expected coupling degradation factor is 0.85 in J- and H-bands.}
\label{adc3}
\end{figure}

Combining the manufacturing residual errors leads to the result shown on Figure\,\ref{adc3}. Appendix~A shows that the maximum image shift due to combined optical manufacturing and positioning errors is less than 1~$\mu$m on the entrance fiber heads. The resulting coupling efficiency is not affected by more than 15\% in J- and H-bands, compliant with the specifications given in Sect.\,\ref{23}. 

\subsection{Tolerance analysis of the instrument from the pupil to fiber entrance}
\label{43}
ZEMAX is used to define a tolerance on all the elements and surfaces. 
As this software also models atmospheric refraction and fiber coupling \citep{wagner}, we could compute directly some ZEMAX coupling coefficients in order to compare them to Figure\,\ref{fig_rho} results. The coupling values are in accordance.
ZEMAX gives an estimation of the fiber coupling efficiency and performs the tolerance analysis of the elements located before the fiber entrance such that the coupling stays very close to a maximum value previously defined. This defines a Merit Function (MF).
The parameters considered here are: surface peak to valley irregularities, transverse positioning errors of the optics and residual inclinations relative to the optical axis. They are complementary to those of the previous Section.

\begin{figure}
\begin{center}
\includegraphics[width=84mm]{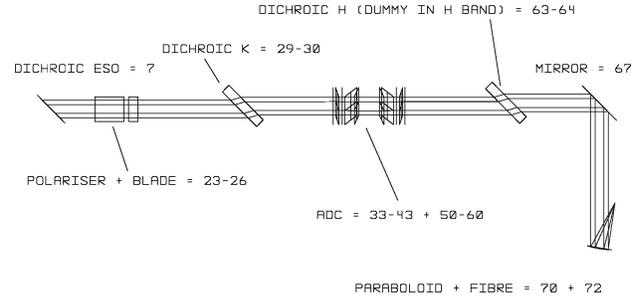}
\end{center}
\caption{Optical set-up for the VLTI injection tolerance analysis (J- and H-bands).}
\label{zem}
\end{figure}

Before entering the spatial filters, the VLTI beam travels towards the fiber heads through a calcite polarizer (to control the polarization direction parallel to the fiber neutral axis), a wedge (to compensate the transverse shift and residual inclination of the wavefront at the polarizer exit), one dichroic for H (reflection of K and transmission of H), two dichroics for J (reflection of K and H and transmission of J) and the ADC. The injection of light in the fiber is performed with an off-axis mirror. The focal length $F$ is related to the fiber numerical aperture $\rm\it NA$ and to the beam diameter $d$ at the entrance of the instrument. The best coupling efficiency is given by ZEMAX for $\rm\it NA.F/d=$0.46, taking into account the telescope obstruction and the spider arms. 
The analyzed optical set-up is described by Figure\,\ref{zem}, with increasing numbers defining in ZEMAX the surfaces from the entrance optics to the fiber head. 

For this study, the central obstruction of the UT is considered through the $k$ factor of the obstruction over the entrance pupil, and a pupil spider arm width of about 0.03 over the entrance pupil diameter. The unperfect VLTI optical quality is also taken into account.
The results provided by ZEMAX are a respective final degradation factor of the coupling efficiency of 93\% in H-band and 86\% in J-band. These values are taken into account in the instrumental throughput budget.


\subsection{VLTI expected fiber coupling with AMBER equipped with the ADC}
\label{44}
From the previous analysis, we can estimate that the global VLTI expected fiber coupling, observing with the instrument AMBER equipped with the ADC, is 48\% in J-band and 45\% in H-band, taking into account:

- The maximal coupling of the fibers (VLTI obstruction included): 69\% in J-band and 61\% in H-band, with the current optical fibers (Figure\,\ref{fig_rho}), and in the single-bandwidth observation mode.

- The coupling degradation error due to the manufacturing errors of the optical pieces: 85\% (Sect.\,\ref{42}).

- The coupling degradation error due to the positioning of the elements, and to the surface quality (VLTI included): 93\% in H-band and 86\% in J-band (Sect.\,\ref{43}).

- Additional coupling degradation factor due to residual errors on the relative control between the fiber and the injected light: 95\%\citep{robbe}.

These values lead to the expected global throughput \citep{robbe} meeting the specifications of AMBER to observe stars with magnitude up to 11 in the J- and H-bands in low resolution mode.

\subsection{Short view of the mechanical system}
\label{45}
\begin{figure}
\begin{center}
\includegraphics[width=50mm]{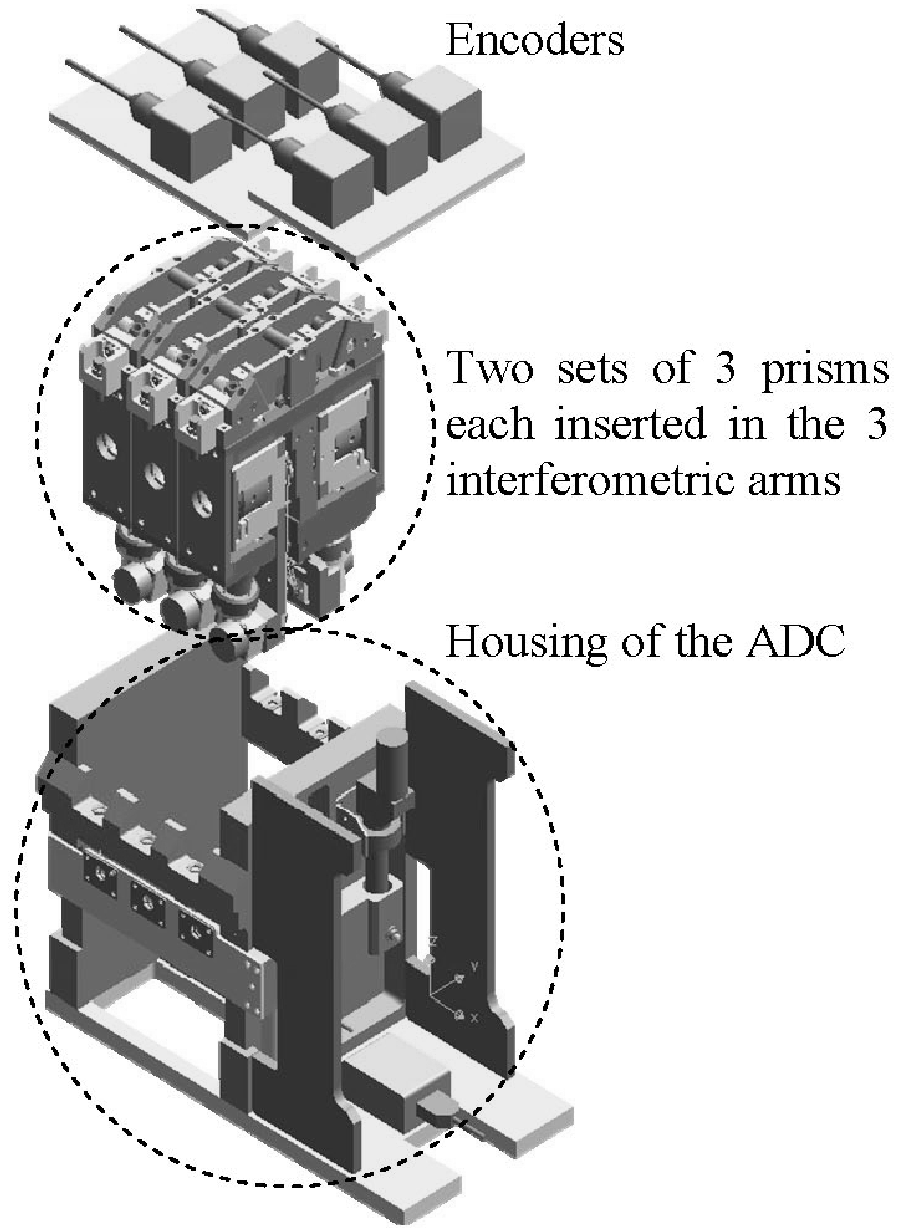}
\includegraphics[width=50mm]{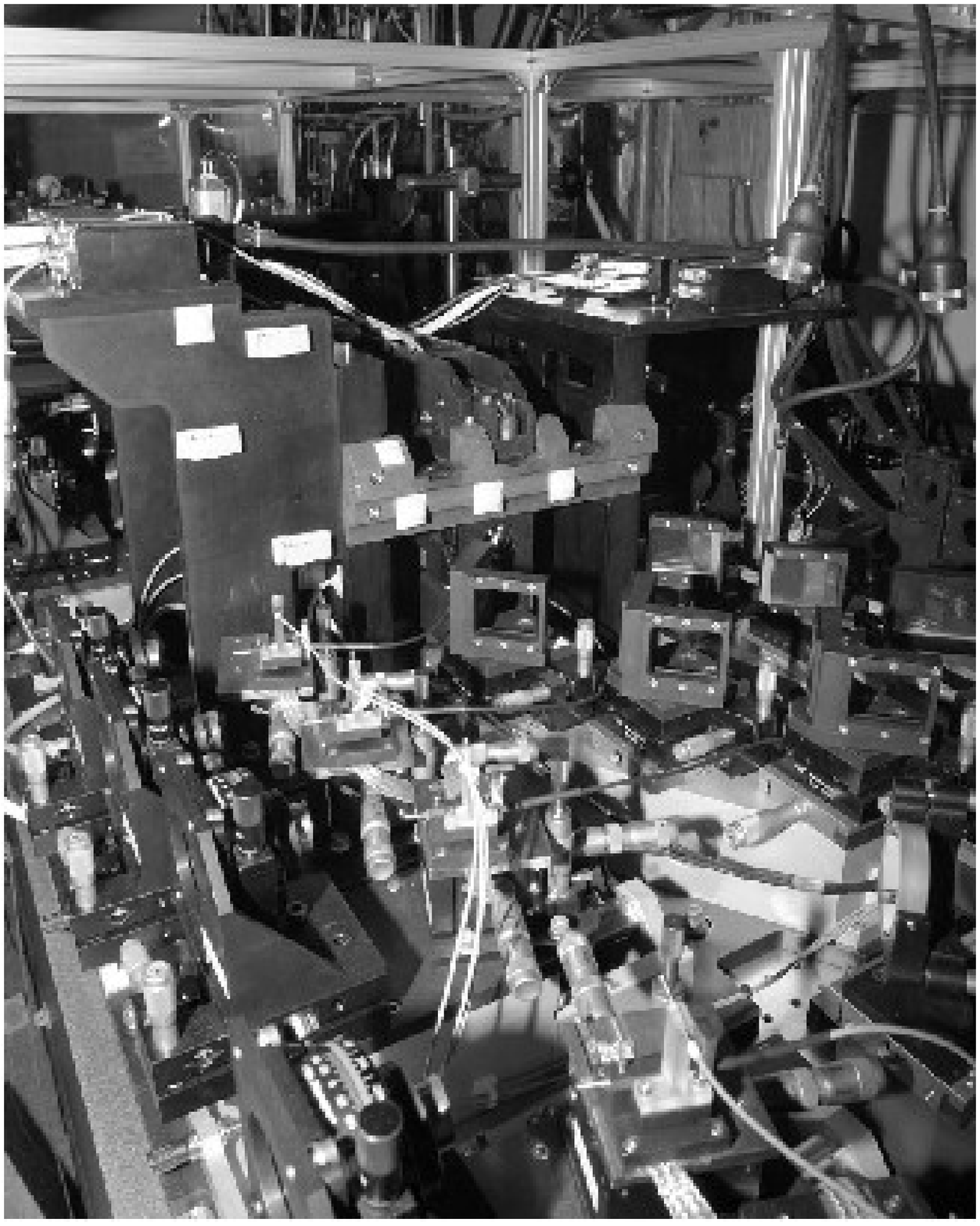}
\end{center}
\caption{Mechanical plan of the ADC housing (top) and picture of the ADC located before the J and H spatial filters on the AMBER instrument at Paranal, Chile (bottom).}
\label{cath}
\end{figure}

One mechanical assembly, with the 2 sets of 3 prisms (one single prism and one composite prism made of two prisms), is inserted in each individual interferometric arm. The three assemblies are installed in an unique structure (housing of the ADC) as shown in Figure\,\ref{cath}. 

Let us now describe the ADC monitoring and validate the performance.

\section{ADC rotation monitoring}
\label{5}
This paragraph describes the operation of the ADC. Observations with UTs and with ATs are considered, even if the correction is not so essential with the latter.
\subsection{Calculation of the rotation angle}\label{51}
\begin{figure}
\begin{center}
\includegraphics[width=80mm]{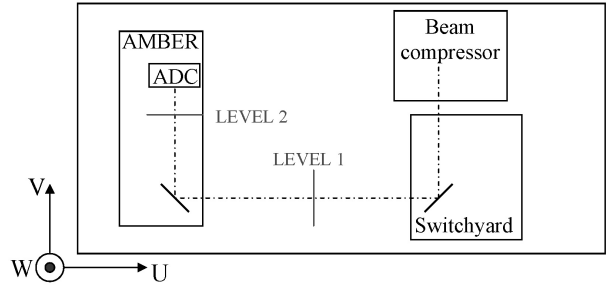}
\end{center}
\caption{Reference frame in the Paranal laboratory. W: upward axis.}
\label{vv}
\end{figure}

The direction of the dispersion from low to high wavelengths is described by the atmospheric dispersion angle $v$ which depends on the type of telescope (UT/AT) and its position in the case of the ATs. The angle $v$ is calculated by the ESO Software in the (V,W) plane of the reference frame of the Paranal laboratory (assuming no further reflection after the VLTI beam switchyard) (see Fig.\,\ref{vv}). 
\citet{ESO} give the following expressions for $v$:\newline
$v = -a + A -12.98$ for the UTs,\newline
$v = a - A + 12.98$ for an AT located north of the delay line tunnel,\newline
$v = a - A - 167.02$ for an AT located south of the delay line tunnel,\newline
where $a = 90^\circ-z$ is the altitude angle and $A$ is the azimuth angle.
This means that an angle $v$ of 0$^\circ$ corresponds to the $+V$ direction and an angle of +90$^\circ$ corresponds to the $+W$ direction. 

\begin{figure}
\begin{center}
\includegraphics[width=84mm]{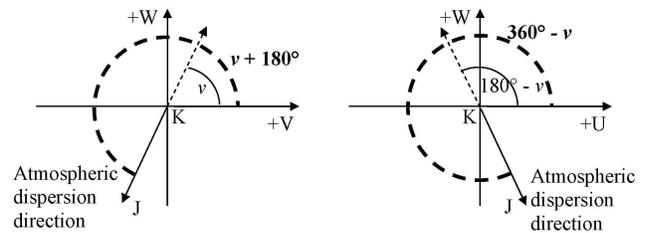}
\end{center}
\caption{Atmospheric dispersion direction (from K to J), after the switchyard (level 1) and after the feeding optics of AMBER or at the ADC level (level 2).}
\label{vdir}
\end{figure}

Figure\,\ref{vdir} gives the atmospheric dispersion direction (from the high to the low wavelengths), after the switchyard (level 1) and after the feeding optics of AMBER or at the ADC level (level 2).

\begin{figure}
\begin{center}
\includegraphics[width=84mm]{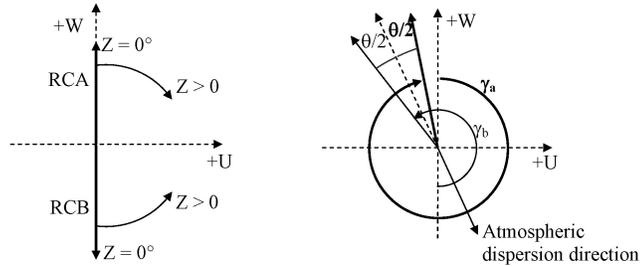}
\end{center}
\caption{Rotation of the two sets of prisms RCA and RCB by some angles $\gamma_{a}$ and $\gamma_{b}$ to correct for the atmospheric dispersion at $z~\neq~0^\circ$.}
\label{rc}
\end{figure}

Let us call RCA and RCB the two sets of 3 prisms. At $z~=~0^\circ$, RCA and RCB are positioned in opposite directions as shown in Figure\,\ref{adc}. During the observations, as $z$ increases, RCA and RCB rotate together (see Fig.\,\ref{rc}) in opposite directions (for mechanical reasons), with two angles $\gamma_{a}$ and $\gamma_{b}$ depending on $v$:\newline
$\gamma_{a} = 270^\circ + v + \theta/2$,\newline
$\gamma_{b} = 270^\circ - v + \theta/2$,\newline
where $\theta = 2\;\arccos(\alpha/2 \Delta D)$, $\alpha = R(\lambda)-R(2.2 \mu m)$ being the differential atmospheric dispersion and 2$\Delta D$ the maximum differential dispersion (maximum possible value for $\alpha$) that each set of prisms can correct. $\Delta D$ is 160 mas  for the UTs and 0.7$\arcsec$ for the ATs between 1.0~$\mu$m and 2.2~$\mu$m. These values are defined by the magnification between the telescopes and the size of the beam in the focal laboratory (see Appendix~B).

\subsection{Example}\label{52}
Let us consider the following parameters:\newline
- zenith angle: $z = 60^\circ$ (maximal direction required for the AMBER observations).\newline
- Altitude: $a = 90^\circ-z =30^\circ$.\newline
- Azimuth angle: $A = 22^\circ$.\newline
The $z$ value gives the differential atmospheric dispersion (from Eq.\,\ref{angleR}) between 1.0~$\mu$m and 2.2~$\mu$m:
$\alpha~=~R(1.0 \mu m)-R(2.2 \mu m)~=~320$~mas.\newline

Table\,\ref{rot} gives the atmospheric dispersion angle $v$ and the angles describing the ADC rotation from its initial position (at $z = 0^\circ$) for the UTs and the ATs.
\begin{table}
	\caption{Atmospheric dispersion angle $v$ and the angles describing the ADC rotation for the UTs and the ATs. $z$~=~60$^\circ$; $A$~=~22$^\circ$; $\alpha$~=~R(1.0~$\mu$m)~-~R(2.2~$\mu$m)~=~320~mas.}
	\begin{center}
		\begin{tabular}{c c c c c} \hline
\scriptsize &\scriptsize $v$  &\scriptsize $\theta /2$	&\scriptsize	$\gamma_{a}$ &\scriptsize	$\gamma_{b}$\\\hline
\scriptsize UT  &\scriptsize 338.53$^\circ$  &\scriptsize 1$^\circ$	&\scriptsize 249.53$^\circ$	&\scriptsize 292.47$^\circ$	\\
\scriptsize AT north &\scriptsize 21.47$^\circ$ &\scriptsize 77$^\circ$ &\scriptsize 8.47$^\circ$&\scriptsize	325.53$^\circ$ \\
\scriptsize AT south &\scriptsize 201.47$^\circ$ &\scriptsize 77$^\circ$&\scriptsize 188.47$^\circ$&\scriptsize 145.53$^\circ$ \\\hline
		\end{tabular}
		\end{center}
	\label{rot}
\end{table}

Next Section gives the results of the ADC operation in laboratory, then on sky. 

\section{Validation of the ADC on the VLTI}
\label{6}
This Section gives the results of the tests performed in laboratory on the ADC and concludes with the validation of its performance on sky. The tests in laboratory, as the alignment procedure, are described in the Appendix~B.

\subsection{Results of the tests performed in laboratory}
\label{61}
The alignment of the three ADCs in laboratory resulted in the measurement of a maximum image shift in K-band of about 0.7~$\mu$m at the fiber head level, leading to less than 10\% of coupling loss in J-band and 5\% in H-band.
Measurements in J-band allowed to deduce the maximum value of $\Delta D$ between 1.0~$\mu$m and 2.2~$\mu$m: 160~mas for the UTs and 0.7$\arcsec$ for the ATs. The deducted maximum value of $\alpha$ is equal to 320~mas for the UTs and 1.4$\arcsec$ for the ATs between 1.0~$\mu$m and 2.2~$\mu$m (Eq.\,\ref{angleR}). This confirms the feasibility of the dispersion correction for observations at $z$ up to 60$^\circ$ (Sect.\,\ref{51} and Sect.\,\ref{52}).

\subsection{Validation on sky}
\label{62}

The purpose of the sky observations was to verify the rotation of the ADC depending on the rotation of the pupil in the laboratory. We have done this by observing the $\alpha$~Cir star (A7Vp spectral type, m$_V$~=~3.2) with one UT at more than
30$^\circ$ from zenith in order to be sensitive enough to the atmospheric refraction but with a not too high airmass so that a good AO correction in J-band could still be achieved. 
The seeing of the night was 0.65$\arcsec$ on average. The altitude and azimuth angles were $a~=~52.914^\circ$
($z~\approx~37^\circ$) and $A~=~207.978^\circ$. The differential atmospheric dispersion between 1.0~$\mu$m and 2.2~$\mu$m was then $\alpha~=~142.6$~mas (Eq.\,\ref{angleR}). These inputs gave two theoretical values for the RCA and RCB angles, 
$\gamma_a~=~115.61^\circ$ and $\gamma_b~=~191.44^\circ$ (see the illustration on Fig.\,\ref{rc}, right).

\begin{figure}
\begin{center}
\includegraphics[width=75mm]{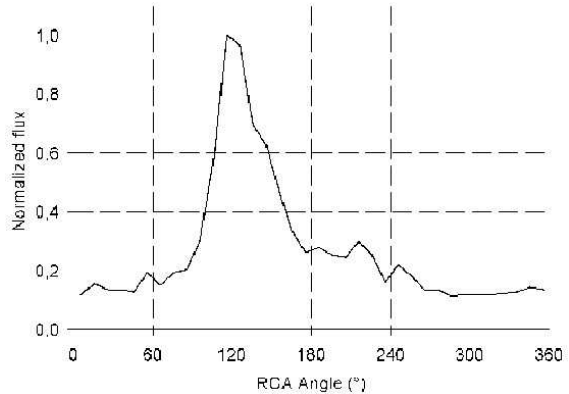}
\end{center}
\caption{Normalized flux in the image recorded at the focal point of AMBER as a function of the $\gamma_a$ angle while keeping the differential angle $\theta$ to its nominal value. Maximal value reached for the expected angle $\gamma_a$ of 116$^\circ$.}
\label{c1_etienne}
\end{figure}

A script was executed that permitted simultaneous rotation of the two sets RCA and RCB while keeping the differential angle $\theta$ at its nominal value. This is equivalent to varying the direction of the dispersion correction direction. Figure\,\ref{c1_etienne} shows the normalized flux in the image recorded at the focal point of AMBER as a function
of $\gamma_a$, incremented every 10$^\circ$. The flux reached its maximal value for the expected angle $\gamma_a$ of 116$^\circ$.

\begin{figure}
\begin{center}
\includegraphics[width=75mm]{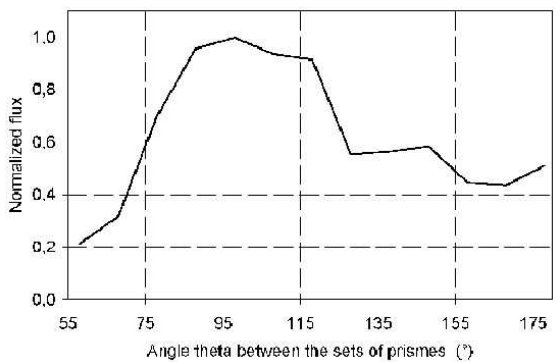}
\end{center}
\caption{Normalized flux in the image recorded at the focal point of AMBER as a function of the angle $\theta$ while maintaining the dispersion correction direction to its nominal value. Maximal value reached for the expected angle $\theta$ of 97$^\circ$.}
\label{c2_etienne}
\end{figure}

Having confirmed the expected dispersion correction direction, we studied the impact of the angle $\theta$. 
The new altitude and azimuth angles while tracking the star were $a~=~41.502^\circ$ ($z\approx48.5^\circ$) and
$A~=~263.091^\circ$, while external atmospheric conditions were similar to the ones when changing the direction of the dispersion correction.  
Under these conditions, the expected angles are $\alpha~= 213.2~$mas, $\theta~=~96.46^\circ$, $\gamma_a~=~166.84^\circ$
and $\gamma_b~=~109.62^\circ$. Figure\,\ref{c2_etienne} shows the normalized flux in the image recorded at the focal point of AMBER as a function of the angle $\theta$. The flux reaches its maximal value for the expected angle $\theta$ of 97$^\circ$. This represents a gain of 50\% compared with the flux recorded when the sets of prisms are 180$^\circ$ apart from each other, so when the system does not correct the dispersion.

These two tests on sky validated the expected direction for the dispersion correction, and the angle calculated for the ADC. 
They showed the flux gain in J-band at this nominal direction by correctly positioning the ADC.

\section{Conclusion}
The need to correct the atmospheric refraction in a single-mode instrument is now assessed for observations with UTs in J- and H-bands. With no correction, numerical simulations showed that the optical fiber coupling efficiency drops by more than 50\% for zenith angles higher than 40$^\circ$ in the totality of J-band, drops by more than 20\% of its initial value for zenith angles above 50$^\circ$ in the totality of H-band, and is almost null at $z$~=~60$^\circ$ at extreme wavelengths. This is obtained while observing in the individual spectral bandwidths. These results are different if the observations are simultaneously performed in the three spectral bands, while the VLTI is optimized at one wavelength. 

Taking into account the atmospheric turbulence, it can be noted that the atmospheric refraction is predominant for $z$ larger than 30$^\circ$ in J-band, and 50$^\circ$ in H-band. It was also shown that the refraction correction is perfectly justified in spite of the Strehl degradation with $z$.

The study performed for the AMBER near infrared instrument of the VLTI resulted in an ADC system made of 2 sets of 3 prisms rotating with respect to each other. This system allows to reach a global fiber coupling of 48\% in J-band and 45\% in H-band, including the maximum coupling of the fibers presently installed on the instrument, the degradation error due to the imperfect manufacturing and assembly of the ADC, the degradation error due to element positioning and surface quality, and the additional error due to an unperfect light injection in the fiber. This study confirms also the acceptability of the uncorrected dispersion in K-band for which the global fiber coupling remains acceptable (in the [35 +/- 4] \% level).

First observations with the ADC showed a flux improvement by a factor of 2 for the expected ADC functioning.
Up to now, most of the observations on AMBER are carried out in K-band and in H-band, spectral bandwidths for which the instrument is optimized today. The updated values on limiting magnitudes are K~=~H~=~7 in low-resolution and medium-resolution ($R$~=~1500), and K~=~H~=~6 in high-resolution ($R$~=~12000) with the UTs (seeing~$\leq$~0.8$\arcsec$).
In a near future, AMBER will benefit of the real advantage brought by the ADC. 


\section*{Acknowledgments}
The AMBER project
has been funded by the French Centre National de la Recherche Scientifique (CNRS), the Max Planck Institute f\"{u}r Radioastronomie (MPIfR) in Bonn, the Osservatorio Astrofisico di Arcetri (OAA) in Firenze, the French Region Provence Alpes C\^{o}te d'Azur and  the European Southern Observatory (ESO). The CNRS funding has been made through the Institut National des Sciences de l'Univers (INSU) and its Programmes Nationaux (ASHRA, PNPS, PNP).

\appendix
\section{Tolerance analysis for the optical components and mechanical compensation}
To follow the paragraph let us call the First PRism (BK7) with the acronym FPR, the Second PRism (F2) with the acronym SPR and the Third PRism (SF14) with the acronym TPR.

\subsection{Prism angles}
\begin{figure}
\begin{center}
\includegraphics[width=84mm]{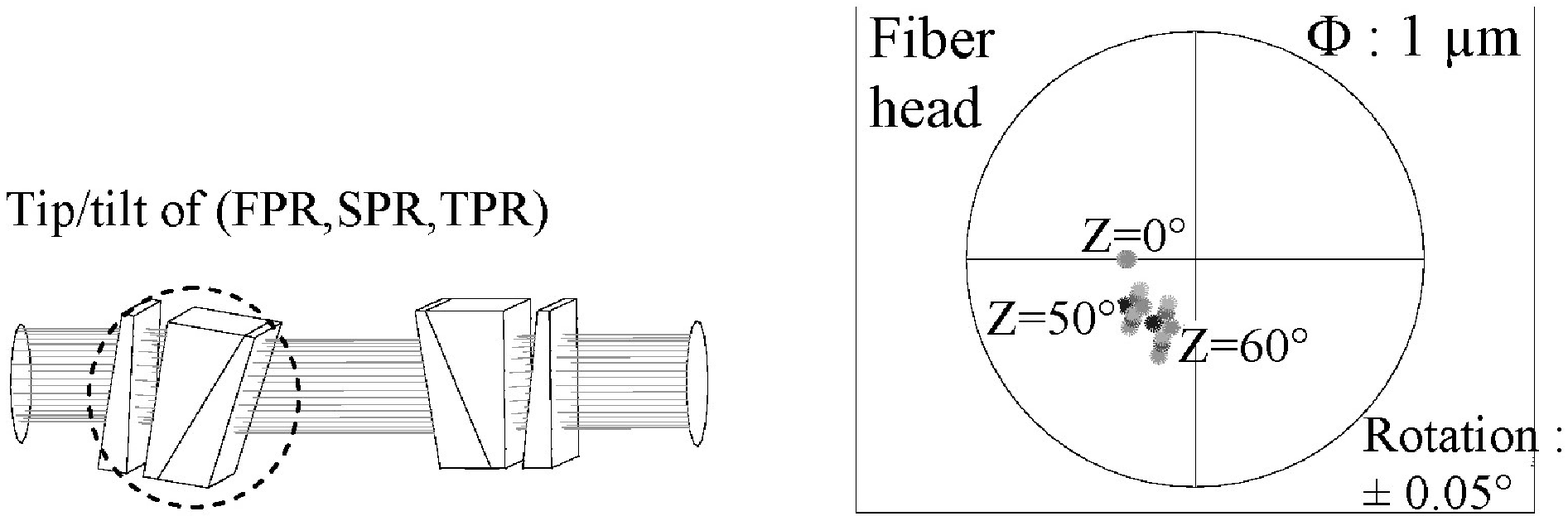}
\end{center}
\caption{Left: tip-tilt of (FPR, SPR, TPR) of 8$^\circ$ to compensate for the errors on the prism angles. Right: effect on the spectrum generated by an error on this adjustment of $\pm0.05^\circ$ (larger than the requirement). The spots represent the dispersion of the beam from 1.1~$\mu$m to 1.82~$\mu$m.}
\label{prism1}
\end{figure}

The prism angle manufacturing error of $\pm$0.01$^\circ$ is compensated by a tip-tilt of one of the 3-prism system (FPR, SPR, TPR) of up to 8$^\circ$ (see Fig.\,\ref{prism1}). The sensitivity of this compensation is $\pm$36$\arcsec$. 
 
\subsection{Cemented angles}
\begin{figure}
\begin{center}
\includegraphics[width=84mm]{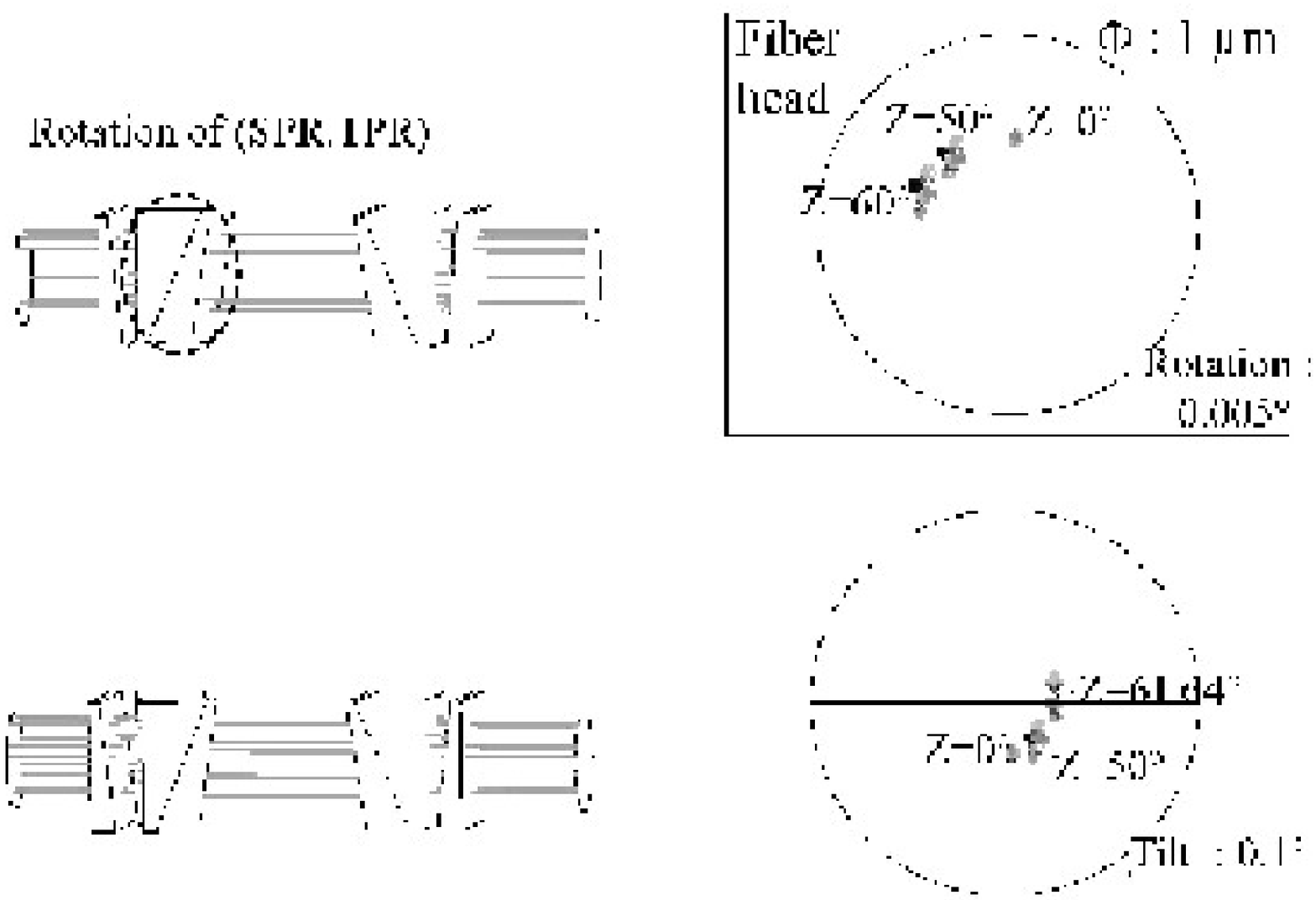}
\end{center}
\caption{Top: rotation of (SPR,TPR) and results on the spectrum position at the fiber heads for different $z$. Bottom: effects of a 0.1$^\circ$ inclination of the rotation axis of (SPR,TPR).}
\label{prism2_3}
\end{figure}

The two prisms SPR and TPR are cemented together with a $\pm0.5^\circ$ accuracy. The generated error is compensated by a rotation of up to 5$^\circ$ of the 2-prism assembly (SPR, TPR) respective to the first prism (FPR) (see Fig.\,\ref{prism2_3}). The sensitivity of this rotation is $\pm$18$\arcsec$. During this motion, an inclination of $\pm0.1^\circ$ of the rotation axis of the assembly (SPR, TPR) is acceptable.

Taking into account the errors on the prism angles, due to the cementing operation, the compensation leads to an image decentering on the fiber head of less than 0.5 $\mu$m.

\subsection{Positioning errors during the correction of the refraction}
\begin{figure}
\begin{center}
\includegraphics[width=84mm]{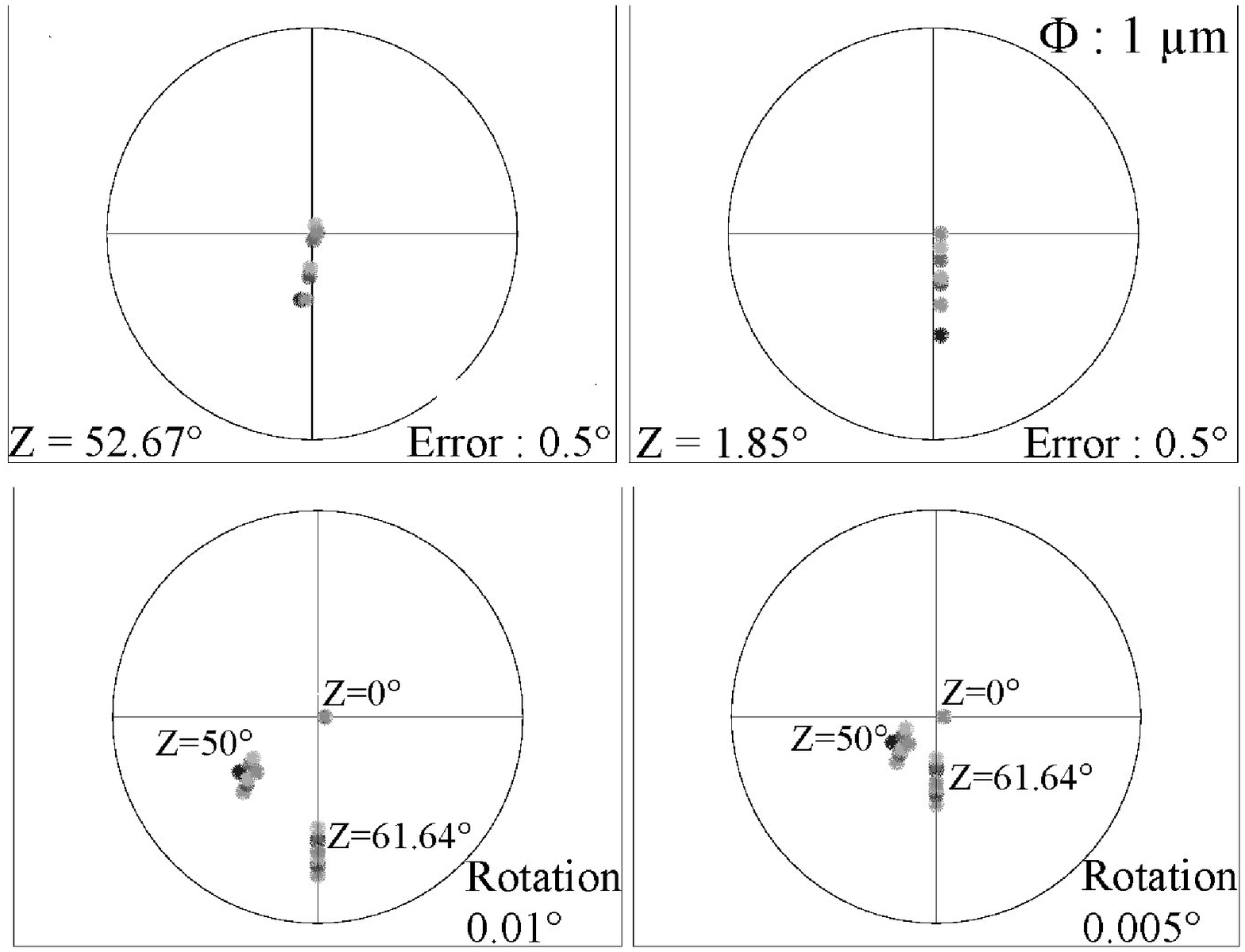}
\end{center}
\caption{Top: effect of a 0.5$^\circ$ error on the rotation of the ADC at z=52.67$^\circ$ (left) and z=1.85$^\circ$ (right). This error includes a mechanical sensitivity 
and an imprecision in the order of a few \% on the atmospheric parameters (air pressure, temperature, density, refraction index). Bottom: effect of  an inclination of the ADC rotation axis: 0.01$^\circ$ (left) and 0.005$^\circ$ (right)}
\label{prism4}
\end{figure}

During an observation of objects located from 0$^\circ$ up to 60$^\circ$ from zenith, the correction of the refraction is performed by the axial rotation of the two sets of 3 prisms. The two systems are positioned after each exposure with a $\pm0.2^\circ$ accuracy. Typical time between corrections is a few minutes. The duration of positioning is about 1\,s. The rotation is performed with a 0.5$^\circ$ accuracy (see Fig.\,\ref{prism4}). The inclination (tip/tilt) of the ADC rotation axis relative to the optical axis is controlled to be much less than 1.8$\arcmin$ with a $\pm$20$\arcsec$ sensitivity.

\section{Tests in laboratory}
\begin{figure}
\begin{center}
\includegraphics[width=84mm]{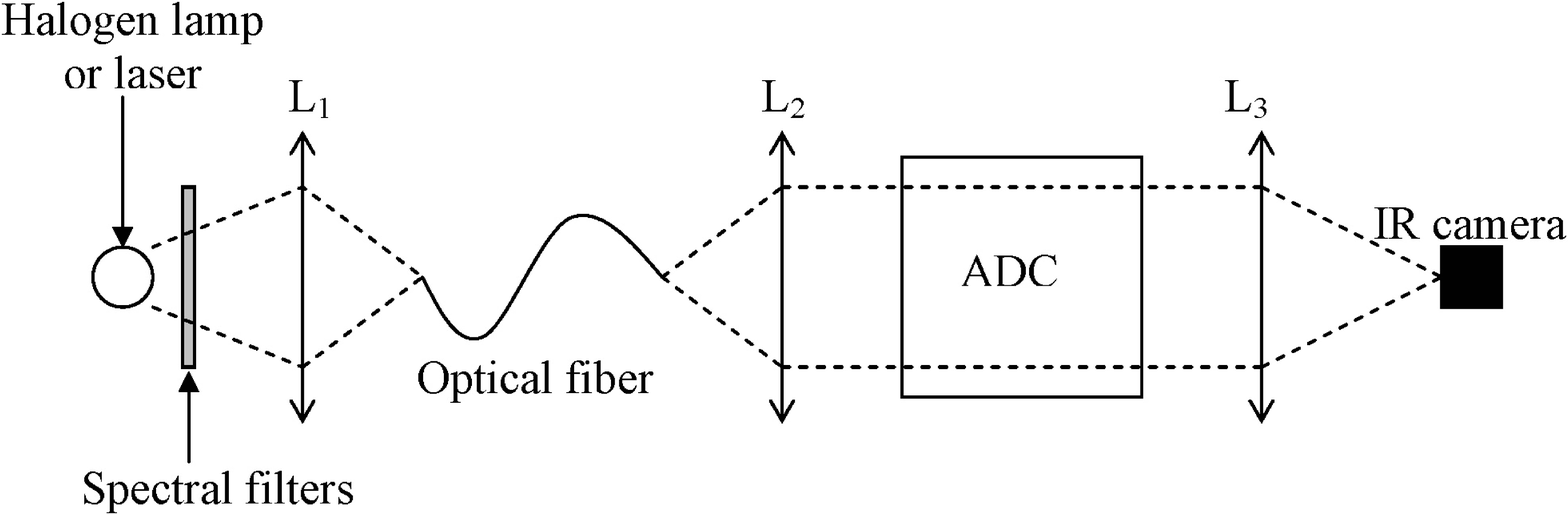}
\end{center}
\caption{Scheme of the optical set-up used for the ADC tests in laboratory.}
\label{test}
\end{figure}

The ADC tests in laboratory were performed at the Observatoire de la C\^ote d'Azur (OCA) in Nice. The temperature inside the room underwent fluctuations between 17$^\circ$C and 21$^\circ$C. The experiments were realized on a Newport optical table (RS4000). 
Two light sources were used (see Fig.\,\ref{test}):\newline
-	a laser source for the rough alignment.\newline
-	a white halogen lamp with spectral filters for the fine adjustment and the measurements,\newline
The characteristics of the spectral filters were:\newline
-	J-filter: $\lambda$ = 1.2954 $\mu$m, $\Delta\lambda$ = 10.2 nm, \newline
-	K-filter: $\lambda$ = 2.2055 $\mu$m, $\Delta\lambda$= 63 nm. \newline
The light is transported by an optical fiber at the focal point of a collimating optics (L$_2$). This optics, equipped with a mask, produces a collimated beam with a diameter of 18 mm which feeds the ADC.
The control system is composed of a lens (L$_3$) of focal length 780~mm, and of an IR camera (JADE SWIR from CEDIP) with a pixel size of 30x30~$\mu$m.\newline
One pixel of the camera corresponds to 1.85 $\mu$m at the fiber head level. 
Each image position was estimated from the mean value of several measurements. 
The corresponding image centroid estimation accuracy (0.025~pixel) allows the extrapolation of the image position on a fiber head with an accuracy of 0.05~$\mu$m.

A specific mechanical device was built in order to perform the first step of the ADC alignment described in the next Section.


\subsection{Internal alignment of the sets of prisms}
The alignment consists in adjusting the tip-tilt angle of the overall set of prisms in their mounts and the relative axial rotation angle between the BK7 prism (FPR) and the two others (SPR, TPR).
A tilt of the 3-prisms assembly or an axial rotation of the FPR relative to the system (SPR, TPR) translates the image in the focal plane of the optical set-up: 0.1$^\circ$ tilt generates 5.4 $\mu$m translation at the fiber head level; 0.01$^\circ$ rotation generates a 7.3 $\mu$m tangential translation.

\begin{figure}
\begin{center}
\includegraphics[width=40mm]{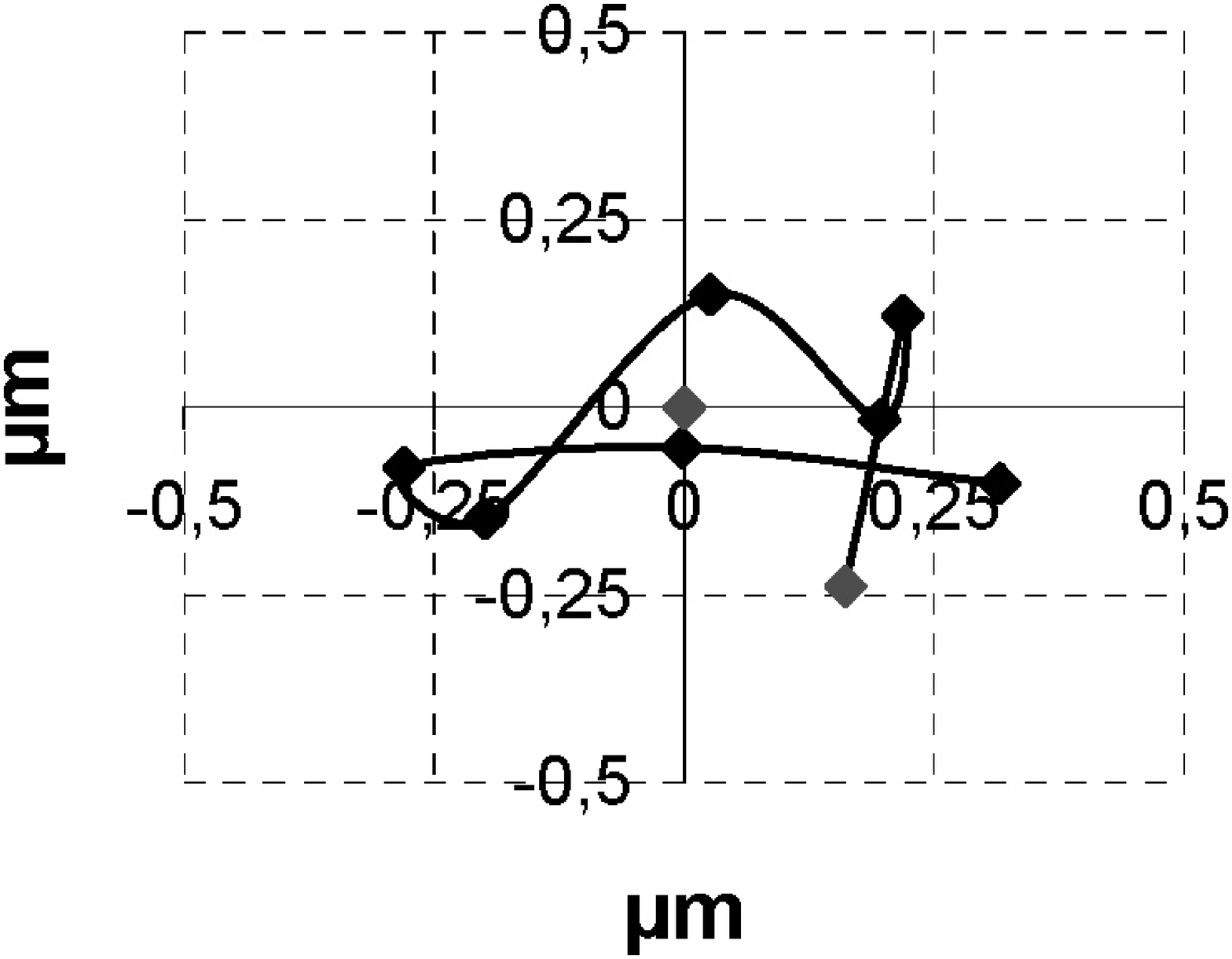}
\end{center}
\caption{Image shift in micrometers at the fiber head level for a rotation of 45$^\circ$ of the 3-prisms set.}
\label{test1}
\end{figure}

The alignment is performed with the K-filter and the specific mechanical device.
Figure\,\ref{test1} displays the result of the internal adjustment of one set of 3 prisms. Each point corresponds to a rotation of 45$^\circ$ of the set (FPR, SPR, TPR). The scale represents the image shift in micrometers at the fiber head level. It is about 0.25 $\mu$m, ensuring a minimal fiber coupling loss. The alignment of the 6 other sets of 3 prisms resulted in the same performance.

\subsection{Overall adjustment}
\begin{figure}
\begin{center}
\includegraphics[width=40mm]{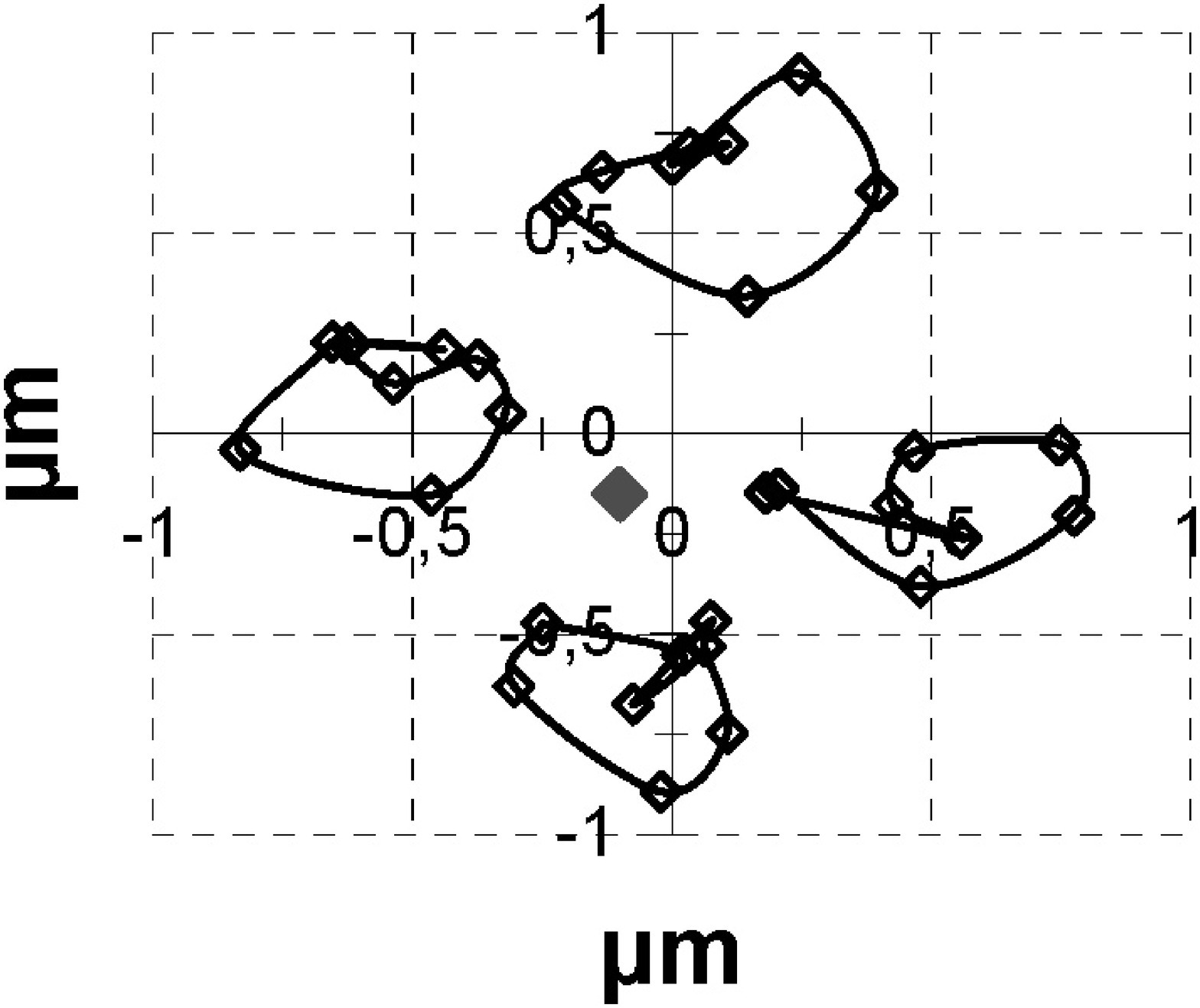}
\includegraphics[width=40mm]{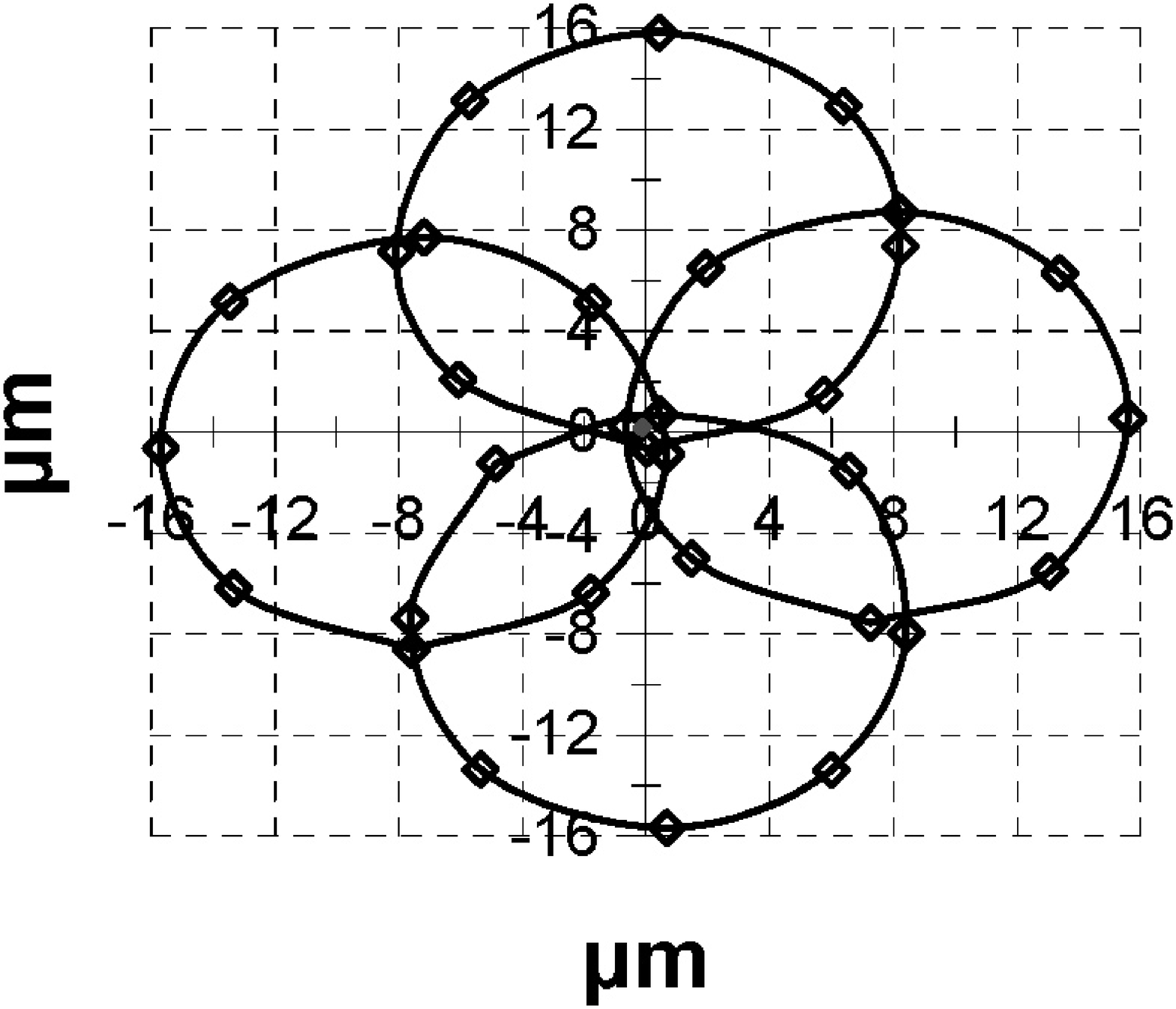}
\end{center}
\caption{Image shift at the fiber head level recorded during the rotation of a FPR prism relative to the system (SPR, TPR) for 4 angular positions of one set (FPR, SPR, TPR) relative to the other set, in K-band (left) and in J-band (right).}
\label{test2}
\end{figure}

The optics and their individual mounts are now mounted in the final overall ADC struture. The alignment is performed using the K-filter. It consists in adjusting the rotation axis of the two sets of 3 prisms relatively to the optical axis.
Figure\,\ref{test2} (left) displays the image shift at the fiber head level recorded during the rotation of a FPR prism relative to the system (SPR, TPR) for 4 angular positions of one set (FPR, SPR, TPR) relative to the other set: 0$^\circ$ (bottom), 90$^\circ$ (right), 180$^\circ$ (top), 270$^\circ$ (left). 
The image describes a circle during the rotation. The circles are all tangent to the non-dispersed point (position for $z = 0^\circ$). The final ADC mounting is correct if the non-dispersed points obtained at the four positions of the two triplet assemblies are superimposed within 1~$\mu$m.
If not, the ADC should be dismounted and the optical adjustment of the individual 3-prisms assemblies performed again as in the previous Section.

For the three ADCs, the maximum image shift in K-band of about 0.7 $\mu$m at the fiber head level leads to less than 10\% of coupling loss in J-band and 5\% in H-band.

\subsection{Maximal correction of the ADC}
The alignment is verified using the J spectral filter (see Fig.\,\ref{test2}, right). The diameter value of 16 $\mu$m of the circles described by the images in the fiber head plane allows to calculate the maximum differential dispersion (maximum possible value for $\alpha$) that the ADC can correct. This diameter value corresponds to an angle of about 70$\arcsec$ for the wavefront received by the fiber injection optics with a $\approx$48-mm focal length, the beam diameter being 18~mm. The maximum value $\Delta D$ is then 160~mas for the UTs and 0.7$\arcsec$ for the ATs between 1.0~$\mu$m and 2.2~$\mu$m. The deduced maximum value of $\alpha$ is equal to 320~mas for the UTs and 1.4~$\arcsec$ for the ATs between 1.0 $\mu$m and 2.2 $\mu$m (Eq.\,\ref{angleR}). This confirms the feasibility of the dispersion correction for observations at $z$ up to 60$^\circ$ (Sect.\,\ref{51} and Sect.\,\ref{52}).

\bsp

\label{lastpage}

\end{document}